\documentclass[10pt,twocolumn,letterpaper]{article}

\usepackage[pagenumbers]{iccv} 

\usepackage[acronym, shortcuts]{glossaries}
\usepackage{amsmath,amssymb,amsfonts}
\usepackage{algorithmic}
\usepackage{textcomp}
\usepackage{xcolor}
\usepackage{glossaries}
\usepackage{tabularx,ragged2e}
\usepackage{booktabs}
\usepackage{subfloat}
\usepackage{subcaption}
\usepackage{array,multirow,graphicx}
\usepackage{float}
\usepackage{mathtools}
\usepackage{footmisc}
\usepackage{siunitx}
\usepackage{colortbl} 
\usepackage{transparent}

\definecolor{blue}{rgb}{0,0,0}

\newacronym{emd}{EMD}{Earth's Mover Distance}
\newacronym{auc}{AUC}{Area Under the Curve}
\newacronym{aucd}{$\text{AUC}_{\text{d}}$}{Area Under the ROC Curve}
\newacronym{aucl}{$\text{AUC}_{\text{l}}$}{Area Under the Curve}
\newacronym{ba}{BA}{Balanced Accuracy}
\newacronym{bad}{$\text{BA}_{\text{d}}$}{Balanced Accuracy}
\newacronym{bal}{$\text{BA}_{\text{l}}$}{Balanced Accuracy}
\newacronym{ba_th0}{BA@0}{BA with 0 threshold}
\newacronym{ba05}{BA@0.5}{Balanced accuracy with 0.5 threshold}
\newacronym{tpr}{TPR}{True Positive Rate}
\newacronym{fnr}{FNR}{False Negative Rate}
\newacronym{fpr}{FPR}{False Positive Rate}
\newacronym{roc}{ROC}{Receiver Operating Characteristic}
\newacronym{bpp}{BPP}{Bits Per Pixel}
\newacronym{qf}{QF}{Quality Factor}
\newacronym{mss}{MS-SSIM}{Multi-scale SSIM}
\newacronym{vif}{VIF}{Visual Information Fidelity}
\newacronym{cnn}{CNN}{Convolutional Neural Network}
\newacronym{gan}{GAN}{Generative Adversarial Network}
\newacronym{vit}{ViT}{Vision Transformer}
\newacronym{svm}{SVM}{Support Vector Machine}
\newacronym{nn}{NN}{Neural Network}
\newacronym{nic}{NIC}{Neural Image Compression}
\newacronym{ifosn}{IFOSN}{ImageForensicsOSN}
\newacronym{npr}{NPR}{Neighboring Pixel Relationships}

\definecolor{iccvblue}{rgb}{0.21,0.49,0.74}
\usepackage[pagebackref,breaklinks,colorlinks,allcolors=iccvblue]{hyperref}

\def\paperID{13844} 
\def\confName{ICCV}
\def\confYear{2025}

\title{Is JPEG AI going to change image forensics?}

\author{Edoardo Daniele Cannas$^{\dagger}$,
Sara Mandelli$^{\dagger}$,
Nataša Popović$^{\dagger}$,
Ayman Alkhateeb$^{\star}$\\
Alessandro Gnutti$^{\star}$,
Paolo Bestagini$^{\dagger}$,
Stefano Tubaro$^{\dagger}$\\
$^{\dagger}$ \textit{Politecnico di Milano},
$^{\star}$ \textit{Università degli Studi di Brescia}
}

\begin{document}
\maketitle

\sloppy
\begin{abstract}
In this paper, we investigate the counter-forensic effects of the new JPEG AI standard based on neural image compression, focusing on two critical areas: deepfake image detection and image splicing localization. Neural image compression leverages advanced neural network algorithms to achieve higher compression rates while maintaining image quality. However, 
it introduces artifacts that closely resemble those generated by image synthesis techniques and image splicing pipelines, complicating the work of researchers when discriminating pristine from manipulated content.
We comprehensively analyze JPEG AI’s counter-forensic effects through extensive experiments on several state-of-the-art detectors and datasets. Our results demonstrate a reduction in the performance of leading forensic detectors when analyzing content processed through JPEG AI. 
By exposing the vulnerabilities of the available forensic tools 
we aim to raise the urgent need for multimedia forensics researchers to include JPEG AI images in their experimental setups and develop robust forensic techniques to distinguish between neural compression artifacts and actual manipulations.
\end{abstract}   
\vspace{-10pt}
\section{Introduction}
\label{sec:intro}

\begin{figure}[t!]
\centering
        \includegraphics[width=\columnwidth]{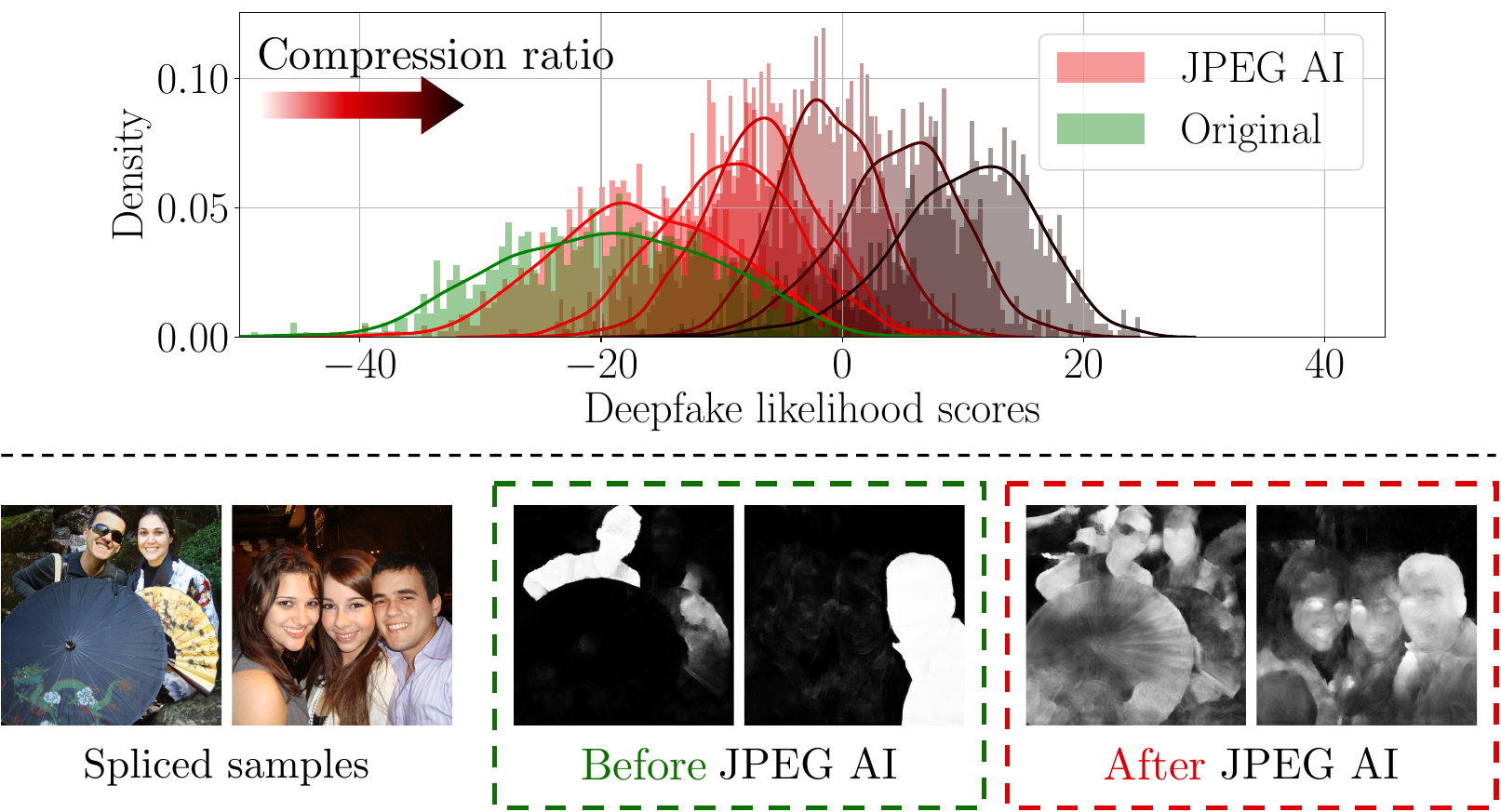}
        \caption{Main findings of our experimental campaign on the counter-forensic effect of JPEG AI. On the top row, we report the deepfake detection scores of pristine images and their JPEG AI-compressed versions as the compression ratio increases (darker red tones), proving that pristine images compressed with this new standard can be mistaken for deepfakes. On the bottom row, we show that, after JPEG AI compression, localization maps produced by state-of-the-art algorithms can no longer distinguish genuine from manipulated content.}
\label{fig:teaser}
\vspace{-15pt}
\end{figure}

Neural image compression is an emerging technology that leverages deep learning techniques to compress images more efficiently than traditional methods~\cite{toderici2017, minnen2018, mentzer2020, minnen2020channel, hoogeboom2023high, yang2023}. This approach is needed to address the growing demand for high-quality image storage and transmission in an era of proliferating visual content. Neural image compression is gaining traction due to its ability to achieve superior compression rates while maintaining visual quality, making it an attractive solution for various applications~\cite{Lieberman2023}.

JPEG AI is a new standard in image compression developed to harness \glspl{nn}' power~\cite{JPEGAI_whitepaper, ascenso2023jpeg}. 
Published as an International Standard on February 2025~\cite{jpegai_pub},
JPEG AI aims not only to achieve superior rate-distortion performance for enhanced visual quality but also to enable faster coding speeds and provide the flexibility to optimize human viewing and machine analysis~\cite{ascenso2023jpeg, JPEGAI_usecase}.

However, the introduction of JPEG AI poses new challenges for image forensics~\cite{hofer2024}. JPEG AI, as a novel standard, introduces unique forensic footprints that require dedicated study to understand and exploit~\cite{Bergmann2023, tsereh2024}. Additionally, existing literature indicates that neural image compression schemes
leave distinctive upsampling traces similar to those found in images generated by \glspl{nn}, such as deepfakes~\cite{Bergmann2024}.
Deepfake image detection is particularly active, with numerous contributions highlighting that \gls{nn}-based image generation techniques leave artifacts due to upsampling operators~\cite{wang2020, durall2020upconv, gragnaniello2021, Corvi_2023_CVPR, Mandelli2024, Tan_2024_CVPR}. 
Furthermore, it is well-documented that locally manipulated (i.e., spliced)
images may exhibit artifacts related to upsampling~\cite{Stamm2013}, too.

All these elements raise a significant issue: the potential for detectors to erroneously classify genuine content as malicious, either as deepfakes or locally manipulated images, 
due to the presence of JPEG AI compression traces. 
While the forensic literature is currently analyzing the implications of adopting JPEG AI, there are no general studies on its counter-forensic effects to our knowledge. 
Indeed, previous investigations have either focused on analyzing the reconstruction artifacts left by neural compression methods~\cite{hofer2024} or the traces left by neural compression techniques that are practical for other forensic purposes~\cite{Bergmann2023, Bergmann2024, tsereh2024}. Other contributions examined how image splicing detectors exploiting standard JPEG compression artifacts lower their performances whenever a user relies on neural compression~\cite{Berthet2022} to save images. 

This paper investigates whether state-of-the-art algorithms designed for deepfake image detection and image splicing localization can mistakenly identify 
JPEG AI-compressed images as malicious content. 
To achieve our goal, we consider several state-of-the-art detectors from both tasks.
We first collect their test datasets and compress them using the official JPEG AI Reference Software~\cite{jpegai_vm}. Then, we evaluate if the detectors present a pronounced tendency to misclassify JPEG AI compressed samples as malicious, varying the compression ratio. 
We also examine the detector's performance on images encoded with traditional JPEG compression, confirming that artifacts introduced by JPEG AI 
make detectors behave in a sort of opposite way to conventional compression methods.
Finally, we preliminary show that incorporating JPEG AI images into the training pipeline of detectors may help mitigate this phenomenon, and also experiment with double JPEG AI compression.

The main findings of our experimental campaign are reported in~\cref{fig:teaser}. We prove that deepfake image detectors tend to classify 
images compressed with JPEG AI as synthetic samples (top row), while image splicing localization techniques risk being fooled by JPEG AI, as many missed detections and false alarms may arise in the estimated tampering masks (bottom row). 
All these effects prove to be stronger as the JPEG AI compression ratio increases. 


In summary, the main contributions of our paper are: i) we are the first to investigate deepfake image detection and image splicing localization tasks focusing on the official implementation of JPEG AI;
ii) despite the exceptional visual quality of JPEG AI images, we demonstrate that state-of-the-art detectors are misled by their interpolation artifacts; 
iii) we reveal that this effect is incomparable with standard compression formats like JPEG; 
iv) we introduce JPEG AI samples into training to reduce the effects of this phenomenon; v) we investigate the counter-forensic effects of double JPEG AI compression.
\section{JPEG AI background}
\label{sec:background}

\begin{figure}[t]
\centering
        \includegraphics[width=\columnwidth]{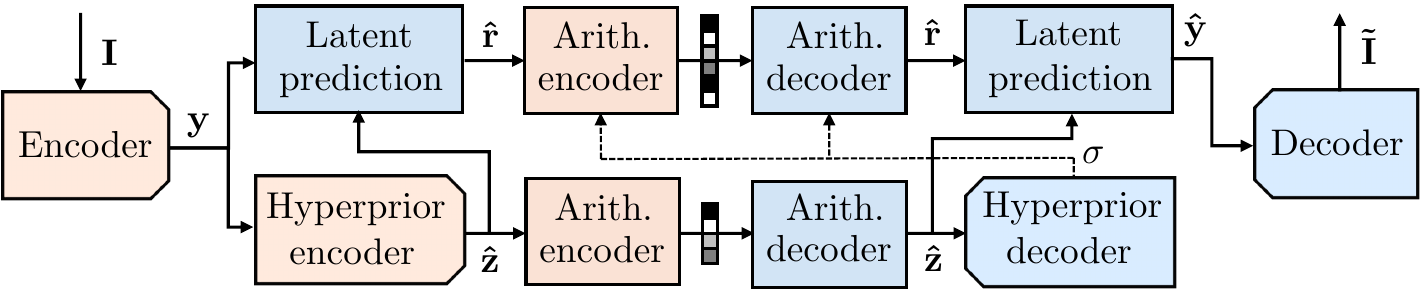}
        \caption{JPEG AI framework.}
\label{fig:architecture}
\vspace{-15pt}
\end{figure}

\noindent \textbf{General framework. } 
In February 2025, the JPEG Standardization Committee published JPEG AI as the first international image coding standard based on end-to-end learning~\cite{jpegai_pub}. 
The general architecture of JPEG AI is depicted in Fig.~\ref{fig:architecture}~\cite{JPEGAI_overview}.
The basic functioning follows the same of many other neural image compression algorithms: it first transforms an image into a latent tensor, which is then compressed and transmitted to the receiver, allowing the decoder to either reconstruct the image or carry out additional operations directly in the compressed domain exploiting the latent tensor~\cite{ascenso2023jpeg}. 
The entire JPEG AI codec is trained end-to-end by minimizing a rate-distortion loss function. By adjusting the weight of the rate term, different models can be trained to achieve variable-rate compression, corresponding to various \gls{bpp}. 
We refer the reader to the supplemental material, \cref{sec:supp:jpeagi}, for more details on each architectural module. 

While many neural image compression techniques exist~\cite{Mishra2022} (please refer to the supplemental material for more discussion), 
in our analysis we specifically focus on JPEG AI given its forthcoming impact. Indeed, JPEG AI presents 
interesting characteristics that are likely to spread its adoption among users, including support for a wide array of devices with different characteristics~\cite{jpegai_pub} and the direct involvement of technology companies~\cite{jpegai_draft}. 

It is noteworthy to mention that JPEG AI, as a standardized format, enforces strict compliance requirements on manufacturers, limiting customization options within the compression pipeline. In particular, the blue modules in Fig.~\ref{fig:architecture}, among which we can notice the decoder, are formally normative.
This adherence to a fixed architecture suggests that images decoded with JPEG AI, even when produced by different manufacturers, will likely display similar characteristics and forensic artifacts. 


\begin{figure*}[t]
\centering
        \includegraphics[width=\textwidth]{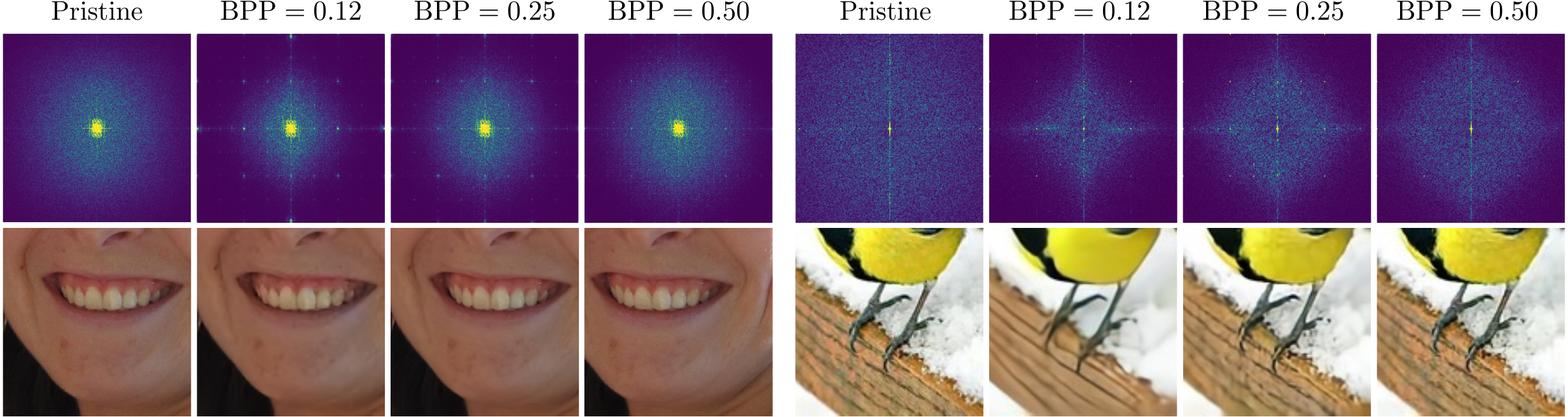}
\caption{Example of JPEG AI artifacts left at different \gls{bpp} values in the Fourier and spatial domains. Left block: FFHQ dataset; right block: LSUN dataset.
In the first row, we show the average Fourier spectra (magnitude) of pristine and JPEG AI images; all spectra are centered in the spatial frequencies $(0, 0)$. In the second row, we show the close-up of one image per dataset. Best viewed in electronic format.}
\label{fig:jpeg_ai_artifacts}
\vspace{-15pt}
\end{figure*}
\noindent \textbf{JPEG AI compression artifacts. }Recently, it has been shown that neural compression codecs can leave frequency artifacts similar to those of synthetic generation techniques~\cite{Bergmann2023, Bergmann2024}.
These artifacts presumably stem from the use of upsampling operators in the decoding process, 
and can be misleading for forensic detectors that rely on similar forensic footprints to accomplish their task. 
In this vein, we perform a similar analysis for JPEG AI compression, following a well-known forensic procedure, i.e., we compute the average Fourier spectrum of noise residuals extracted from images under analysis~\cite{Corvi_2023_ICASSP, Mandelli2024}. 

The first row of~\cref{fig:jpeg_ai_artifacts} shows the average Fourier spectra of residuals computed from the JPEG AI versions of $10K$ images of the FFHQ dataset~\cite{ffhq} (left block) and of $13.5K$ LSUN images~\cite{LSUN} (right block). We consider \gls{bpp} target values of $0.12, 0.25$ and $0.5$, as suggested by the JPEG AI Call For Evidence~\cite{jpegai_ce}.
Interestingly, JPEG AI spectra present peaks that resemble those already noticed for synthetic images~\cite{Mandelli2024, Corvi_2023_ICASSP}, though being less pronounced. The peaks vary in intensity and frequency index according to the \gls{bpp} used.
These results drive us to our proposed analysis of the counter-forensic effects of JPEG AI. 

Moreover, if LSUN samples compression at lower \glspl{bpp} shows a prominent low-pass filtering effect, this behavior is much less pronounced on FFHQ. As an example, in the second row of~\cref{fig:jpeg_ai_artifacts}, we report the close-up of one image per dataset. On FFHQ samples (left), few artifacts are visible on specific face regions that are well known to be hard to reconstruct by neural decoders, like teeth, eyes, and ears~\cite{verdoliva2020media}. LSUN samples (right) look more low-pass filtered at low \glspl{bpp}, even though $0.50$ \gls{bpp} are already enough to reconstruct high-frequency details.




\section{Case studies}
\label{sec:studies}



In this paper, we aim to verify if forensic detectors 
fail to work correctly when dealing with JPEG AI-compressed images. 
We focus mainly on two forensics tasks: i) deepfake image detection and ii) image splicing localization. In the following, we illustrate each of them and motivate our proposed experiments.

\noindent \textbf{Deepfake image detection. }With deepfake image detection, we refer to identifying images created through synthetic generation techniques, such as \glspl{gan} or diffusion models. 
Given an image $\mathbf{I}$, state-of-the-art detectors implement a binary classifier $D(\cdot)$ returning a detection score $s=D(\mathbf{I})$ for the image being a deepfake. 
Higher scores usually indicate a higher likelihood of the image being synthetically generated, with practitioners using a fixed threshold, e.g., $0$, to separate the two classes (if the score is greater than $0$, the image is classified as fake, otherwise it is detected as pristine).

Our study investigates whether state-of-the-art detectors produce high scores when processing images compressed with JPEG AI. 
We aim to check if the artifacts introduced by JPEG AI (like the ones previously shown in \cref{sec:background}) are misinterpreted by these detectors 
leading to the misclassification of pristine samples as deepfakes or to alterations of the scores of deepfake images. To do so, we compare the distributions of the scores obtained from JPEG AI-uncompressed images and their JPEG AI-compressed versions.

\noindent \textbf{Image splicing localization. }With image splicing localization, we refer to the pixel-level identification of local manipulations in an image. 
For instance, this general manipulation manifests in the so-called ``Photoshopped'' or ``cheapfake''~\cite{paris2019} pictures, i.e., images where a malicious actor altered some pixels by copying content from another picture. 
To assess the integrity of the image under analysis, state-of-the-art detectors output an estimated real-valued tampering mask $\mathbf{\tilde{M}}$, whose values indicate the likelihood of manipulation for each pixel. 
Similarly to deepfake image detection, imposing a threshold on the likelihood values generates a binary tampering mask that classifies each pixel as manipulated or genuine.

We aim to verify whether image splicing localization tools alter their output masks $\mathbf{\tilde{M}}$ when the input samples are compressed with JPEG AI, i.e., we compare the masks $\mathbf{\tilde{M}}$ obtained from spliced samples before and after JPEG AI compression. In particular, we check if manipulated pixels are still correctly reported in the final estimated binary mask, and we inspect if JPEG AI artifacts cause some pristine pixels to be incorrectly classified as manipulated.
\section{Experimental setup}
\label{sec:setup}

\subsection{Deepfake image detection}
\label{subsec:setup_deepfake_detection}

\noindent \textbf{Forensic detectors. }To have an exhaustive picture of the effects of JPEG AI on deepfake image detection, we select several state-of-the-art detectors ranging from \glspl{cnn} to transformers-based architectures, trained with different paradigms on diverse datasets. In particular, we choose the detector proposed in~\cite{wang2020}, available in two variants, i.e., \cite{wang2020}-A and \cite{wang2020}-B, \textcolor{blue}{the detectors proposed in \cite{gragnaniello2021, Corvi_2023_ICASSP, Ojha_2023_CVPR, Mandelli2024, Tan_2024_CVPR}} and, finally, the two approaches presented in \cite{cozzolino2023raising}, i.e., \cite{cozzolino2023raising}-A and \cite{cozzolino2023raising}-B. 
All these tools are publicly available, with both test code and model weights. We rely on the original authors' implementation for each detector to execute our experiments~\footnote{Complete experimental details in the Supplementary Material \cref{sec:supp:details}. 
All code for replicating our experiments is available \href{https://github.com/polimi-ispl/jpeg-ai-antifor}{here}.
\label{footnote_details}
}.

\noindent \textbf{Test set. }To avoid any possible bias due to data distribution mismatching, we test each detector on the same test set used by the original authors. 
Our experiments' final set of images comprehends pristine samples from CelebA~\cite{celeba}, COCO~\cite{coco}, FFHQ~\cite{ffhq}, Imagenet~\cite{imagenet}, LAION~\cite{laion}, LSUN~\cite{LSUN}, and RAISE~\cite{raise} datasets. For every pristine test set, we include a deepfake counterpart, selected by the original authors to depict similar content to the pristine one\footref{footnote_details}. We process all these images through JPEG AI compression, following a specific paradigm we detail in~\cref{subsec:setup_compression}.

\noindent \textbf{Evaluation metrics. } 
To quantify the counter-forensic effects of JPEG AI, we compare pristine and deepfake scores' distributions 
considering two scenarios, i.e., \textit{before} and \textit{after} JPEG AI compression is applied to our test images.
Our goal is to quantify how much JPEG AI alters these scores' distribution.

As previously conjectured, we expect JPEG AI to move the score distribution of pristine samples to the ``deepfake'' area, i.e., above the $0$ threshold, potentially producing more false alarms. Moreover, we later show that JPEG AI compression does not only modify pristine samples' distributions, but it can also affect those of synthetic images. To quantify these modifications, we 
use standard binary classification metrics, namely the \gls{auc}, the \gls{ba}, the \gls{fpr}, and the \gls{fnr} evaluated at $0$ threshold. 
The \gls{fpr} quantifies the ratio of pristine samples misclassified as synthetic (i.e., the false alarms), while the \gls{fnr} represents the ratio of synthetic samples misclassified as pristine. 
If the JPEG AI has no effect, these metrics should remain the same before and after the compression.

\subsection{Image splicing localization}

\noindent \textbf{Forensic detectors. }
We focus on the latest and top-performing state-of-the-art techniques based on deep learning models. In particular, we choose three detectors: i) TruFor~\cite{Guillaro_2023_CVPR}, a method based on a denoising \gls{cnn}; ii) MMFusion~\cite{triaridis2024exploring}, a framework for the multi-modal fusion of different image splicing localization techniques; iii) \gls{ifosn}~\cite{Wu2022}, a technique robust to social network compression schemes.
Code and model weights are available online, and we rely on the original author's implementation for our experiments.

\noindent \textbf{Test set. }We consider standard benchmarks in the field, i.e., CASIA~\cite{Dong2013}, Columbia~\cite{hsu06crfcheck}, Coverage~\cite{wen2016}, COCOGlide~\cite{Guillaro_2023_CVPR}, and DSO-1~\cite{Carvalho2013, Carvalho2016}. All detectors were originally tested with excellent performances on these datasets, which comprise uncompressed samples (DSO-1, Columbia, Coverage, COCOGlide) and JPEG images (CASIA). We compress all these images through JPEG AI by following the specifications provided in~\cref{subsec:setup_compression}.

\noindent \textbf{Evaluation metrics. }In this case as well, we evaluate the performances of all detectors on the JPEG AI-uncompressed version of the datasets and their JPEG AI-compressed version. In specific, we compare the ground truth mask $\mathbf{M}$ against the estimated real-valued mask $\mathbf{\tilde{M}}$ for each sample \textit{before} and \textit{after} the compression.

Without counter-forensic effects, we expect a correct separation between pristine and tampered with pixels. On the contrary, in the case of the counter-forensic effect of JPEG AI, we expect a potential performance drop since JPEG AI compression might hinder or introduce new artifacts. 
To measure this behavior, we consider standard metrics for the image splicing localization task, namely the pixel level \gls{auc}, the \gls{ba} and the F1-score. The \gls{ba} and the F1-score are evaluated at the $0.5$ threshold. 

\subsection{Dataset compression}
\label{subsec:setup_compression}
We compress the datasets illustrated above using the official JPEG AI Reference Software~\cite{jpegai_vm}. 
In particular, we use version 7.0 of the JPEG AI Verification Model~\cite{jpegai_vm} with the High Operation Point (HOP) configuration and enable all tools to achieve the best coding performance~\cite{jpegai_vm}.
We rely on six different \gls{bpp} values, i.e., $0.12, 0.25, 0.5, 0.75, 1.0$, and $2.0$. We select these values from the official target rates of the JPEG AI Call for Evidence Challenge~\cite{jpegai_ce}. 
With this approach, we have a believable overview of how the compression ratio affects the presence of artifacts and, in turn, the performance of forensic detectors. 
\section{Results}
\label{sec:results}

\subsection{Deepfake image detection}
\label{subsec:results_synth_img_det}

\noindent \textbf{Preliminary analysis. }Before analyzing the counter-forensic effect of JPEG AI, we first evaluate the various detectors on pristine and synthetic images \textit{not compressed with JPEG AI}. This approach allows us to verify if detectors achieve acceptable performances for the standard real-vs-deepfake task. 
The first column of~\cref{tab:jpegai_metrics_all_detectors_all_datasets_final} reports the results averaged across all datasets.
On average, all detectors show satisfactory performances\footnote{Complete results for all detectors and all datasets are available in \cref{sec:supp:deepfake} and \cref{sec:supp:splicing} of the Supplementary Material.\label{footnote_deepfake}}.

\begin{table*}
\caption{\gls{ba}, \gls{fpr}, and \gls{fnr} evaluated at $0$ threshold for the deepfake image detection case study before and after JPEG AI compression. 
}
\centering
\resizebox{\textwidth}{!}{
\begin{tabular}{lccc|ccc|ccc|ccc|ccc|ccc}
\toprule
 & \multicolumn{3}{c |}{Original format} & \multicolumn{3}{c|}{BPP = $0.12$} & \multicolumn{3}{c|}{BPP = $0.25$} & 
 \multicolumn{3}{c|}{BPP = $0.5$} &
 \multicolumn{3}{c|}{BPP = $0.75$} & \multicolumn{3}{c}{BPP = $1.0$} \\
 \midrule
& $\textrm{BA}_{\uparrow}$ & $\textrm{FPR}_{\downarrow}$ & $\textrm{FNR}_{\downarrow}$ & $\textrm{BA}_{\uparrow}$ & $\textrm{FPR}_{\downarrow}$ & $\textrm{FNR}_{\downarrow}$ & $\textrm{BA}_{\uparrow}$ & $\textrm{FPR}_{\downarrow}$ & $\textrm{FNR}_{\downarrow}$ & $\textrm{BA}_{\uparrow}$ & $\textrm{FPR}_{\downarrow}$ & $\textrm{FNR}_{\downarrow}$ & $\textrm{BA}_{\uparrow}$ & $\textrm{FPR}_{\downarrow}$ & $\textrm{FNR}_{\downarrow}$ & $\textrm{BA}_{\uparrow}$ & $\textrm{FPR}_{\downarrow}$ & $\textrm{FNR}_{\downarrow}$ \\
\midrule
\cite{wang2020}-A & $0.82$ & $0.04$ & $0.31$ & $0.62$ & $0.56$ & $0.20$ & $0.71$ & $0.35$ & $0.22$ & $0.76$ & $0.16$ & $0.32$ & $0.80$ & $0.09$ & $0.32$ & $0.79$ & $0.06$ & $0.36$ \\
\cite{wang2020}-B & $0.74$ & $0.01$ & $0.51$ & $0.64$ & $0.16$ & $0.56$ & $0.69$ & $0.06$ & $0.57$ & $0.71$ & $0.02$ & $0.56$ & $0.73$ & $0.01$ & $0.53$ & $0.73$ & $0.01$ & $0.53$ \\
\cite{gragnaniello2021} & $0.72$ & $0.03$ & $0.54$ & $0.50$ & $0.96$ & $0.04$ & $0.52$ & $0.86$ & $0.11$ & $0.72$ & $0.38$ & $0.18$ & $0.72$ & $0.22$ & $0.33$ & $0.76$ & $0.08$ & $0.41$ \\
\cite{Corvi_2023_ICASSP} & $0.77$ & $0.00$ & $0.46$ & $0.50$ & $0.97$ & $0.02$ & $0.59$ & $0.52$ & $0.29$ & $0.65$ & $0.03$ & $0.66$ & $0.67$ & $0.01$ & $0.66$ & $0.69$ & $0.00$ & $0.63$ \\
\cite{Ojha_2023_CVPR} & $0.91$ & $0.02$ & $0.15$ & $0.71$ & $0.42$ & $0.17$ & $0.80$ & $0.19$ & $0.21$ & $0.84$ & $0.08$ & $0.24$ & $0.86$ & $0.05$ & $0.22$ & $0.87$ & $0.04$ & $0.21$ \\
\cite{cozzolino2023raising}-A & $0.77$ & $0.29$ & $0.16$ & $0.59$ & $0.71$ & $0.11$ & $0.60$ & $0.66$ & $0.14$ & $0.60$ & $0.64$ & $0.17$ & $0.64$ & $0.56$ & $0.16$ & $0.65$ & $0.54$ & $0.17$ \\
\cite{cozzolino2023raising}-B & $0.74$ & $0.30$ & $0.23$ & $0.60$ & $0.72$ & $0.07$ & $0.63$ & $0.65$ & $0.09$ & $0.64$ & $0.61$ & $0.11$ & $0.67$ & $0.54$ & $0.13$ & $0.68$ & $0.51$ & $0.14$ \\
\cite{Tan_2024_CVPR} & $0.92$ & $0.13$ & $0.02$ & $0.50$ & $0.99$ & $0.00$ & $0.51$ & $0.99$ & $0.00$ & $0.51$ & $0.98$ & $0.00$ & $0.51$ & $0.98$ & $0.00$ & $0.51$ & $0.97$ & $0.00$ \\
\cite{Mandelli2024} & $0.84$ & $0.16$ & $0.15$ & $0.50$ & $0.96$ & $0.05$ & $0.53$ & $0.78$ & $0.15$ & $0.58$ & $0.61$ & $0.22$ & $0.67$ & $0.41$ & $0.24$ & $0.71$ & $0.32$ & $0.26$ \\
\bottomrule
\end{tabular}
}
\label{tab:jpegai_metrics_all_detectors_all_datasets_final}
\vspace{-10pt}
\end{table*}

Given our goal of studying the effect of JPEG AI on misleading deepfake image detectors,
in the following, we discard results from datasets where detectors' performances do not exceed an \gls{auc} of $0.75$. Indeed, we want to analyze the counter-forensic effects of JPEG AI only when the initial dataset (\textit{before} JPEG AI compression) is classified with acceptable performances by the detectors.

\begin{figure}[t]
\centering
        \includegraphics[width=\columnwidth]{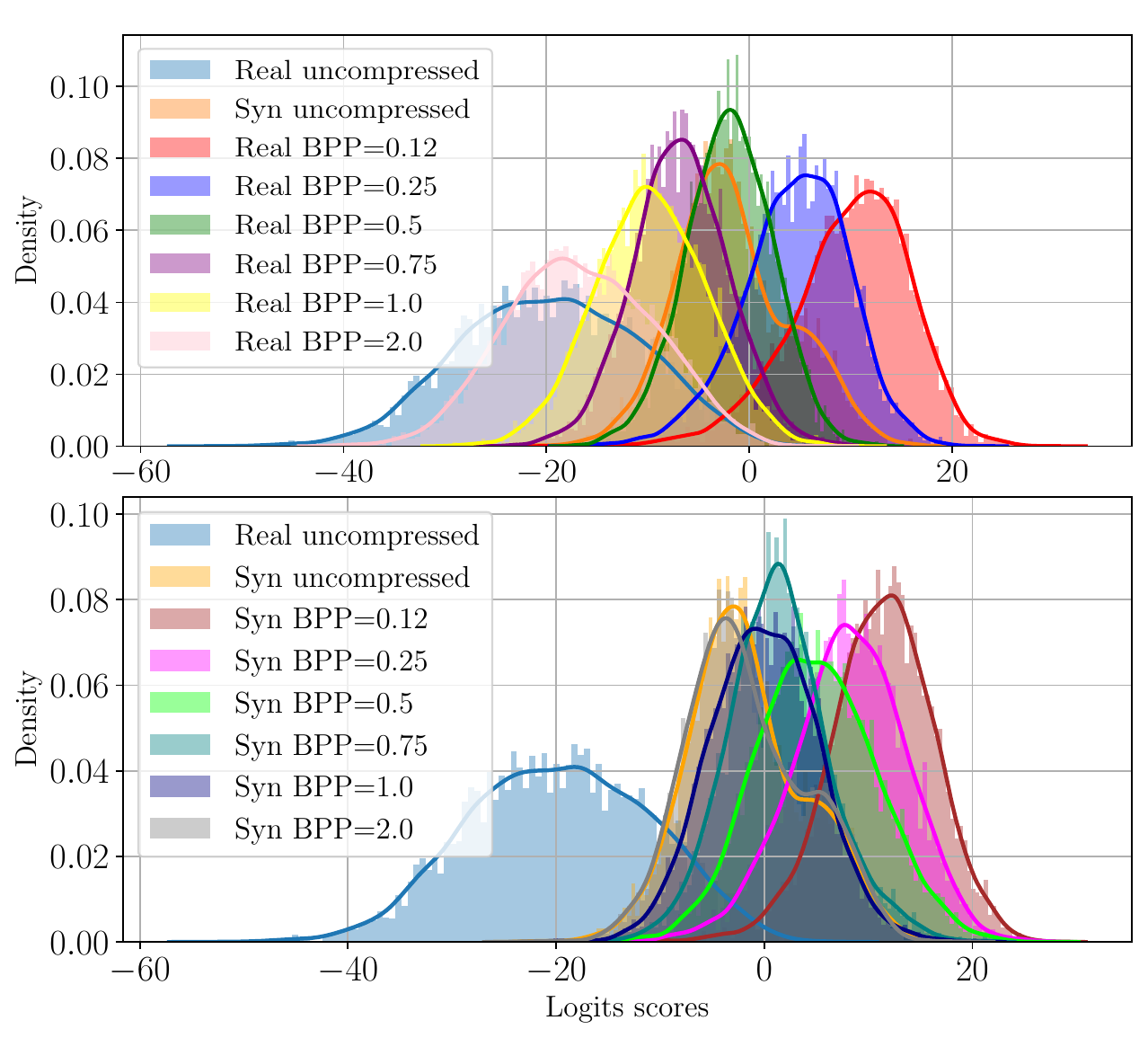}
        \caption{Scores distribution for~\cite{gragnaniello2021} over COCO \textit{pristine} samples (top row) and their \textit{synthetic} counterpart samples (bottom row) compressed at different \gls{bpp} values.}
\label{fig:jpeg-ai_dist}
\vspace{-17pt}
\end{figure}

\noindent \textbf{Impact of compression ratio. }
We begin our analysis by illustrating the effect of JPEG AI 
on the score distributions.
Figure~\ref{fig:jpeg-ai_dist} depicts the logit scores distribution of detector \cite{gragnaniello2021} when processing original format and JPEG AI compressed samples from the COCO dataset\footref{footnote_deepfake}. 

Interestingly, as the compression ratio increases (i.e., when we lower the \glspl{bpp}), the distribution of the scores moves its center of gravity to the right (i.e., above the $0$ threshold), almost independently from the class of the samples. This behavior has different implications. For pristine samples, we expect an increase in false positives. This right shift may cause fewer false negatives for deepfake samples instead. 
To confirm this hypothesis quantitatively, \cref{tab:jpegai_metrics_all_detectors_all_datasets_final} (starting from the second left column) reports the average results obtained by testing each detector on all its datasets\footnote{We omit the \gls{auc} results and \gls{bpp}$=2.0$ scenario for clarity. Complete results are available in the Supplementary Material  \cref{sec:supp:deepfake}.\label{auc_footnote}}.
For extremely low \glspl{bpp}, all detectors exhibit a drop in the \gls{ba}, especially for~\cite{Tan_2024_CVPR, Corvi_2023_ICASSP, gragnaniello2021, Mandelli2024},  
and they all experience an increase in the \gls{fpr}.
All these elements tend to confirm that, in low-rate settings, 
the score distributions of JPEG AI pristine samples move above the $0$ threshold. Hence, these images can be mistaken for deepfakes.
On the contrary, for all detectors except~\cite{wang2020}-B and~\cite{Ojha_2023_CVPR}, the \gls{fnr} decreases for aggressive compression settings. This phenomenon seems to corroborate our intuition: JPEG AI artifacts might ``overwrite'' synthetic generation traces at very low \gls{bpp} values. However, most detectors still mistake these artifacts as deepfake generation clues, thus reducing the \gls{fnr} and ending up improving the classification performances of synthetic images.

As might be expected, when increasing the \gls{bpp}, JPEG AI score distributions tend to align more closely with the original ones. Nonetheless, the \gls{ba} can still be less than $0.7$, and the \gls{fpr} risks to be high even when the JPEG AI compression ratio gets lower
(see the case in which \gls{bpp} is equal to $1$ for detectors~\cite{cozzolino2023raising}-A, \cite{cozzolino2023raising}-B and~\cite{Tan_2024_CVPR}).
Moreover, when the \gls{bpp} increases, 
also a left-shift of the scores of synthetic samples can occur, thus resulting in a slight enhancement of \gls{fnr} and reducing the detection performances (see, for instance, detectors~\cite{wang2020}-A, \cite{gragnaniello2021},~\cite{Corvi_2023_ICASSP},~\cite{Ojha_2023_CVPR} and~\cite{Mandelli2024}).


\begin{table}
\centering
\caption{JPEG AI MS-SSIM and VIF quality metrics.}
\resizebox{.85\columnwidth}{!}{
\begin{tabular}{l S[table-format=1.2] S[table-format=1.2] S[table-format=1.2] S[table-format=1.2] S[table-format=1.2] S[table-format=1.2]}
\toprule
BPP & 0.12 & 0.25 & 0.5 & 0.75 & 1.0 & 2.0 \\
\midrule
MS-SSIM & 0.97 & 0.98 & 0.99 & 1.00 & 1.00 & 1.00 \\
VIF & 38.15 & 40.28 & 42.47 & 44.33 & 45.89 & 48.08 \\
\bottomrule
\end{tabular}}
\label{tab:jpegai_quality_metrics}
\vspace{-10pt}
\end{table}

\begin{table*}[t]
\caption{\gls{ba}, \gls{fpr}, and \gls{fnr} evaluated at $0$ threshold for the deepfake image detection case study before and after JPEG compression. 
\gls{auc} results available in the Supplementary Material \cref{sec:supp:deepfake}.
}
\centering
\resizebox{.9\textwidth}{!}{
\begin{tabular}{lccc|ccc|ccc|ccc|ccc}
\toprule
 & \multicolumn{3}{c|}{Original format} & \multicolumn{3}{c}{QF = 65} & \multicolumn{3}{c}{QF = 75} & 
 \multicolumn{3}{c}{QF = 85} &
 \multicolumn{3}{c}{QF = 95} \\
 \midrule
& $\textrm{BA}_{\uparrow}$ & $\textrm{FPR}_{\downarrow}$ & $\textrm{FNR}_{\downarrow}$ & $\textrm{BA}_{\uparrow}$ & $\textrm{FPR}_{\downarrow}$ & $\textrm{FNR}_{\downarrow}$ & $\textrm{BA}_{\uparrow}$ & $\textrm{FPR}_{\downarrow}$ & $\textrm{FNR}_{\downarrow}$ & $\textrm{BA}_{\uparrow}$ & $\textrm{FPR}_{\downarrow}$ & $\textrm{FNR}_{\downarrow}$ & $\textrm{BA}_{\uparrow}$ & $\textrm{FPR}_{\downarrow}$ & $\textrm{FNR}_{\downarrow}$ \\
\midrule
\cite{wang2020}-A & $0.82$ & $0.04$ & $0.31$ & $0.69$ & $0.02$ & $0.59$ & $0.69$ & $0.01$ & $0.60$ & $0.69$ & $0.01$ & $0.60$ & $0.73$ & $0.01$ & $0.54$ \\
\cite{wang2020}-B & $0.74$ & $0.01$ & $0.51$ & $0.69$ & $0.00$ & $0.62$ & $0.69$ & $0.00$ & $0.61$ & $0.69$ & $0.00$ & $0.61$ & $0.71$ & $0.00$ & $0.59$ \\
\cite{gragnaniello2021} & $0.72$ & $0.03$ & $0.54$ & $0.57$ & $0.00$ & $0.85$ & $0.58$ & $0.00$ & $0.84$ & $0.60$ & $0.00$ & $0.80$ & $0.63$ & $0.01$ & $0.73$ \\
\cite{Corvi_2023_ICASSP} & $0.77$ & $0.00$ & $0.46$ & $0.78$ & $0.11$ & $0.33$ & $0.78$ & $0.07$ & $0.37$ & $0.77$ & $0.03$ & $0.43$ & $0.76$ & $0.00$ & $0.48$ \\
\cite{Ojha_2023_CVPR} & $0.91$ & $0.02$ & $0.15$ & $0.80$ & $0.03$ & $0.36$ & $0.82$ & $0.03$ & $0.33$ & $0.82$ & $0.01$ & $0.34$ & $0.87$ & $0.02$ & $0.24$ \\
\cite{cozzolino2023raising}-A & $0.77$ & $0.29$ & $0.16$ & $0.68$ & $0.09$ & $0.55$ & $0.66$ & $0.07$ & $0.60$ & $0.64$ & $0.11$ & $0.61$ & $0.77$ & $0.21$ & $0.24$ \\
\cite{cozzolino2023raising}-B & $0.74$ & $0.30$ & $0.23$ & $0.68$ & $0.28$ & $0.35$ & $0.68$ & $0.27$ & $0.36$ & $0.64$ & $0.25$ & $0.48$ & $0.70$ & $0.32$ & $0.27$ \\
\cite{Tan_2024_CVPR} & $0.92$ & $0.13$ & $0.02$ & $0.50$ & $0.00$ & $1.00$ & $0.50$ & $0.00$ & $1.00$ & $0.50$ & $0.00$ & $0.99$ & $0.51$ & $0.03$ & $0.95$ \\
\cite{Mandelli2024} & $0.84$ & $0.16$ & $0.15$ & $0.76$ & $0.14$ & $0.34$ & $0.77$ & $0.15$ & $0.32$ & $0.78$ & $0.16$ & $0.27$ & $0.80$ & $0.18$ & $0.22$ \\
\bottomrule
\end{tabular}
}
\label{tab:jpeg_metrics_all_detectors_all_datasets_final}
\vspace{-12pt}
\end{table*}

\noindent \textbf{JPEG AI image quality analysis. }
Given the performance drop of all detectors when tested against JPEG AI images, we now investigate if this loss might be due to the poor visual quality of JPEG AI compressed images with respect to the original ones. 
To do this, we compute the \gls{mss} and the \gls{vif} for all the JPEG AI images in the datasets, comparing them to their
JPEG AI-uncompressed
versions. The JPEG AI committee proposes both metrics to evaluate the quality of JPEG AI implementations~\cite{jpegai_ce, JPEGAI_whitepaper}~\footnote{Additional results available in \cref{sec:supp:deepfake}.}.

\cref{tab:jpegai_quality_metrics} shows the average results. It is worth noticing that all the considered \gls{bpp} achieve high quality metrics. Extremely low \gls{bpp}, as previously shown in \cref{sec:background}, might carry a few low-pass filtering effects, but they do not introduce important losses in the visual quality of the images. 
This suggests that other artifacts than only low-pass frequency effects drive the detectors 
when classifying JPEG AI images as synthetic.

\begin{table*}[t]
\color{blue}
\caption{Evaluation metrics for detector~\cite{Mandelli2024} before and after retraining or fine-tuning with $20\%$ of JPEG AI samples. All metrics are computed with a $0$ threshold.}
\centering
\resizebox{\textwidth}{!}{
\begin{tabular}{lccc|ccc|ccc|ccc|ccc|ccc}
\toprule
 & \multicolumn{3}{c |}{Original format} & \multicolumn{3}{c|}{BPP = $0.12$} & \multicolumn{3}{c|}{BPP = $0.25$} & 
 \multicolumn{3}{c|}{BPP = $0.5$} &
 \multicolumn{3}{c|}{BPP = $0.75$} & \multicolumn{3}{c}{BPP = $1.0$} \\
 \midrule
& $\textrm{BA}_{\uparrow}$ & $\textrm{FPR}_{\downarrow}$ & $\textrm{FNR}_{\downarrow}$ & $\textrm{BA}_{\uparrow}$ & $\textrm{FPR}_{\downarrow}$ & $\textrm{FNR}_{\downarrow}$ & $\textrm{BA}_{\uparrow}$ & $\textrm{FPR}_{\downarrow}$ & $\textrm{FNR}_{\downarrow}$ & $\textrm{BA}_{\uparrow}$ & $\textrm{FPR}_{\downarrow}$ & $\textrm{FNR}_{\downarrow}$ & $\textrm{BA}_{\uparrow}$ & $\textrm{FPR}_{\downarrow}$ & $\textrm{FNR}_{\downarrow}$  & $\textrm{BA}_{\uparrow}$ & $\textrm{FPR}_{\downarrow}$ & $\textrm{FNR}_{\downarrow}$\\
 \midrule
Before & $0.84$ & $0.16$ & $0.15$ & $0.50$ & $0.96$ & $0.05$ & $0.53$ & $0.78$ & $0.15$ & $0.58$ & $0.61$ & $0.22$ & $0.67$ & $0.41$ & $0.24$ & $0.71$ & $0.32$ & $0.26$ \\
Retraining & $0.83$ & $0.15$ & $0.19$ & $0.61$ & $0.65$ & $0.14$ & $0.69$ & $0.41$ & $0.21$ & $0.74$ & $0.25$ & $0.27$ & $0.76$ & $0.19$ & $0.29$ & $0.77$ & $0.16$ & $0.30$ \\
Fine-tuning & $0.81$ & $0.07$ & $0.31$ & $0.58$ & $0.63$ & $0.21$ & $0.64$ & $0.43$ & $0.28$ & $0.73$ & $0.24$ & $0.31$ & $0.76$ & $0.17$ & $0.32$ & $0.76$ & $0.14$ & $0.33$ \\
\bottomrule
\end{tabular}
}
\label{tab:retraining_jpegai_final}
\vspace{-12pt}
\color{black}
\end{table*}

\noindent \textbf{Comparisons with standard JPEG. } 
In the forensic community, it is well-known that post-processing operations, especially image compression like standard JPEG, might alter the performance of detectors~\cite{wang2020, mandelli2020training, Corvi_2023_ICASSP}. 
Thus, the results obtained using JPEG AI, as previously presented, could be considered somewhat predictable.
Nonetheless, we argue that standard JPEG, while undermining the performance of detectors, 
behaves fundamentally differently from JPEG AI.

\begin{figure}[t]
\centering
        \includegraphics[width=\columnwidth]{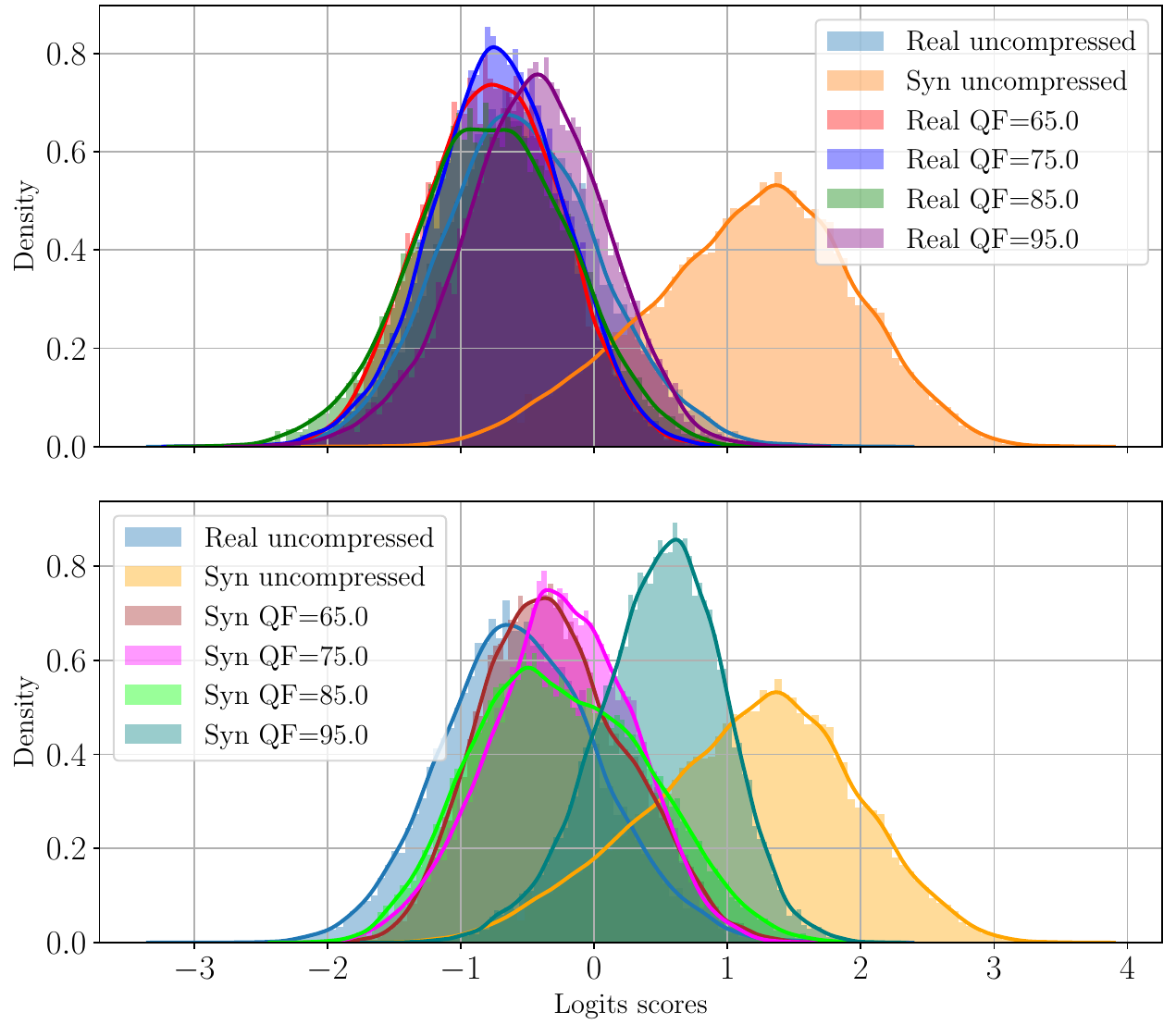}
        \caption{Scores distribution for~\cite{cozzolino2023raising}-A over LSUN \textit{pristine} samples (top row) and their \textit{synthetic} counterpart (bottom row) compressed with JPEG at different \gls{qf} values.}
\label{fig:jpeg_dist}
\vspace{-17pt}
\end{figure}

To investigate this,
\textcolor{blue}{we compress our datasets using different \glspl{qf}, then compare the distribution of the scores before and after JPEG compression.}
\cref{fig:jpeg_dist} shows an example, i.e., the score distribution for the detector \cite{cozzolino2023raising}-A from the LSUN dataset\footref{footnote_deepfake}.
Notice that, differently from the JPEG AI case, the scores distribution of pristine JPEG-compressed images closely overlaps with those obtained from JPEG-uncompressed samples, meaning the absence of false alarms (both distributions are all centered below the $0$). On the other hand, when compressing deepfake images, the scores distribution moves below the $0$ as we increase the \gls{qf}, meaning a significant increase in the number of false negatives.

To confirm this observation, \cref{tab:jpeg_metrics_all_detectors_all_datasets_final} reports the average classification metrics for all datasets and detectors before (first left column) and after (starting from the second left column) JPEG varying the \gls{qf} used in the compression. 
In this case, the \gls{ba} has a general drop, which is more evident in the case of lower \glspl{qf}. 
However, unlike the JPEG AI scenario, the \gls{fpr} stays constant or even lowers, indicating that the scores distribution of pristine samples remains or moves even below the $0$ threshold.
On the other hand, JPEG compression, apart from detector~\cite{Corvi_2023_ICASSP}, consistently affects the \gls{fnr}, which is generally higher, especially for~\cite{wang2020}-A,~\cite{gragnaniello2021},~\cite{cozzolino2023raising}-A,~\cite{Tan_2024_CVPR} and~\cite{Mandelli2024}. 
This suggests that JPEG compression shifts the scores distribution of deepfake images below the $0$, effectively ``canceling'' out forensic artifacts that are crucial for deepfake detection. 
This behavior is diametrically opposed to that of JPEG AI, which essentially affects the detection of pristine content, increasing the number of false positives.



\noindent \textbf{Introducing JPEG AI into training. }
Given that all the investigated detectors have not been trained with JPEG AI images,
we conduct a last experiment by retraining the forensic detector recently proposed in~\cite{Mandelli2024} with 
JPEG AI samples.
We consider the same data used in the training phase of this detector. However, we compress the $20\%$ of training and validation samples with the \gls{bpp} values considered in our previous analysis, equally distributed among the considered images. 
Then, we retrain this detector in two scenarios: (i) complete retraining over the same dataset used in~\cite{Mandelli2024}, with $20\%$ of its samples compressed with JPEG AI; (ii) a fine-tuning of the already trained detector by exploiting only JPEG AI samples.  
In both cases, we follow the original training procedure~\cite{Mandelli2024}. 

\cref{tab:retraining_jpegai_final} depicts the achieved detection metrics over non-JPEG AI data (first left column) and over JPEG AI images (other columns) before and after retraining and fine-tuning. 
The detector is still capable of classifying original format samples, experiencing a slight performance loss in the retraining case and, as expected, a more significant drop when fine-tuning.
Indeed, the fine-tuning setup includes only JPEG AI-compressed samples;
thus, a subtle loss of generalization is expected.

As for what regards the \gls{ba} and the \gls{fpr}, we can notice a reduction of the JPEG AI counter-forensic effect since these metrics improve upon the initial results.
Nonetheless, the \gls{ba} is still below $0.7$, and the \gls{fpr} is above $0.4$ for lower \glspl{bpp}.
Instead, the effects of retraining on the \gls{fnr} are less visible: as observed before, though the \gls{fnr} tends to reduce for low \glspl{bpp}, this number gradually increases, achieving even higher values than those resulting when processing the original format images. 
These preliminary results show that a huge augmentation campaign should probably be considered to be completely robust to JPEG AI compression.

\begin{figure}[t]
\centering
    \includegraphics[width=\columnwidth]{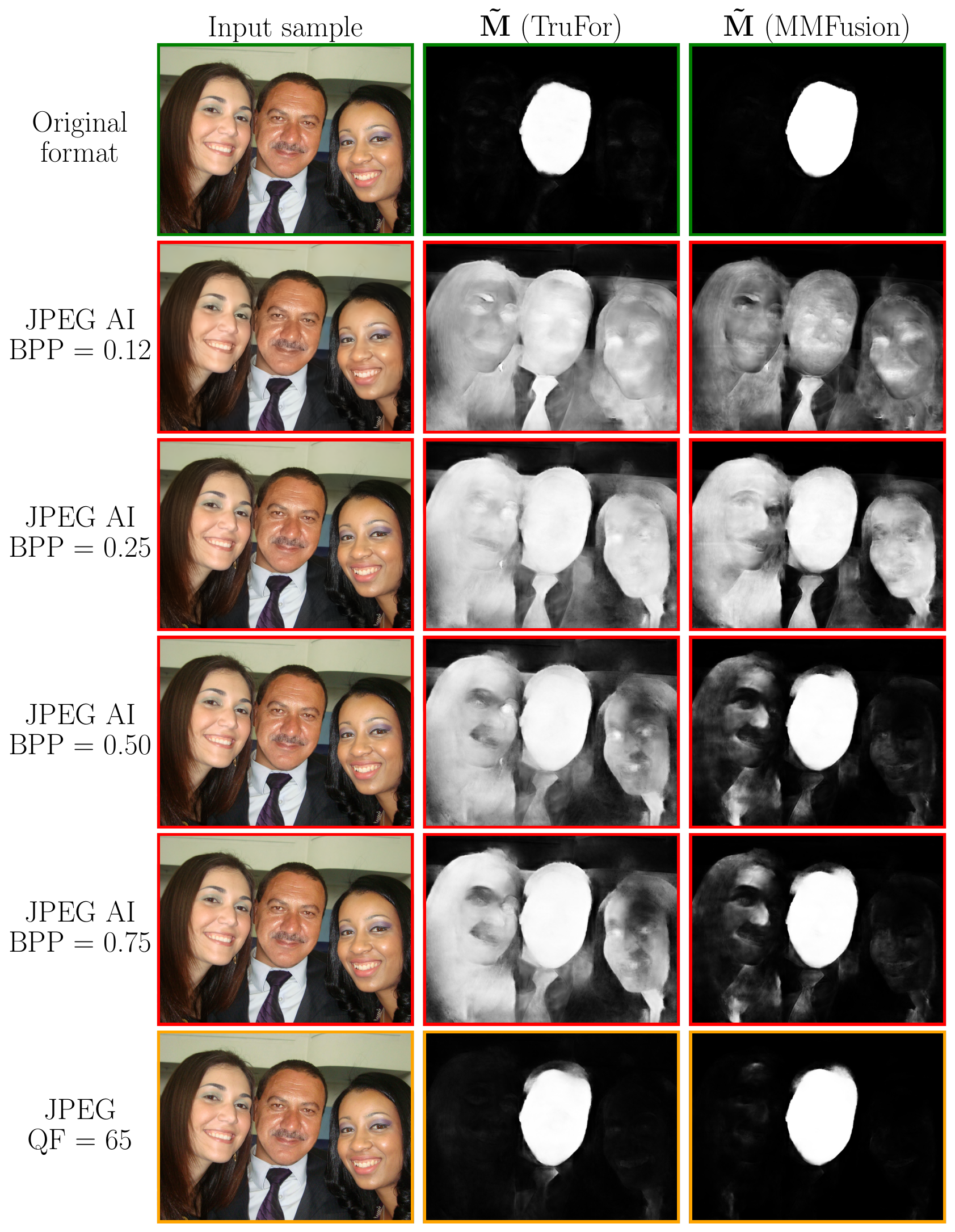}
    \caption{Output of TruFor~\cite{Guillaro_2023_CVPR} and MMFusion~\cite{triaridis2024exploring} for the same DSO-1 input under different JPEG AI compression settings, together with the JPEG $\textrm{QF} = 65$ setting. Other examples available in \cref{sec:supp:splicing} of the supplementary material.
    }
\label{fig:splicing_grid}
\vspace{-17pt}
\end{figure}

\subsection{Image splicing localization}

\noindent \textbf{Visual inspection. }
As introduced in \cref{sec:setup}, we analyze three detectors, TruFor~\cite{Guillaro_2023_CVPR}, MMFusion~\cite{triaridis2024exploring}, and \gls{ifosn}~\cite{Wu2022}. 
We start by visually inspecting their output on a sample from the DSO-1 dataset. Figure~\ref{fig:splicing_grid} shows the output masks $\mathbf{\tilde{M}}$ for the original format 
as well as the masks obtained after compressing the input sample with JPEG AI at different \glspl{bpp} for TruFor and MMFusion. We display the original spliced sample together with its compressed versions as well. 

Even though the quality of the JPEG AI images is visually high, we can observe that the compression operation heavily degrades the output masks $\mathbf{\tilde{M}}$ in two different ways. First, for low \gls{bpp}, the original manipulated area, while still largely identified, may present several false negatives, i.e., manipulated pixels that are incorrectly classified as pristine. Second, we can observe many false positives in the rest of the image, i.e., pristine pixels erroneously classified as manipulated at all \gls{bpp} values. These two separate phenomena suggest that JPEG AI can contemporarily ``hide'' some manipulation traces while introducing artifacts mistaken for tampered with pixels.

\noindent \textbf{Quantitative analysis. }We compute the pixel-level localization metrics illustrated in~\cref{sec:setup} to quantify this visual inspection and report their values in~\cref{tab:splicing} (second and third right column). In particular, we calculate these metrics for all the datasets considered and then average their values. 

As the reader can notice, performance drops sharply as the JPEG AI compression ratio increases for all detectors. 
For the compression at \gls{bpp} $=0.12$, we have 
a drop of more than $0.15$ and $0.13$ points in \gls{auc} and \gls{ba}, respectively. Moreover, the F1 score sharply decreases by more than $0.28$ points. 

It is interesting to notice that, in high \gls{bpp} settings, the F1 score may still show a drop of $0.10$. 
Even if not visible to the naked eye, we conjecture that JPEG AI interpolation artifacts can impair the localization performance of the detectors considered. 
The \gls{ifosn} detector~\cite{Wu2022} is the least affected detector in case of high \glspl{bpp}: we believe this might be due to its specific training process, developed for gaining 
robustness against social network compression. 


\noindent \textbf{Comparison with standard JPEG.} 
The last row of Figure~\ref{fig:splicing_grid} shows the output masks $\mathbf{\tilde{M}}$ obtained after compressing the input sample with JPEG at $\textrm{QF} = 65$ (i.e., the worst considered \gls{qf} in our experiments) for TruFor~\cite{Guillaro_2023_CVPR} and MMFusion~\cite{triaridis2024exploring}. 
Notice the differences to JPEG AI: in the JPEG case, the localization performances are only slightly reduced, and the tampered with area is still well detectable.  

We report in \cref{tab:splicing} (last column) the quantitative results for all detectors when tested against JPEG compression.
It is worth noticing that JPEG impacts the detectors' performance much less for all the \glspl{qf} considered.
In particular, even at the lowest \gls{qf}, all metrics drop by a maximum of $0.09$ points.
We argue that, while JPEG is known for ``hiding'' tampering traces, JPEG AI has a considerably heavier effect due to the various upsampling artifacts introduced.

\begin{table}[t]
\centering
\caption{\textcolor{blue}{Image splicing localization results before and after JPEG AI and JPEG compressions for all detectors. Results are averaged per dataset, considering all datasets at disposal.}}
\resizebox{\columnwidth}{!}{
\begin{tabular}{ll|c|ccccc|cccc}
\toprule
 &  & Orig. & \multicolumn{5}{c|}{JPEG AI compressed (BPP)} & \multicolumn{4}{c}{JPEG compressed (QF)} \\
& Metric& format & 0.12 & 0.25 & 0.5 & 0.75 & 1.0 & 65 & 75 & 85 & 95 \\
\midrule
\multirow{3}{*}{\rotatebox{90}{\cite{Guillaro_2023_CVPR}}} & AUC & $0.91$ & $0.75$ & $0.79$ & $0.83$ & $0.84$ & $0.87$ & $0.85$ & $0.85$ & $0.85$ & $0.86$ \\
& BA & $0.77$ & $0.60$ & $0.62$ & $0.68$ & $0.69$ & $0.72$ & $0.70$ & $0.70$ & $0.70$ & $0.73$ \\
& F1 & $0.66$ & $0.37$ & $0.42$ & $0.51$ & $0.52$ & $0.57$ & $0.58$ & $0.58$ & $0.59$ & $0.63$ \\
\midrule
\multirow{3}{*}{\rotatebox{90}{\cite{triaridis2024exploring}}} & AUC & $0.91$ & $0.72$ & $0.77$ & $0.84$ & $0.85$ & $0.87$ & $0.85$ & $0.85$ & $0.86$ & $0.88$ \\
& BA & $0.80$ & $0.57$ & $0.60$ & $0.68$ & $0.70$ & $0.73$ & $0.72$ & $0.72$ & $0.73$ & $0.76$ \\
& F1 & $0.70$ & $0.31$ & $0.36$ & $0.49$ & $0.52$ & $0.58$ & $0.62$ & $0.62$ & $0.63$ & $0.67$ \\
\midrule
\multirow{3}{*}{\rotatebox{90}{\cite{Wu2022}}} & AUC & $0.86$ & $0.71$ & $0.73$ & $0.80$ & $0.80$ & $0.84$ & $0.77$ & $0.78$ & $0.80$ & $0.82$ \\
& BA & $0.73$ & $0.60$ & $0.62$ & $0.68$ & $0.68$ & $0.71$ & $0.65$ & $0.65$ & $0.67$ & $0.69$ \\
& F1 & $0.56$ & $0.28$ & $0.34$ & $0.44$ & $0.44$ & $0.52$ & $0.48$ & $0.48$ & $0.50$ & $0.53$ \\
\bottomrule
\end{tabular}
}
\label{tab:splicing}
\vspace{-10pt}
\end{table}

\subsection{Double JPEG AI experiments.} 
\label{sec:results:djpegai}
A very common forensic scenario is that represented by double compression~\cite{cozzolino2015splicebuster, barni2017aligned,zampoglou2017large}. In this situation, the input samples might have undergone several compression stages before or after the manipulation. 
For instance, the manipulated area in a spliced sample might present traces of a compression operation performed before the splicing. 
Such traces can be extremely useful for understanding the overall image history and better localizing the tampered with pixels. Moreover, the double compression setup is also valuable for testing image detectors' robustness, i.e., how good they are when analyzing images undergoing several compression stages. 


\color{blue}
In our work, we perform a preliminary analysis of the double compression scenario in the case of JPEG AI. 
To simplify the experimental setting, we consider two opposite setups: 
i) the original-format images have been JPEG AI-compressed with various \gls{bpp} values and then compressed with the lowest \gls{bpp} considered, i.e., \gls{bpp}$ =0.12$; ii) the original-format images have been JPEG AI-compressed with various \gls{bpp} values, and then compressed with the highest \gls{bpp} considered, i.e., \gls{bpp}$ =2.0$.
For the sake of space, we refer the reader to the supplementary material for an in-depth discussion. 
Our findings show that a strong second JPEG AI compression might cancel traces left by the previous compression step, similar to the behavior observed on standard double JPEG compression~\cite{barni2017aligned}.
\color{black}

\section{Discussion and conclusion}
\label{sec:conclusions}

While far from being an exhaustive and complete analysis of the forensic implications of adopting JPEG AI, our results highlight how off-the-shelf forensic detectors need to be prepared to handle the artifacts of this upcoming standard. In both deepfake image detection and image splicing localization tasks, heavy compression leads to inaccurate classification of the investigated content, requiring practitioners to consider JPEG AI images in their work.

As the boundaries between synthetic and authentic media continue to thin~\cite{Moreira2024}, these findings open up new avenues for multimedia forensic researchers on both a general and technical level. On a more general level, the broad adoption of \glspl{nn} in processing media of any kind requires practitioners to reflect on the very nature of the task of deepfake media detection. 
In a scenario where all media are ``deepfakes'', meaning that they went through some \gls{nn} processing, multimedia forensics will have to go deeper in its analysis.
For instance, it will be required to understand if that particular processing was meant for harmful purposes, e.g., to create a deepfake for fake news propaganda, or for innocuous tasks, e.g., for image compression or quality enhancement. 
Recent works on avatar fingerprints~\cite{prashnani2023avatar} go in this direction. In such a scenario, there is also a growing demand for active mechanisms that guarantee the media's provenance, integrity, and reliability. The Content Authenticity Initiative by Adobe~\cite{content_authenticity}, as well as the JPEG Trust standard~\cite{JPEGTrust_whitepaper}, are being developed and adopted in this vein.

On a more technical level, and for the time being, multimedia forensic researchers must develop techniques robust to neural compression artifacts. Robustness to post-processing operations~\cite{Barni2018, wang2020, Bondi2020, mandelli2020training, Wang2024, Mandelli2024} and adversarial attacks~\cite{Carlini2020, Hussain_2021_WACV, derosa2024} are well-known and studied topics in multimedia forensics. However, the nature of neural image compression calls for detectors that distinguish between ``genuine'' neural artifacts and those left by ``malicious'' generation techniques. Correct re-training of data-driven detectors, as shown in~\cref{sec:results} or as proposed in~\cite{Mandelli2024}, are all valid options to further investigate in conjunction with the more general reformulation of the deepfake detection task mentioned above.

In conclusion, the diffusion of JPEG AI fits in a rapidly changing landscape, requiring multimedia forensics to develop more robust instruments and collaborate with social and legal entities to fight against disinformation and re-introduce trust in all media types. 



{
    \small
    \bibliographystyle{ieeenat_fullname}
    \bibliography{main}
}
\clearpage
\setcounter{page}{1}
\maketitlesupplementary


\section{Neural image compression and JPEG AI specification}
\label{sec:supp:jpeagi}
\noindent \textbf{Neural image compression basics. }In \gls{nic}, an input image is transformed into a compact file or bitstream for storage or transmission, which is later used to reconstruct the original image. The basic approach involves mapping an image vector $\mathbf{I}$ to a latent representation $\mathbf{y}$ using a convolutional autoencoder. This latent representation is then quantized to $\hat{\mathbf{y}}$ and compressed into a bitstream using an entropy coding method such as arithmetic coding. During reconstruction, the received bits are decoded via reverse entropy coding, and the resulting $\hat{\mathbf{y}}$ is fed into a \gls{cnn} decoder to produce the reconstructed image $\hat{\mathbf{I}}$. These models can be trained end-to-end with the help of differentiable proxies for entropy coding.

As pointed out in \cref{sec:background}, many techniques exist to perform such operations. In 2017, Theis et al.~\cite{Theis2017} and Ballé et al.~\cite{Balle2017} introduced the first deep-learning approaches for image compression. \cite{Theis2017} proposed compressive autoencoders that learn end-to-end optimized lossy image compression, aiming to outperform traditional codecs, especially at lower bitrates. \cite{Balle2017} presented the first pipeline that jointly optimizes all components of the compression pipeline, achieving superior rate-distortion performance compared to traditional codecs. Building upon their previous work, in \cite{ballé2018variational}, Ballé et al. incorporate a scale hyperprior to model the entropy of the latent representations more accurately, leading to improved compression efficiency. Minnen et al.~\cite{minnen2018} combine autoregressive models with hierarchical priors to better capture spatial dependencies in images, enhancing the performance of learned image compression systems. Choi et al.~\cite{Choi2019} present a conditional autoencoder that enables variable-rate image compression within a single model, adjusting compression rates on the fly without retraining. In 2021, Yang et al.~\cite{Yang2021} proposed slimmable autoencoders that adjust their network width to accommodate different compression rates, making neural compression more adaptable and efficient. Zhu et al.~\cite{Zhu2022} in 2022 presented a unified entropy model using multivariate Gaussian mixtures to better model the distribution of latent representations, improving compression efficiency.

\noindent \textbf{JPEG AI scheme specification. }The astonishing results from the literature pushed in 2019 the JPEG Standardization Committee to launch the first image coding specifications based on end-to-end learning~\cite{JPEGAI_start}, namely the JPEG AI initiative.
At its core, JPEG AI follows a similar structure to standard neural image compression algorithms. In particular, we refer the reader to Fig.~\ref{fig:architecture}~\cite{JPEGAI_overview} for the overall JPEG AI standard scheme specification. 
The encoder consists of a \gls{nn}-based analysis transform that computes the latent information $\mathbf{y}$ of the input image $\mathbf{I}$.
Next, $\mathbf{y}$ is processed by a latent domain prediction module, generating a latent residual $\mathbf{\hat{r}}$, and by a hyperprior encoder, producing a hyperprior $\mathbf{\hat{z}}$, as proposed in the foundational work by Ballé et al.~\cite{ballé2018variational}. 
For entropy decoding, 
a latent prediction network retrieves $\mathbf{\hat{y}}$, which is finally input into the synthesis transform at the decoder for reconstructing the image $\mathbf{\tilde{I}}$.

As explained in \cref{sec:background}, JPEG AI enforces strict compliance requirements on manufacturers as a standardized format, limiting the customization options within the compression pipeline.
Although certain architectural modifications are permitted (indicated by the red modules of~\cref{fig:architecture}), such as adjustments to the analysis transform, to the hyperprior encoder network, or retraining for specific content types to improve performance, these changes are restricted. 
The modified latent space must still function within the official JPEG AI framework to ensure compatibility~\cite{JPEGAI_overview}.

\section{Experimental setup additional details}
\label{sec:supp:details}
\noindent \textbf{Experiments reproducibility.} All code and additional figures and tables for all detectors is available at the following URL \url{https://github.com/polimi-ispl/jpeg-ai-antifor}.

\noindent \textbf{Deepfake image detectors.} We provide some more information regarding the forensic detectors presented in \cref{sec:setup}:
\begin{itemize}
\item \cite{wang2020}: a ResNet-50~\cite{He2016}, pre-trained on the Imagenet dataset and then finetuned on a dataset of real and deepfake images generated through the ProGAN~\cite{progan} architecture. This detector is available in two variants, i.e., \cite{wang2020}-A, trained with an augmentation of Gaussian blurring and JPEG compression applied to each image with a probability of $10\%$, and \cite{wang2020}-B, trained with the same augmentation pipeline but applied with a probability of $50\%$;
\item \cite{gragnaniello2021}: another ResNet-50 trained with the same setup of ~\cite{wang2020}, however avoiding down-sampling of the features extracted in the first layer of the network;
\item \cite{Corvi_2023_ICASSP}: same solution as ~\cite{gragnaniello2021}, but in this case, the authors train the network on a dataset of Latent Diffusion~\cite{rombach2021highresolution} deepfake samples;
\item \cite{Ojha_2023_CVPR}: a frozen \gls{vit}~\cite{vit} pre-trained on the CLIP task, to which a final classification layer is finetuned with same dataset of \cite{wang2020};
\item \cite{cozzolino2023raising}: a similar approach to \cite{Ojha_2023_CVPR}, where the authors substitute the final classification layer with a \gls{svm}. The \gls{svm} is finetuned with only $1000$ real samples from the COCO datasets and $1000$ deepfakes generated with Latent Diffusion. In this case, too, we have two variants of the detector, i.e., \cite{cozzolino2023raising}-A, trained without augmentation, and \cite{cozzolino2023raising}-B, trained with compression and resize as data augmentation;
\item \cite{Mandelli2024}: an EfficientNet-B4~\cite{Tan2019efficientnet} performing a patch-based analysis of the image and aggregating multiple scores to decide the ``fakeness'' of the sample. This detector is trained exclusively on datasets of human faces, with deepfake samples coming from various generators, including \glspl{gan}, transformers, and diffusion models;
\item \cite{Tan_2024_CVPR}: a method focusing on detecting deepfakes by analyzing structural artifacts introduced during the up-sampling operations in generative networks like \glspl{gan} and diffusion models, namely \gls{npr}. These artifacts create local interdependencies among neighboring pixels in synthetic images. \gls{npr} captures and characterizes these relationships to identify forgeries effectively.
\end{itemize}

\noindent \textbf{Deepfake datasets.} To test the performances of every detector in normal conditions, we include for every pristine test set a deepfake ``counterpart''. These images have been selected by the original authors of the detectors to depict content similar to that of the pristine ones but are generated using different deepfake image generation techniques.
In total, we considered $20$ different generation techniques ranging from \glspl{gan} to diffusion models, namely Guided Diffusion~\cite{guided_diffusion}, Latent Diffusion~\cite{rombach2021highresolution}, Taming Transformers~\cite{Esser_2021_CVPR}, DALL·E~\cite{dalle} in various versions (i.e., mini~\cite{Dayma_DALLE_Mini_2021}, 2~\cite{dalle2}, and 3~\cite{dalle3}), GLIDE~\cite{glide}, Stable Diffusion~\cite{rombach2021highresolution} in various versions (i.e., 1.3~\cite{sd13}, 1.4~\cite{sd14}, XL~\cite{sdxl}, and 2~\cite{sd2}), Adobe Firefly~\cite{firefly}, MidJourney v5~\cite{midjourney}, BigGAN~\cite{brock2018large}, GauGAN~\cite{Park2019GauGANSI}, StyleGAN~\cite{Karras_2019_CVPR}, StyleGAN2~\cite{Karras2019stylegan2}, ProGAN~\cite{progan}, and StarGAN~\cite{choi2018stargan}. 
\cref{tab:techniques} report their relationships with the different pristine datasets used in our evaluation.

\begin{table}[t]
\centering
\caption{Deepfake image generation techniques used in our experiments and their relationships with pristine datasets' content.}
\resizebox{\columnwidth}{!}{
\begin{tabular}{l|lllllll}
\toprule
& \rotatebox{90}{CelebA} & \rotatebox{90}{COCO} & \rotatebox{90}{FFHQ} & \rotatebox{90}{Imagenet} & \rotatebox{90}{LAION} & \rotatebox{90}{LSUN} & \rotatebox{90}{RAISE} \\
\midrule
BigGAN~\cite{brock2018large} & & & & \checkmark & & & \\
GauGAN~\cite{Park2019GauGANSI} & & \checkmark & & & & & \\
ProGAN~\cite{progan} & & & & & & \checkmark &\\
StarGAN~\cite{choi2018stargan} & \checkmark & & &  & & & \\
StyleGAN~\cite{Karras_2019_CVPR} & & & & & & \checkmark & \\
StyleGAN2~\cite{Karras2019stylegan2} & & & \checkmark & & & \checkmark & \\
GLIDE~\cite{glide} & & \checkmark & & & \checkmark & & \checkmark \\
Guided Diffusion~\cite{guided_diffusion} & & & & \checkmark & \checkmark & \checkmark & \\
Latent Diffusion~\cite{rombach2021highresolution} & & \checkmark & \checkmark & \checkmark & & \checkmark & \\
Taming Transformers~\cite{Esser_2021_CVPR} & & & \checkmark & \checkmark & & & \\
DALL·E~\cite{dalle, Dayma_DALLE_Mini_2021, dalle2, dalle3} & & \checkmark & & & \checkmark & & \checkmark \\
Stable Diffusion~\cite{rombach2021highresolution, sd13, sd14, sd2, sdxl} & & \checkmark & & & & & \checkmark \\
Adobe Firefly~\cite{firefly} & & & & & & & \checkmark \\
MidJourney v5~\cite{midjourney} & & & & & & & \checkmark \\
\bottomrule
\end{tabular}
}
\label{tab:techniques}
\end{table}

\section{Deepfake image detection additional results}
\label{sec:supp:deepfake}


\begin{table}[t]
\centering
\caption{\gls{auc} results of the various detectors on all datasets considered for the real VS synthetic image detection task. Results below the $0.75$ threshold are reported in red. Entries signed with a $-$ indicate that the detector has not been tested on that particular dataset.}
\resizebox{\columnwidth}{!}{
\begin{tabular}{l|lllllll}
\toprule
& \multicolumn{7}{c}{Dataset} \\
Detector & \rotatebox{90}{LSUN} & \rotatebox{90}{Imagenet} & \rotatebox{90}{COCO} & \rotatebox{90}{CelebA} & \rotatebox{90}{FFHQ} & \rotatebox{90}{LAION} & \rotatebox{90}{RAISE} \\
\midrule
\cite{wang2020}-A      & $0.99$ & $0.88$ & $0.92$ & $0.98$ & $-$    & $-$    & $-$    \\
\cite{wang2020}-B      & $0.98$ & $0.90$ & $0.98$ & $0.95$ & $-$    & $-$    & $-$    \\
\cite{gragnaniello2021}    & $-$    & $0.89$ & $0.96$ & $-$    & $0.93$ & $-$    & $-$    \\
\cite{Corvi_2023_ICASSP} & $-$    & $0.87$ & $0.96$ & $-$    & $0.89$ & $-$    & $-$    \\
\cite{Ojha_2023_CVPR}    & $0.98$ & $0.99$ & $1.00$ & $0.99$ & $-$    & $0.96$ & $-$    \\
\cite{cozzolino2023raising}-A & $0.95$ & $0.97$     & $0.97$ & $-$      & $0.96$ & $0.96$                        & $0.80$  \\
\cite{cozzolino2023raising}-B & $0.79$ & $0.83$     & $0.89$ & $-$      & $0.88$ & \textcolor{red}{$0.74$} & $0.86$  \\
\cite{Tan_2024_CVPR} & $0.94$ & $0.91$ & $1.00$ & $1.00$ & $-$ & $0.99$ & $-$ \\
\cite{Mandelli2024} & $0.86$ & $0.87$ & \textcolor{red}{$0.71$} & $1.00$    & $1.00$ & $0.98$ & $0.83$ \\
\bottomrule
\end{tabular}
}
\label{tab:syn_detection}
\end{table}

\noindent \textbf{Preliminary analysis on synthetic detection.} \cref{tab:syn_detection} reports the results for the simple case of synthetic image detection \textit{without} JPEG AI image compression of the test samples. We report the complete numbers for all detectors on all datasets. 

We tested each detector only on the datasets on which the original authors first proposed the method. \cite{Mandelli2024} represents the only exception since the authors reported good generalization capabilities on various semantic content despite training their solution only on human faces.
As the reader can quickly inspect, all detectors achieve acceptable performance, with only two cases below the $0.75$ threshold for the \gls{auc}, i.e., \cite{cozzolino2023raising}-B on LAION and \cite{Mandelli2024} on COCO.

\noindent \textbf{JPEG AI quality analysis.} \cref{tab:supp:jpegai_quality_full} reports a complete quality analysis of the pristine JPEG AI compressed samples. To achieve this, we computed the following metrics:
\begin{itemize}
    \item Feature Similarity (FSIM)~\cite{Zhang2011FSIM}: a metric based on phase congruency and gradient magnitude to assess quality. The value range for FS is between 0 and 1, with 1 indicating the highest quality;
    \item Multi-Scale SIMilarity index (MS-SSIM)~\cite{Wang2003MSSSIM}: evaluates image quality by comparing details across resolutions. Typical values range between $0$ and $1$, with $1$ indicating the highest quality;
    \item Normalized Laplacian Pyramid (NLP)~\cite{Laparra2016PLP}: based on local luminance and contrast, it uses a Laplacian pyramid decomposition. We normalize the scores in a $0-1$ range, with $1$ indicating higher quality;
    \item Visual Information Fidelity (VIF)~\cite{Sheikh2004IIVQ}: measures the loss of perceived information using natural scene statistics. Values range from $0$ to infinity, with higher scores indicating better quality;
    \item Video Multimethod Assessment Fusion (VMAF)~\cite{LiPracticalPerceptual}: developed by Netflix, it assesses quality by fusing scores from multiple algorithms. Scores range from $0$ to $100$, with high values indicating better quality. 
\end{itemize}
All these metrics have been suggested in the JPEG AI Call For Evidence~\cite{jpegai_ce}, and we computed them using the official JPEG AI Reference Software~\cite{jpegai_vm}. The samples present an overall high perceptual quality as the reader can quickly inspect, even at low \gls{bpp} values.


\begin{table*}
\caption{JPEG AI Call For Evidence~\cite{jpegai_ce} quality metrics for our dataset of JPEG AI images. For clarity's sake, we abbreviate the metrics as follows: FSIM as FS, MS-SSIM as MS, NLP as NL, VIF as VI, and VMAF as VF.}
\centering
\resizebox{\textwidth}{!}{
\begin{tabular}{l|cccccc}
\toprule
 & \multicolumn{1}{c}{BPP = 0.12} & \multicolumn{1}{c}{BPP = 0.25} & 
 \multicolumn{1}{c}{BPP = 0.5} &
 \multicolumn{1}{c}{BPP = 0.75} & \multicolumn{1}{c}{BPP = 1.0} & \multicolumn{1}{c}{BPP = 2.0} \\
 \midrule
Dataset & \multicolumn{1}{c |}{FS / MS / NL / VI / VF} & \multicolumn{1}{c |}{FS / MS / NL / VI / VF} & \multicolumn{1}{c |}{FS / MS / NL / VI / VF} &
\multicolumn{1}{c |}{FS / MS / NL / VI / VF} &
\multicolumn{1}{c |}{FS / MS / NL / VI / VF} & \multicolumn{1}{c |}{FS / MS / NL / VI / VF} \\
\midrule
$\text{Imagenet}$ & $36.57 / 0.96 / 0.42 / 36.27 / 76.65$ & $38.36 / 0.98 / 0.54 / 38.00 / 83.71$ & $40.63 / 0.99 / 0.64 / 39.90 / 88.92$ & $42.30 / 1.00 / 0.69 / 41.28 / 91.25$ & $43.55 / 1.00 / 0.74 / 42.32 / 92.71$ & $46.44 / 1.00 / 0.83 / 44.92 / 95.25$ \\
$\text{COCO}$ & $35.41 / 0.95 / 0.37 / 35.13 / 73.91$ & $36.95 / 0.98 / 0.48 / 36.69 / 81.23$ & $38.76 / 0.99 / 0.58 / 38.36 / 87.13$ & $40.21 / 0.99 / 0.64 / 39.32 / 89.65$ & $41.42 / 1.00 / 0.68 / 40.41 / 91.26$ & $44.00 / 1.00 / 0.78 / 42.64 / 94.26$ \\
$\text{FFHQ}$ & $41.65 / 0.98 / 0.52 / 42.61 / 86.91$ & $44.06 / 0.99 / 0.63 / 44.61 / 91.37$ & $47.31 / 0.99 / 0.72 / 46.96 / 93.27$ & $48.39 / 1.00 / 0.79 / 47.96 / 95.35$ & $49.66 / 1.00 / 0.84 / 49.21 / 96.11$ & $52.37 / 1.00 / 0.91 / 51.65 / 97.52$ \\
$\text{LSUN}$ & $37.39 / 0.96 / 0.39 / 36.95 / 74.96$ & $39.25 / 0.98 / 0.49 / 38.76 / 81.61$ & $42.26 / 0.99 / 0.60 / 41.32 / 87.69$ & $43.79 / 0.99 / 0.65 / 42.51 / 89.92$ & $45.43 / 1.00 / 0.69 / 43.82 / 91.55$ & $48.92 / 1.00 / 0.79 / 47.39 / 94.39$ \\
$\text{RAISE}$ & $39.27 / 0.96 / 0.40 / 37.50 / 78.93$ & $41.04 / 0.98 / 0.52 / 39.35 / 85.44$ & $42.78 / 0.97 / 0.52 / 40.38 / 83.17$ & $44.87 / 1.00 / 0.69 / 42.71 / 92.39$ & $46.01 / 1.00 / 0.73 / 43.69 / 93.67$ & $48.95 / 1.00 / 0.80 / 46.62 / 95.67$ \\
$\text{LAION}$ & $39.82 / 0.98 / 0.50 / 39.83 / 82.71$ & $42.06 / 0.99 / 0.61 / 42.08 / 88.74$ & $45.47 / 1.00 / 0.71 / 44.76 / 92.64$ & $47.93 / 1.00 / 0.77 / 46.76 / 94.39$ & $49.56 / 1.00 / 0.81 / 47.99 / 95.42$ & $53.63 / 1.00 / 0.87 / 52.01 / 96.79$ \\
$\text{CelebA}$ & $38.64 / 0.98 / 0.58 / 39.76 / 85.49$ & $41.63 / 0.99 / 0.72 / 42.45 / 91.35$ & $46.28 / 1.00 / 0.81 / 45.62 / 94.37$ & $48.52 / 1.00 / 0.86 / 47.75 / 95.86$ & $50.56 / 1.00 / 0.90 / 49.81 / 96.67$ & $54.28 / 1.00 / 0.95 / 53.33 / 97.86$ \\
\bottomrule
\end{tabular}
}
\label{tab:supp:jpegai_quality_full}
\vspace{-10pt}
\end{table*}

\begin{figure*}[htb!]
	\centering
    
        \subfloat[\centering JPEG AI scores distribution over the Imagenet dataset.]{
            \includegraphics[width=\columnwidth]{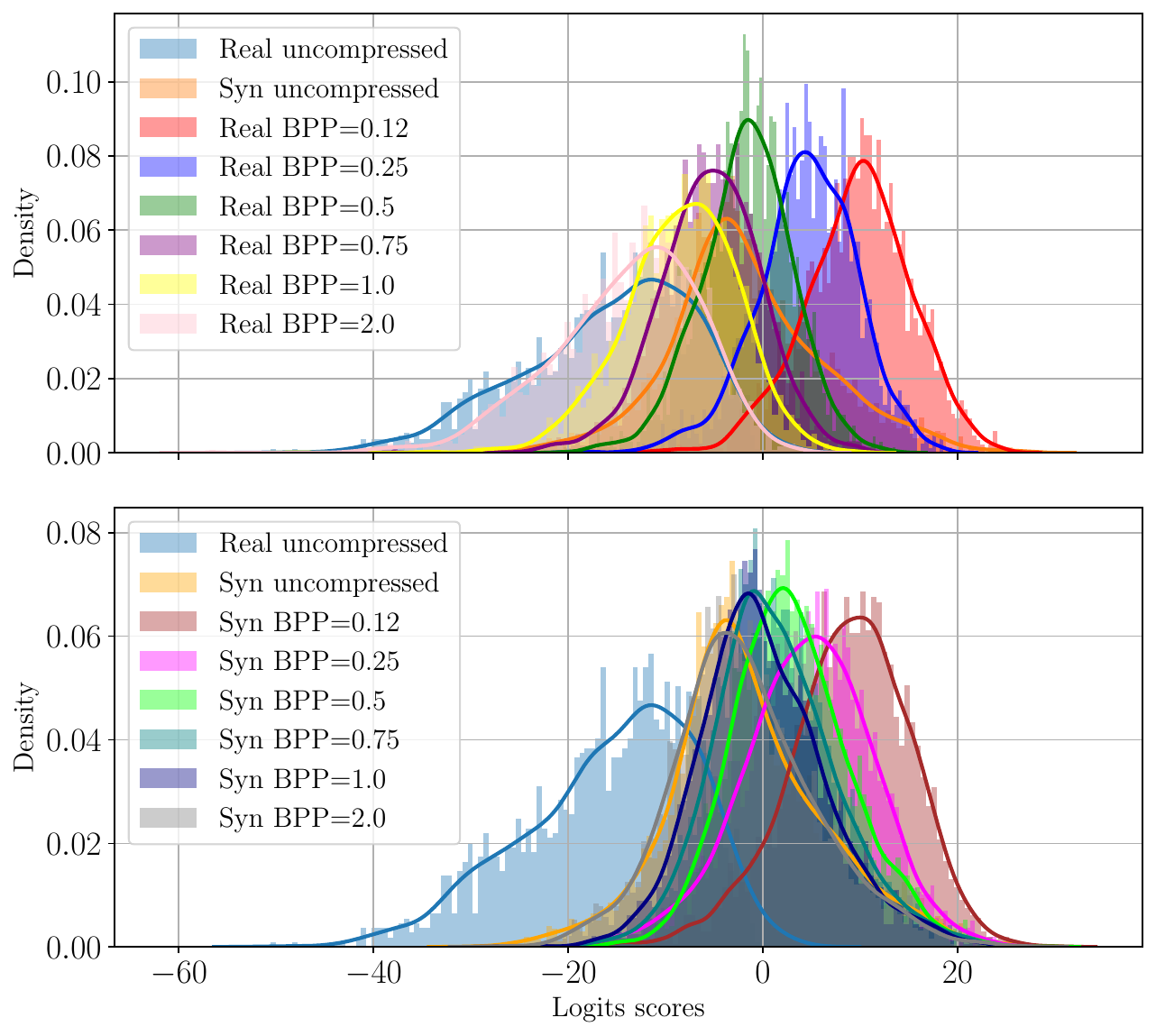}		\label{fig:supp:grag2021:imagenet:jpegai}}
        \hfil
        \subfloat[\centering JPEG scores distribution over the Imagenet dataset.]{
            \includegraphics[width=\columnwidth]{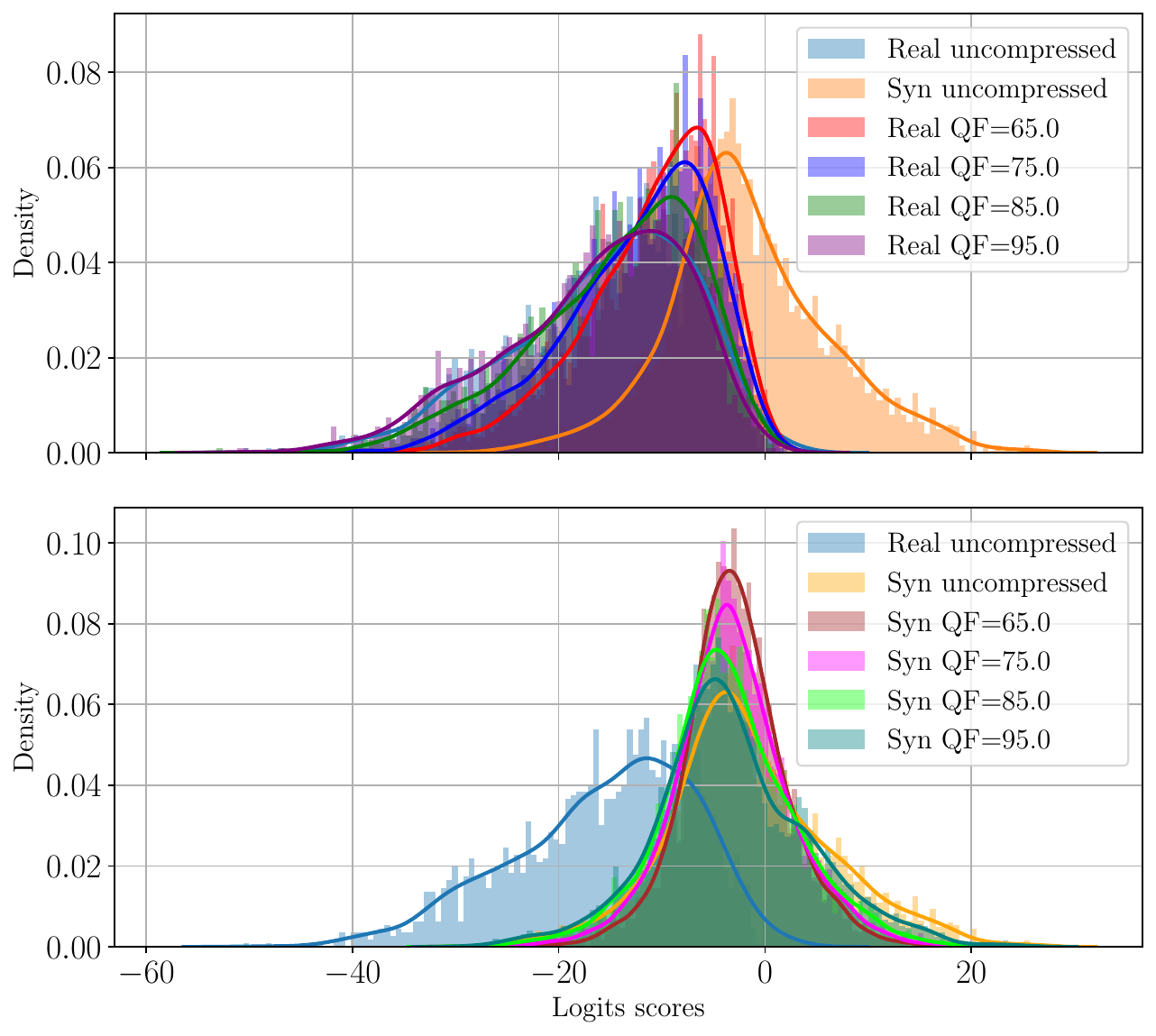}
            \label{fig:supp:grag2021:imagenet:jpeg}}
        
        \subfloat[\centering JPEG AI scores distribution over the COCO dataset.]{
            \includegraphics[width=\columnwidth]{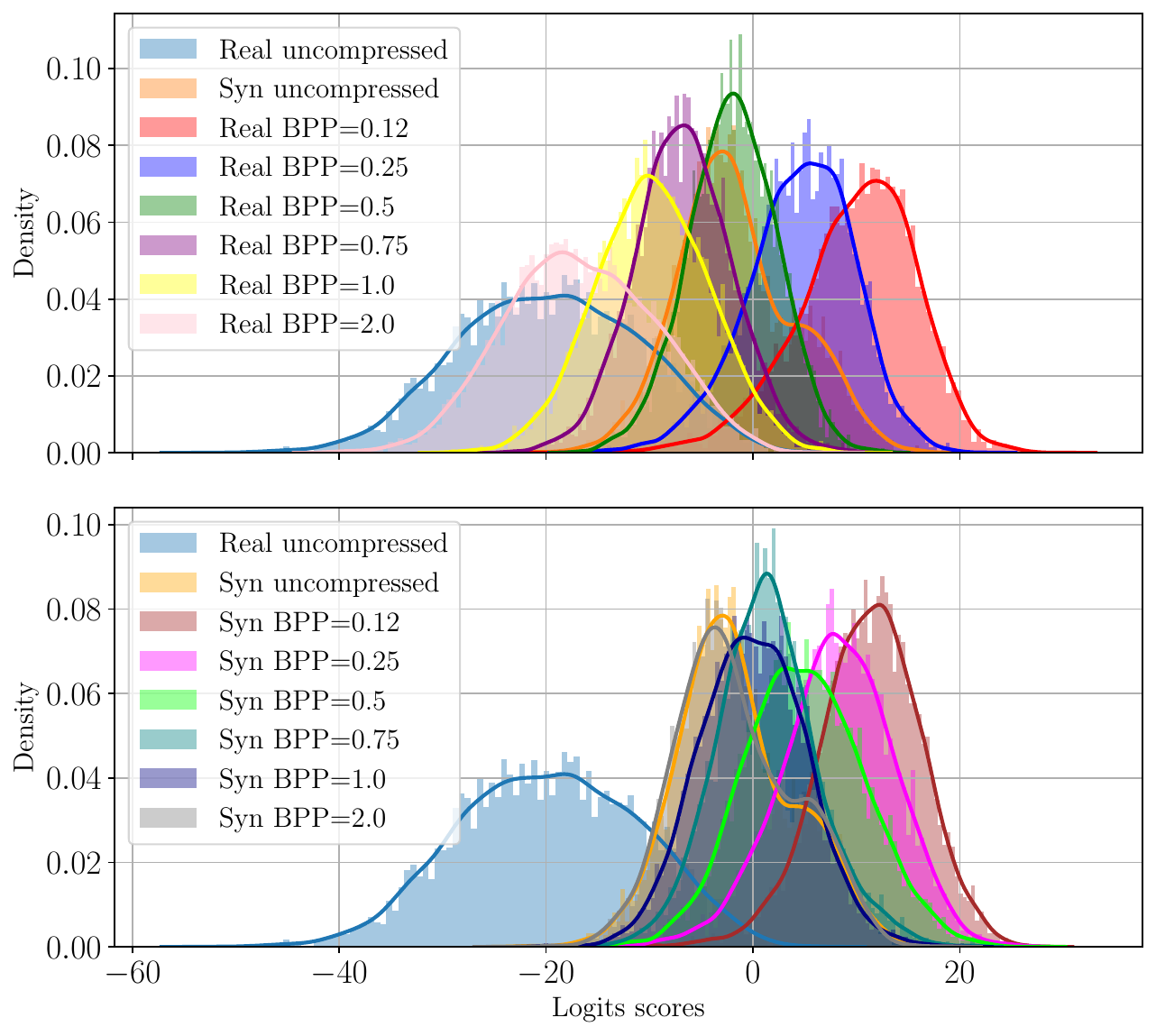}		\label{fig:supp:grag2021:coco:jpegai}}
        \hfil
        \subfloat[\centering JPEG scores distribution over the COCO dataset.]{
            \includegraphics[width=\columnwidth]{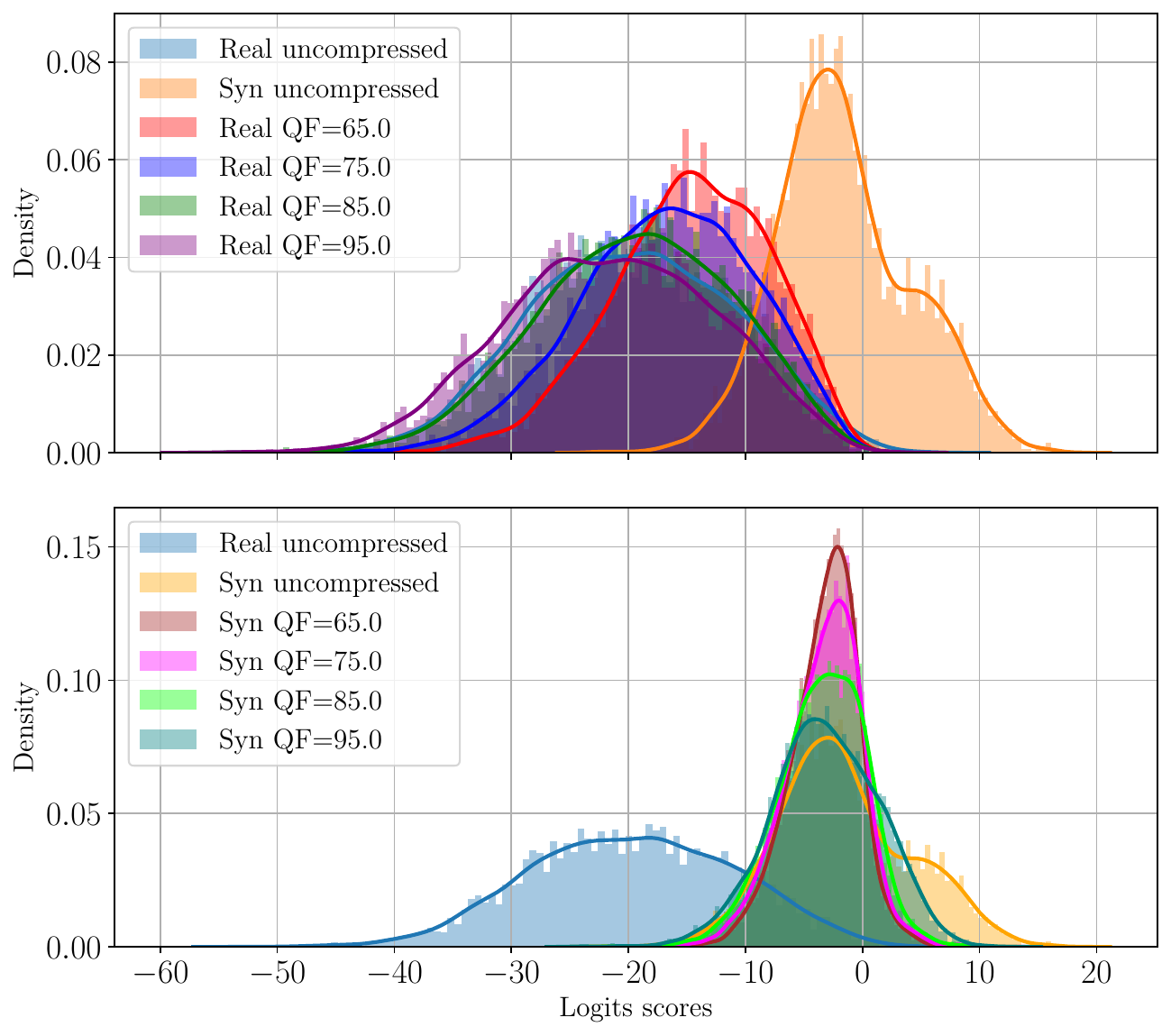}
            \label{fig:supp:grag2021:coco:jpeg}}

        \subfloat[\centering JPEG AI scores distribution over the FFHQ dataset.]{
            \includegraphics[width=\columnwidth]{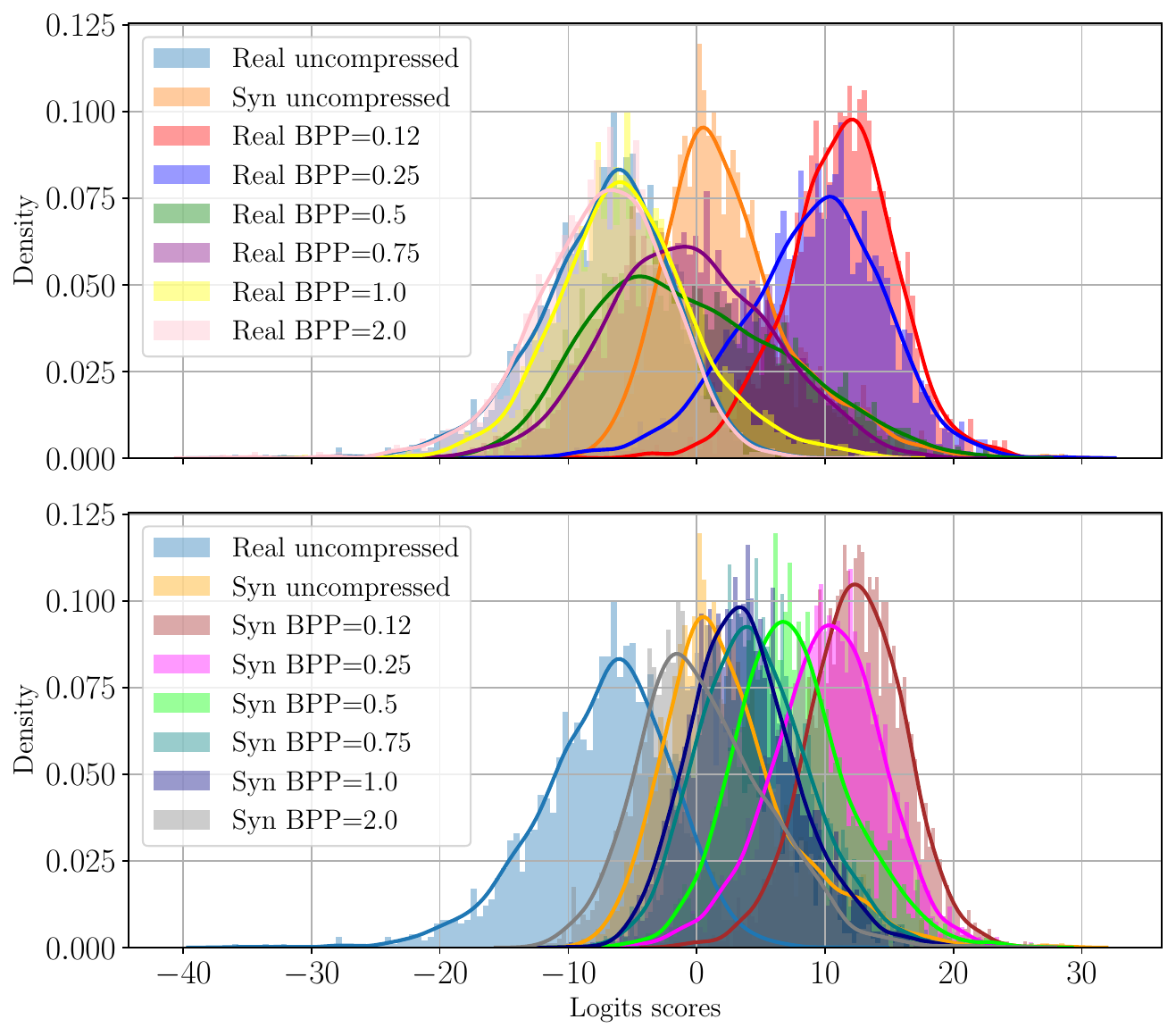}		\label{fig:supp:grag2021:ffhq:jpegai}}
        \hfil
        \subfloat[\centering JPEG scores distribution over the FFHQ dataset.]{
            \includegraphics[width=\columnwidth]{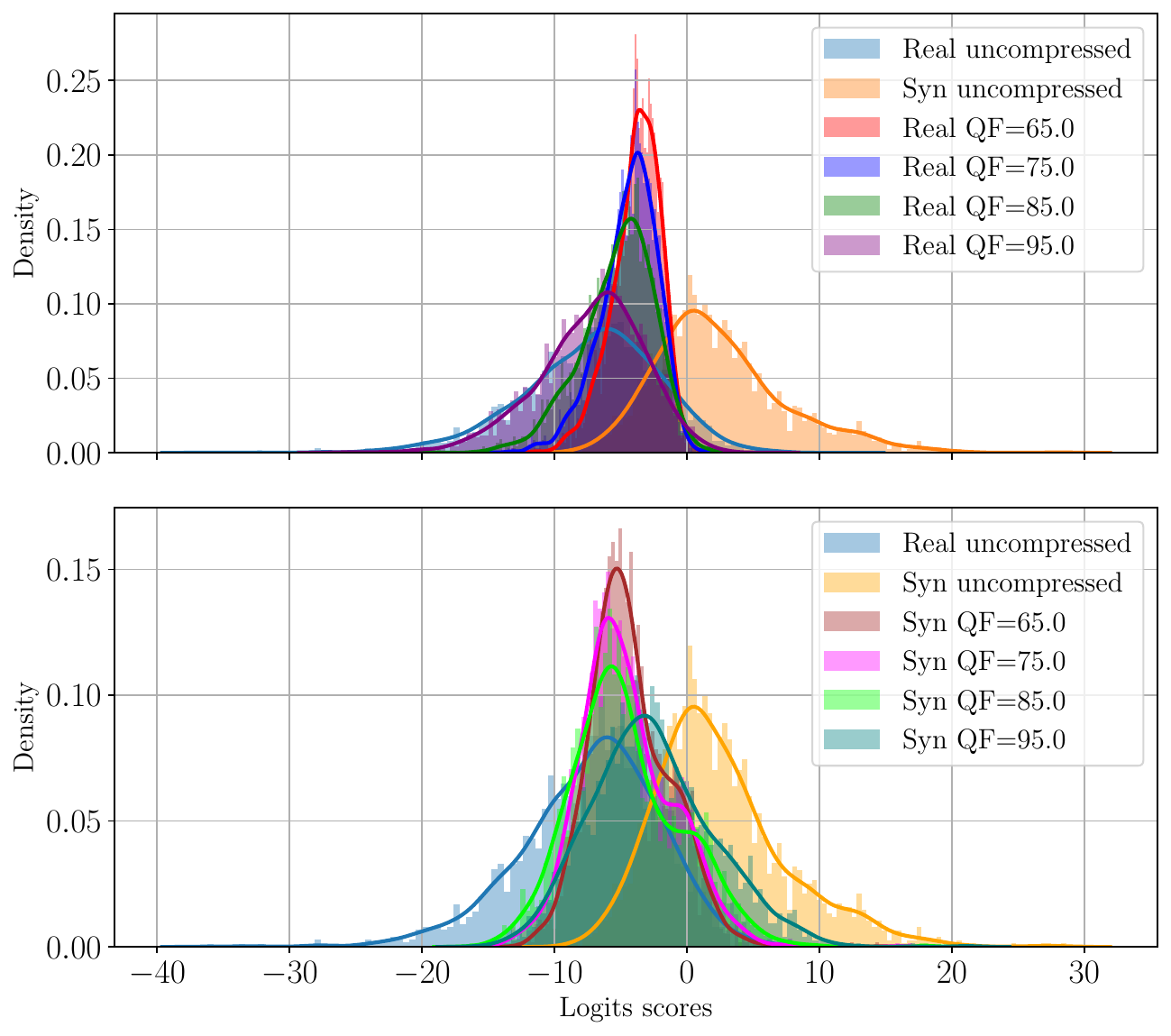}
            \label{fig:supp:grag2021:ffhq:jpeg}}

        \caption{Scores distribution over the different datasets of the \cite{gragnaniello2021} detector.}
	\label{fig:supp:grag2021}
\end{figure*}

\begin{figure*}[htb!]
	\centering
    
        \subfloat[\centering JPEG AI scores distribution over the Imagenet dataset.]{
            \includegraphics[width=\columnwidth]{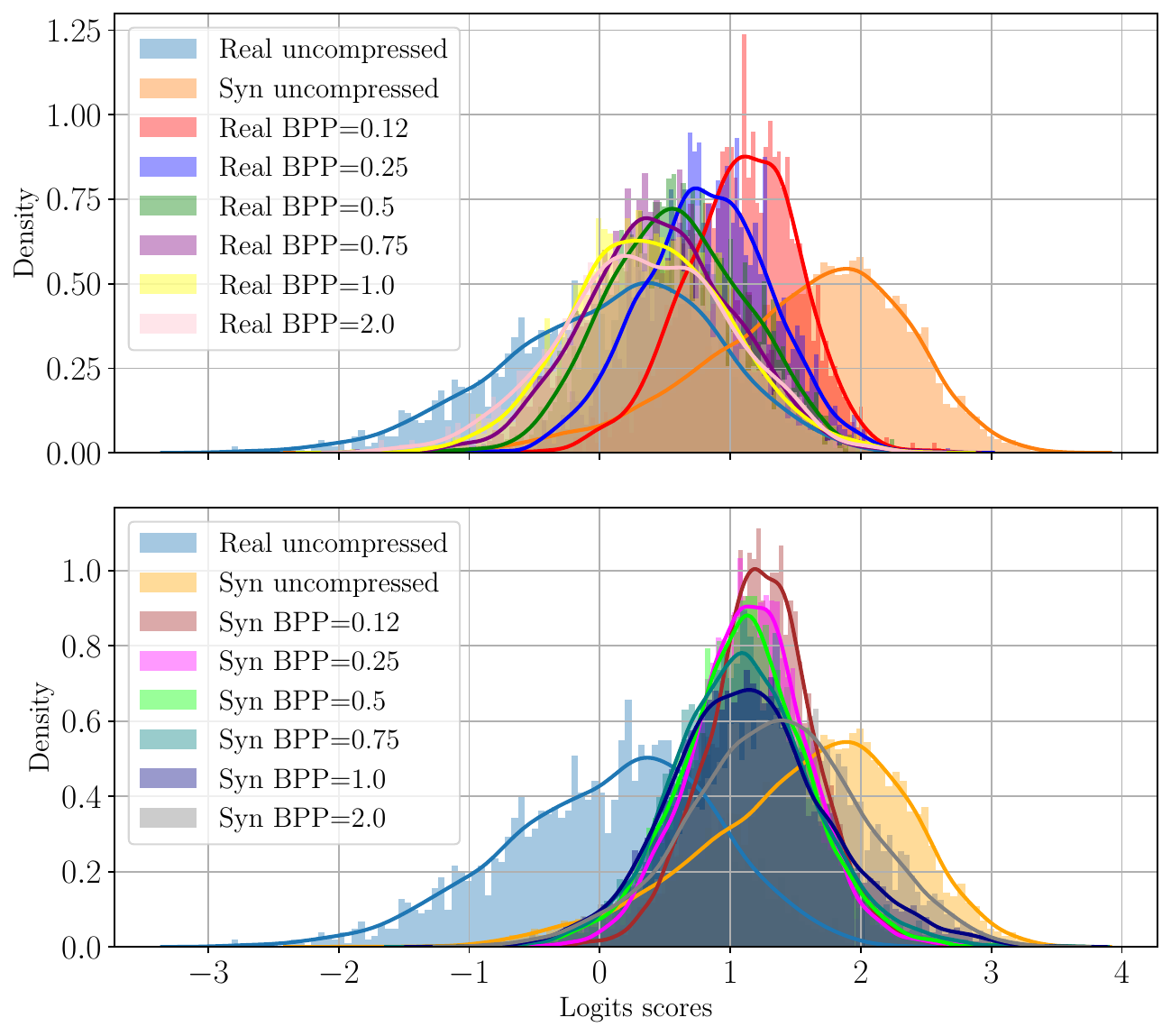}		\label{fig:supp:clip2024a:imagenet:jpegai}}
        \hfil
        \subfloat[\centering JPEG scores distribution over the Imagenet dataset.]{
            \includegraphics[width=\columnwidth]{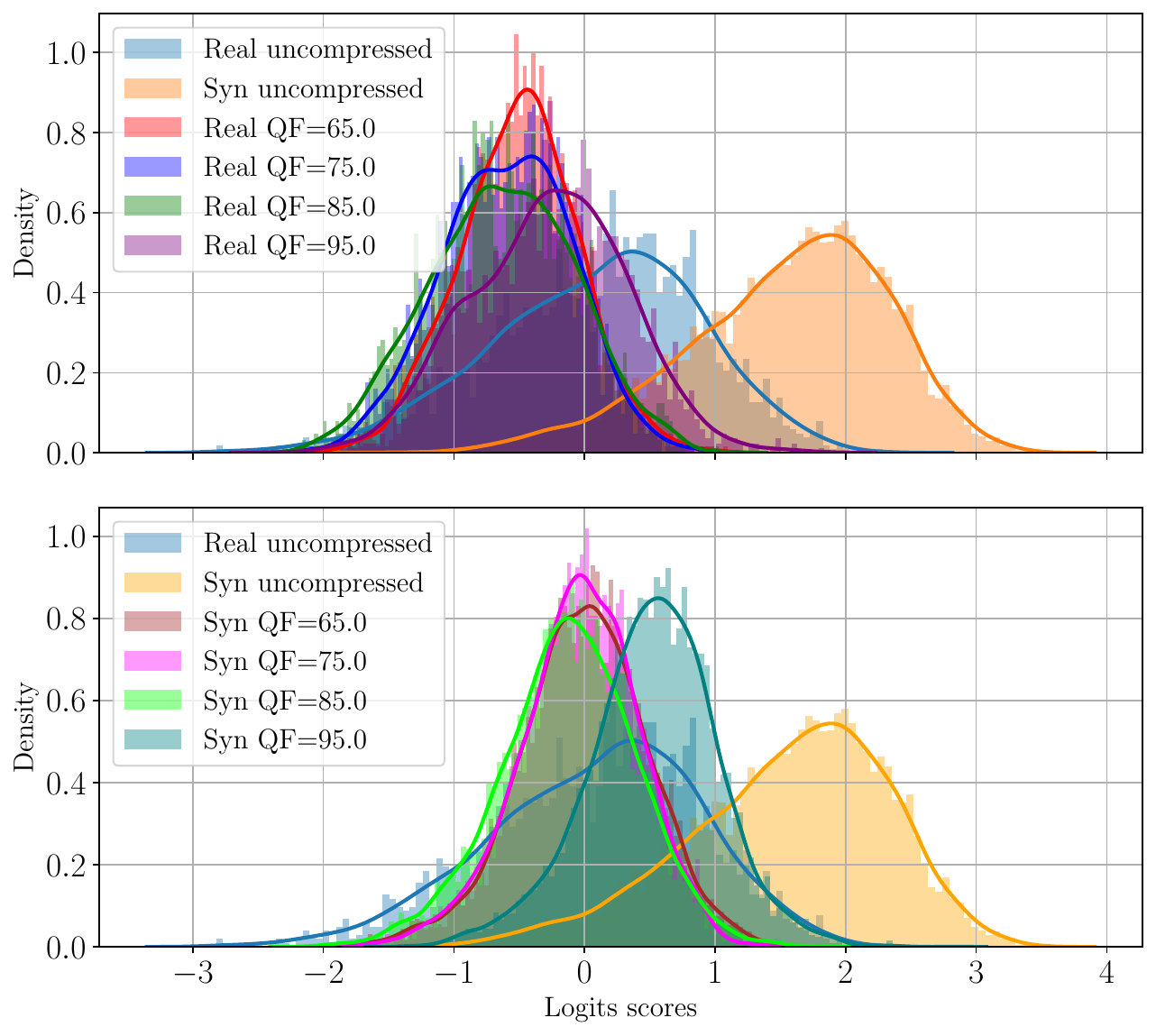}
            \label{fig:supp:clip2024a:imagenet:jpeg}}
        
        \subfloat[\centering JPEG AI scores distribution over the COCO dataset.]{
            \includegraphics[width=\columnwidth]{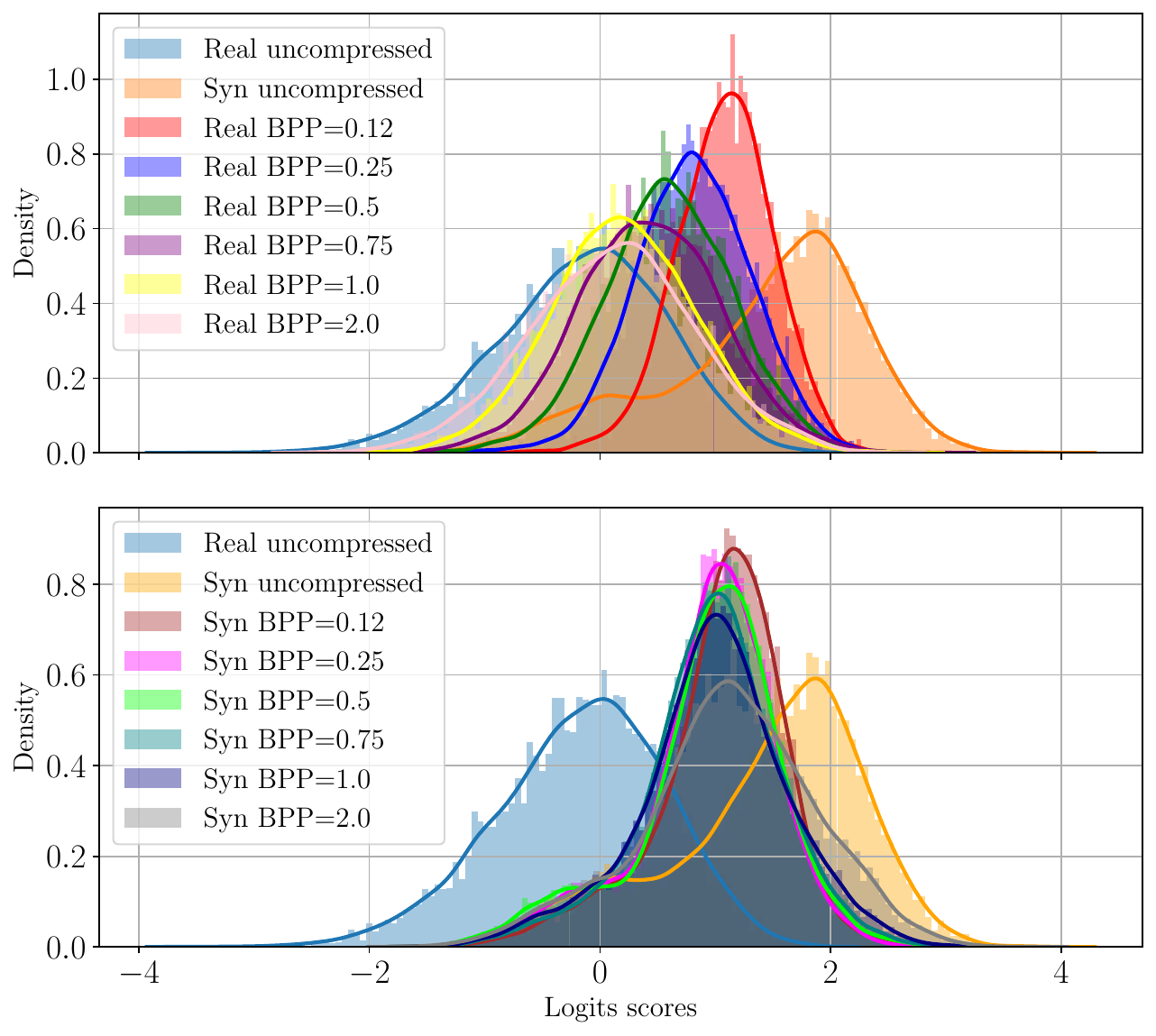}		\label{fig:supp:clip2024a:coco:jpegai}}
        \hfil
        \subfloat[\centering JPEG scores distribution over the COCO dataset.]{
            \includegraphics[width=\columnwidth]{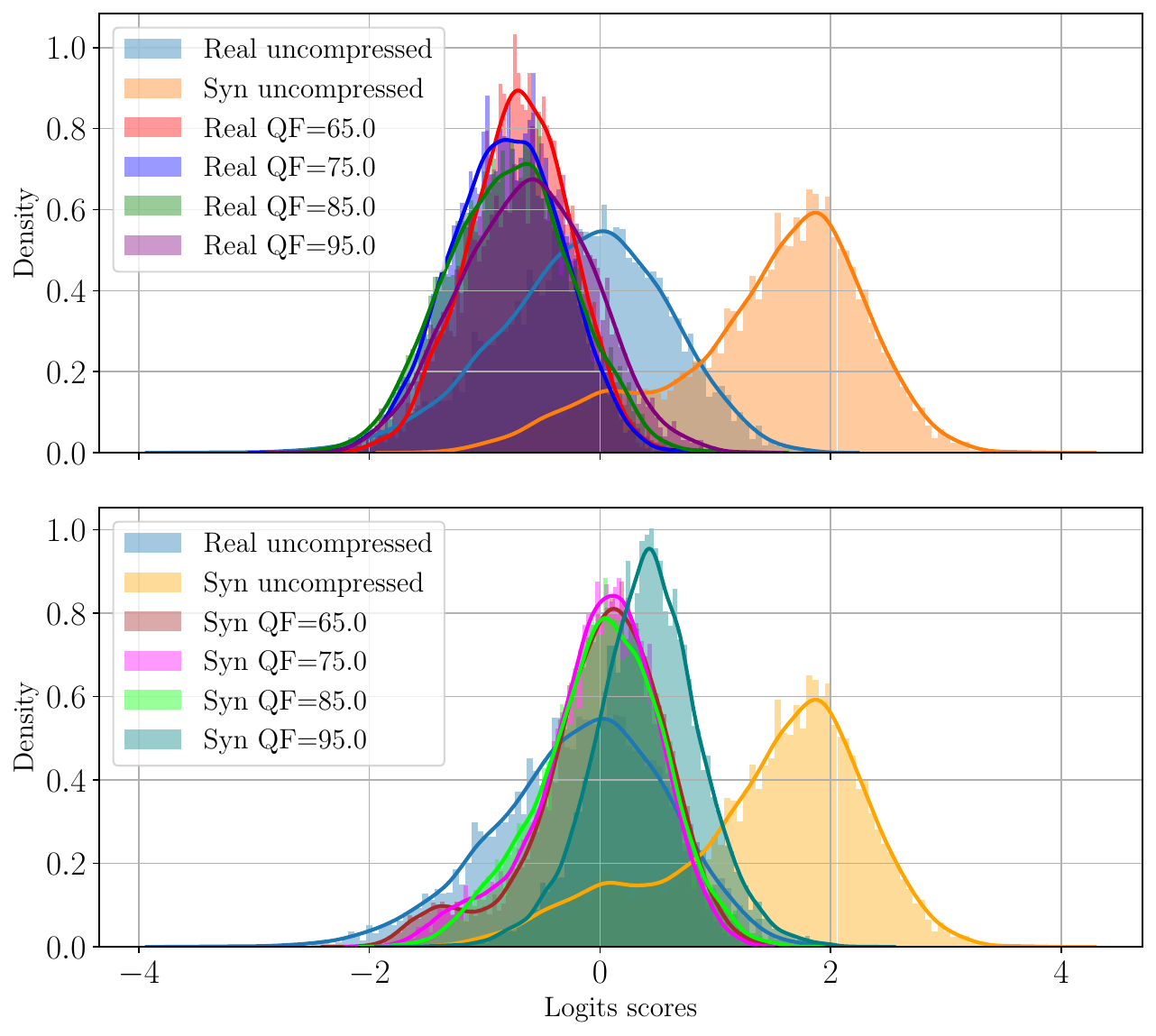}
            \label{fig:supp:clip2024a:coco:jpeg}}

        \subfloat[\centering JPEG AI scores distribution over the FFHQ dataset.]{
            \includegraphics[width=\columnwidth]{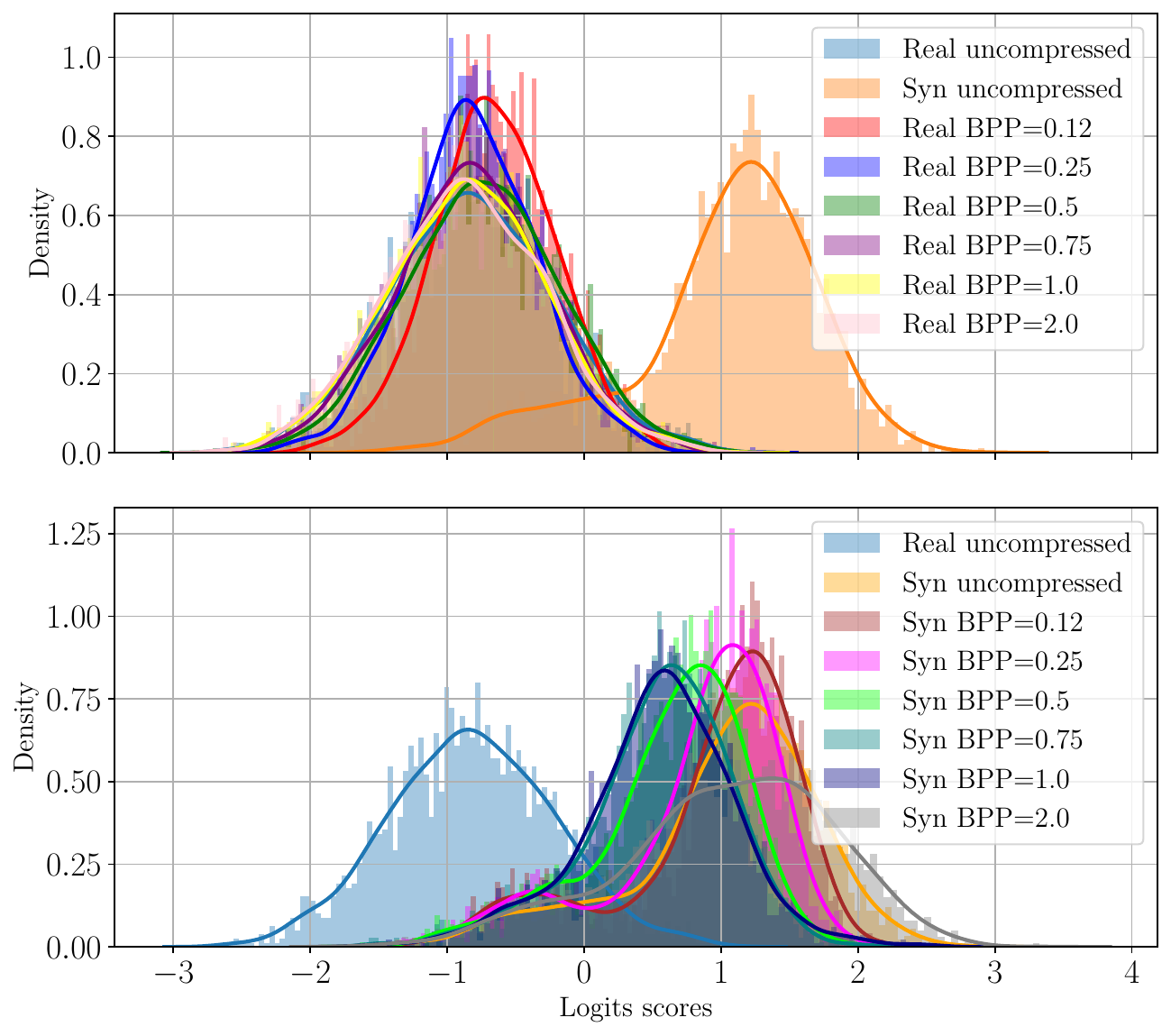}		\label{fig:supp:clip2024a:ffhq:jpegai}}
        \hfil
        \subfloat[\centering JPEG scores distribution over the FFHQ dataset.]{
            \includegraphics[width=\columnwidth]{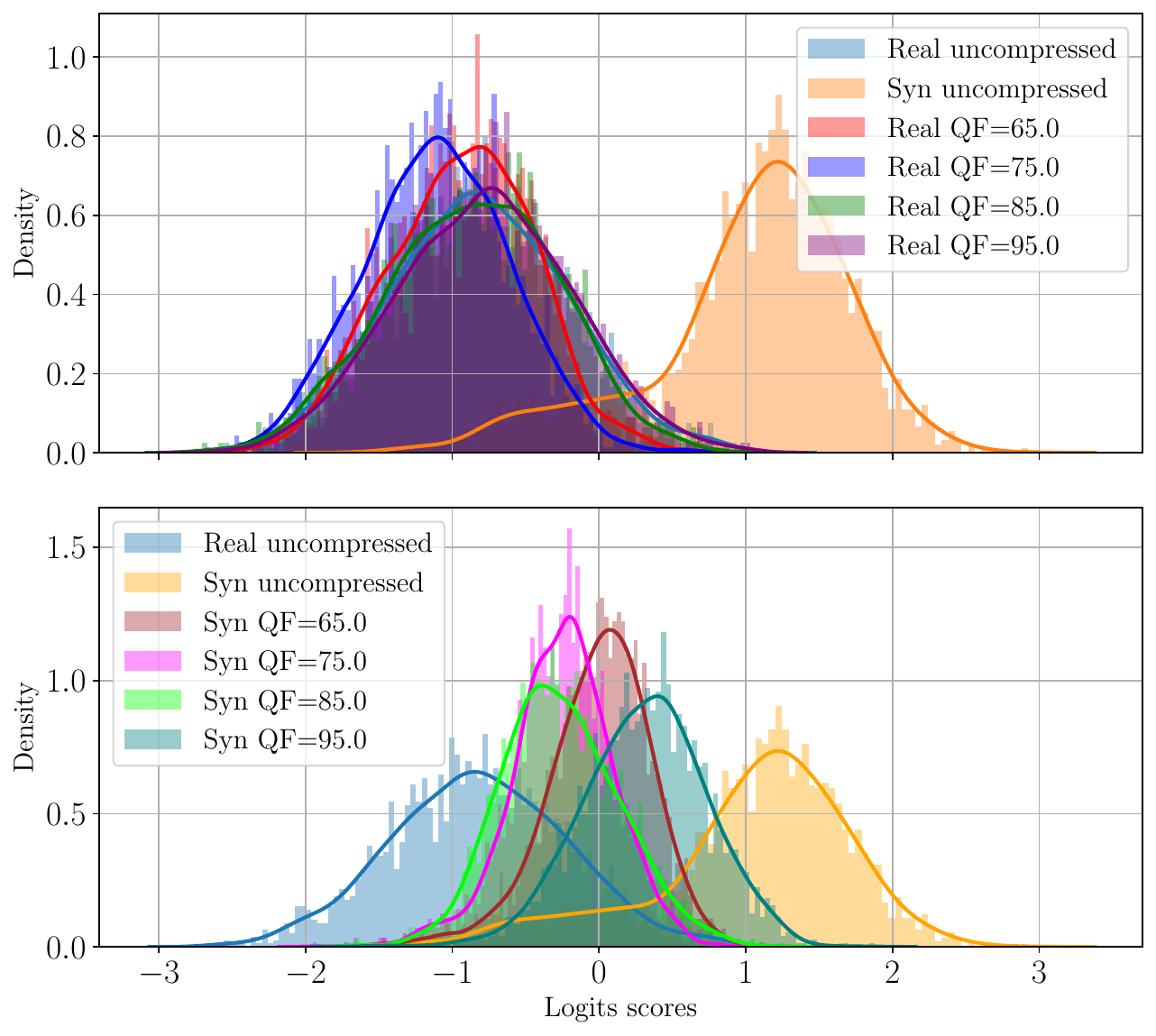}
            \label{fig:supp:clip2024a:ffhq:jpeg}}

    \caption{Scores distribution over the ImageNet, COCO, and FFHQ datasets of the \cite{cozzolino2023raising}-A detector.}
\label{fig:supp:clip2024a:1}

\end{figure*}

\begin{figure*}[htb!]
    \centering

        \subfloat[\centering JPEG AI scores distribution over the LSUN dataset.]{
            \includegraphics[width=\columnwidth]{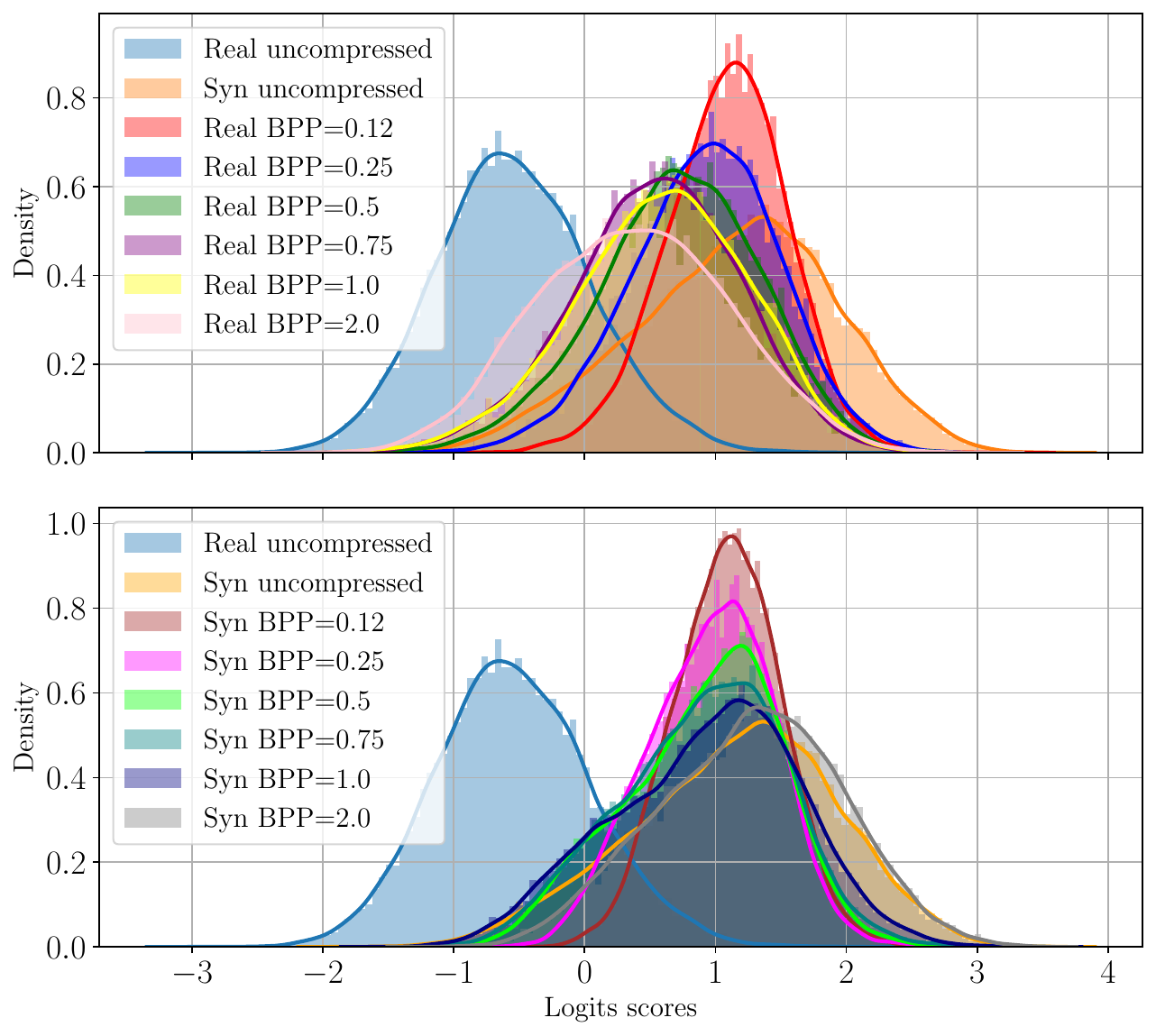}		\label{fig:supp:clip2024a:lsun:jpegai}}
        \hfil
        \subfloat[\centering JPEG scores distribution over the LSUN dataset.]{
            \includegraphics[width=\columnwidth]{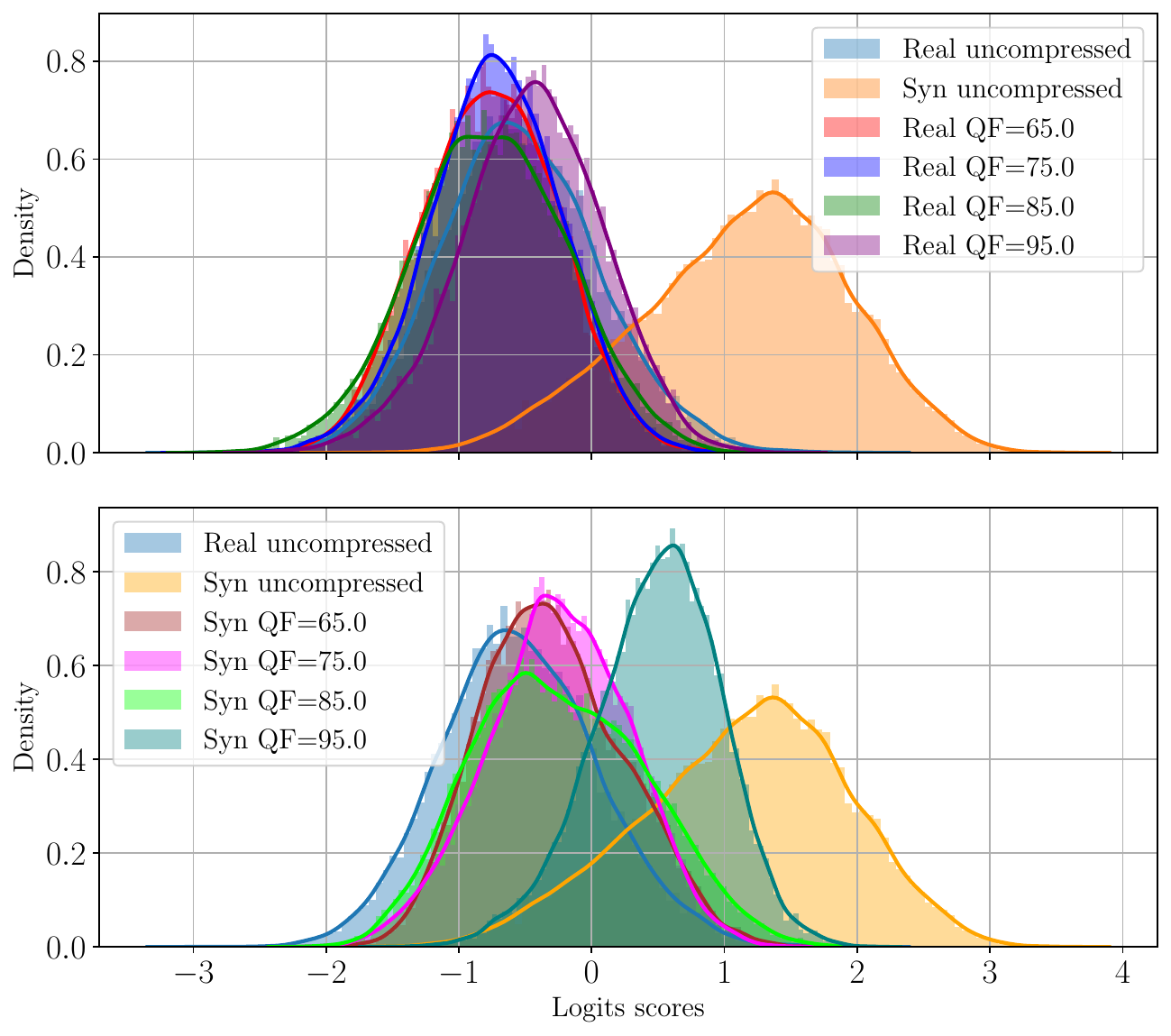}
            \label{fig:supp:clip2024a:lsun:jpeg}}

        \subfloat[\centering JPEG AI scores distribution over the LAION dataset.]{
            \includegraphics[width=\columnwidth]{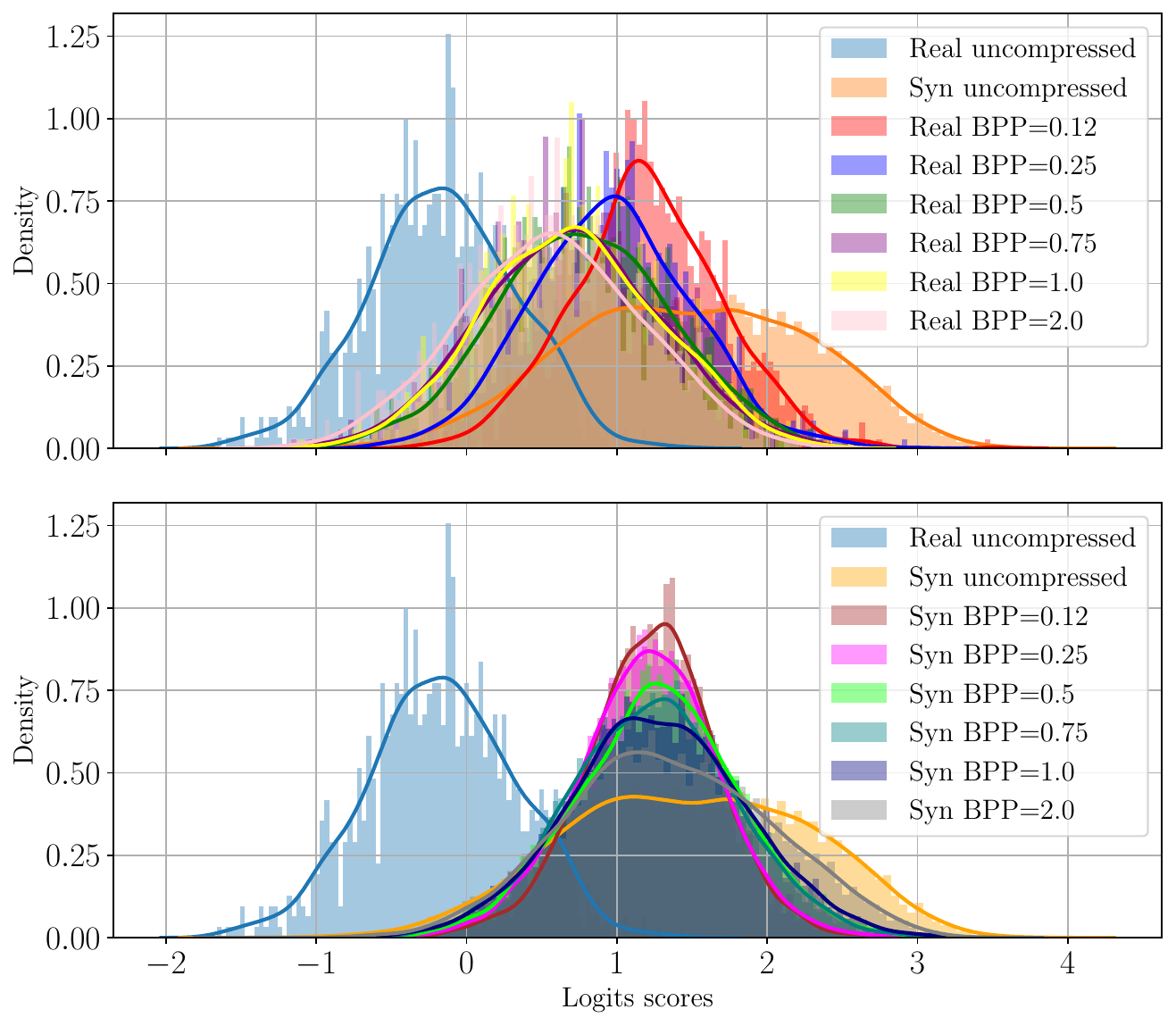}		\label{fig:supp:clip2024a:laion:jpegai}}
        \hfil
        \subfloat[\centering JPEG scores distribution over the LAION dataset.]{
            \includegraphics[width=\columnwidth]{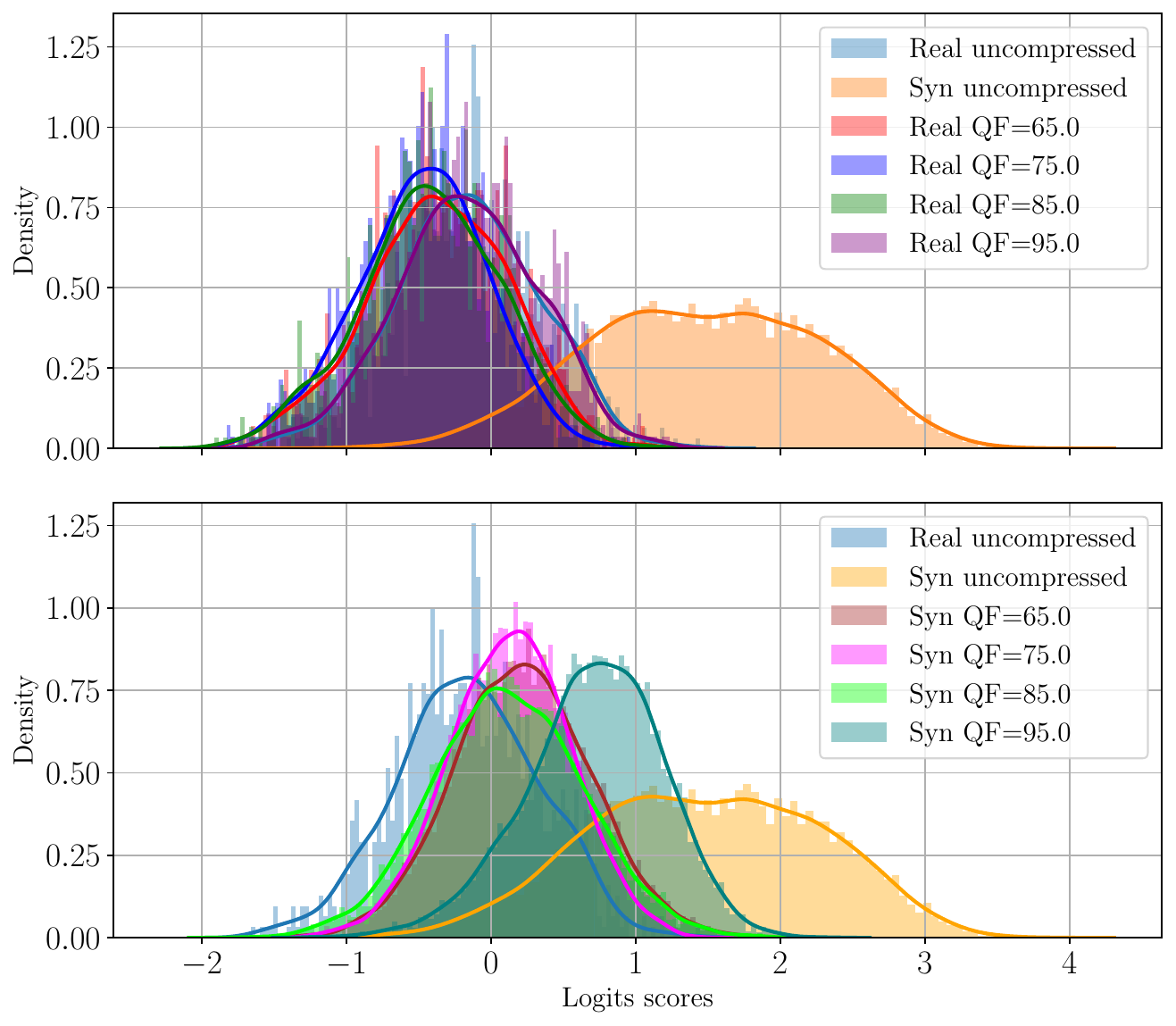}
            \label{fig:supp:clip2024a:laion:jpeg}}

        \subfloat[\centering JPEG AI scores distribution over the RAISE dataset.]{
            \includegraphics[width=\columnwidth]{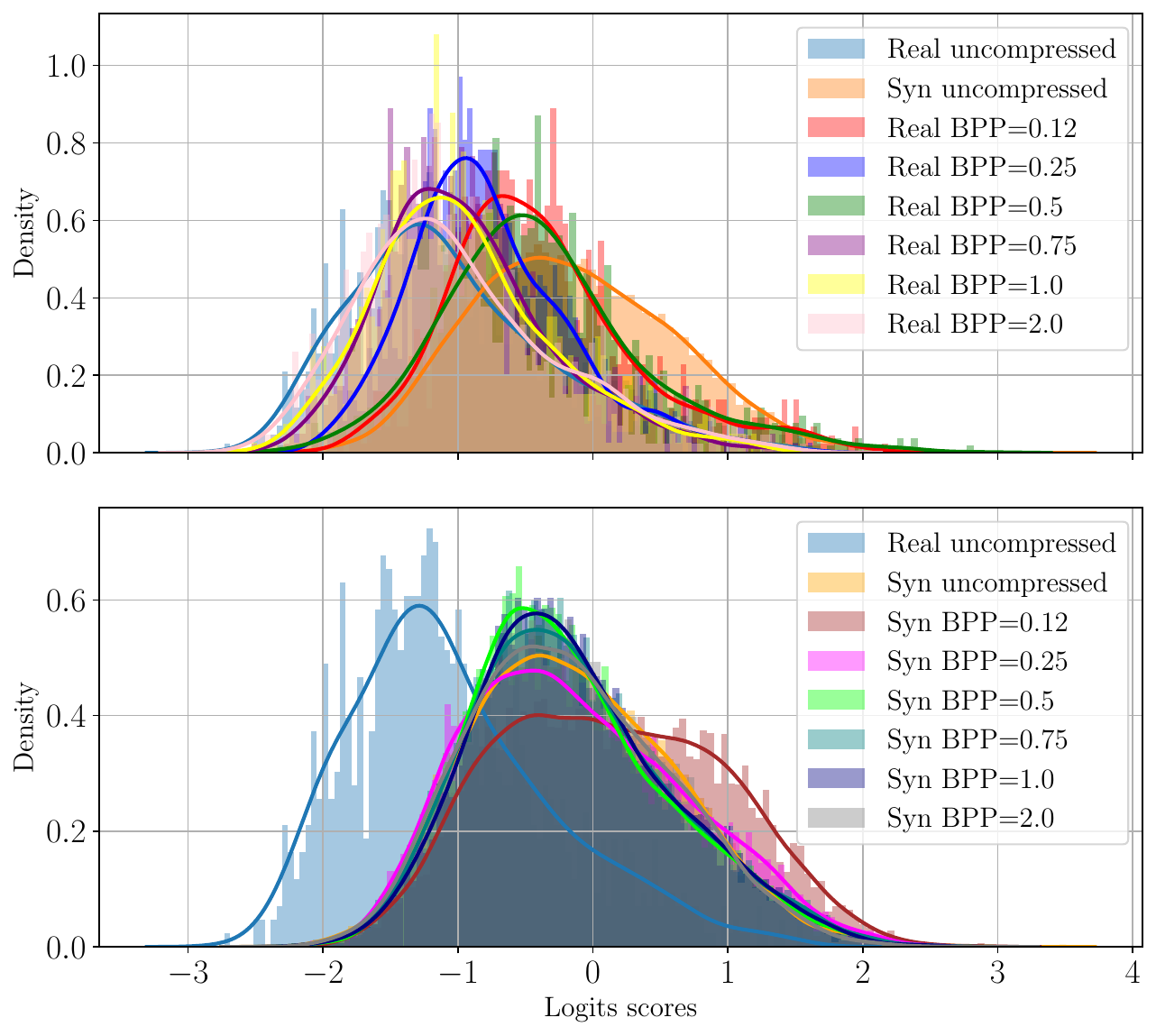}		\label{fig:supp:clip2024a:raise:jpegai}}
        \hfil
        \subfloat[\centering JPEG scores distribution over the RAISE dataset.]{
            \includegraphics[width=\columnwidth]{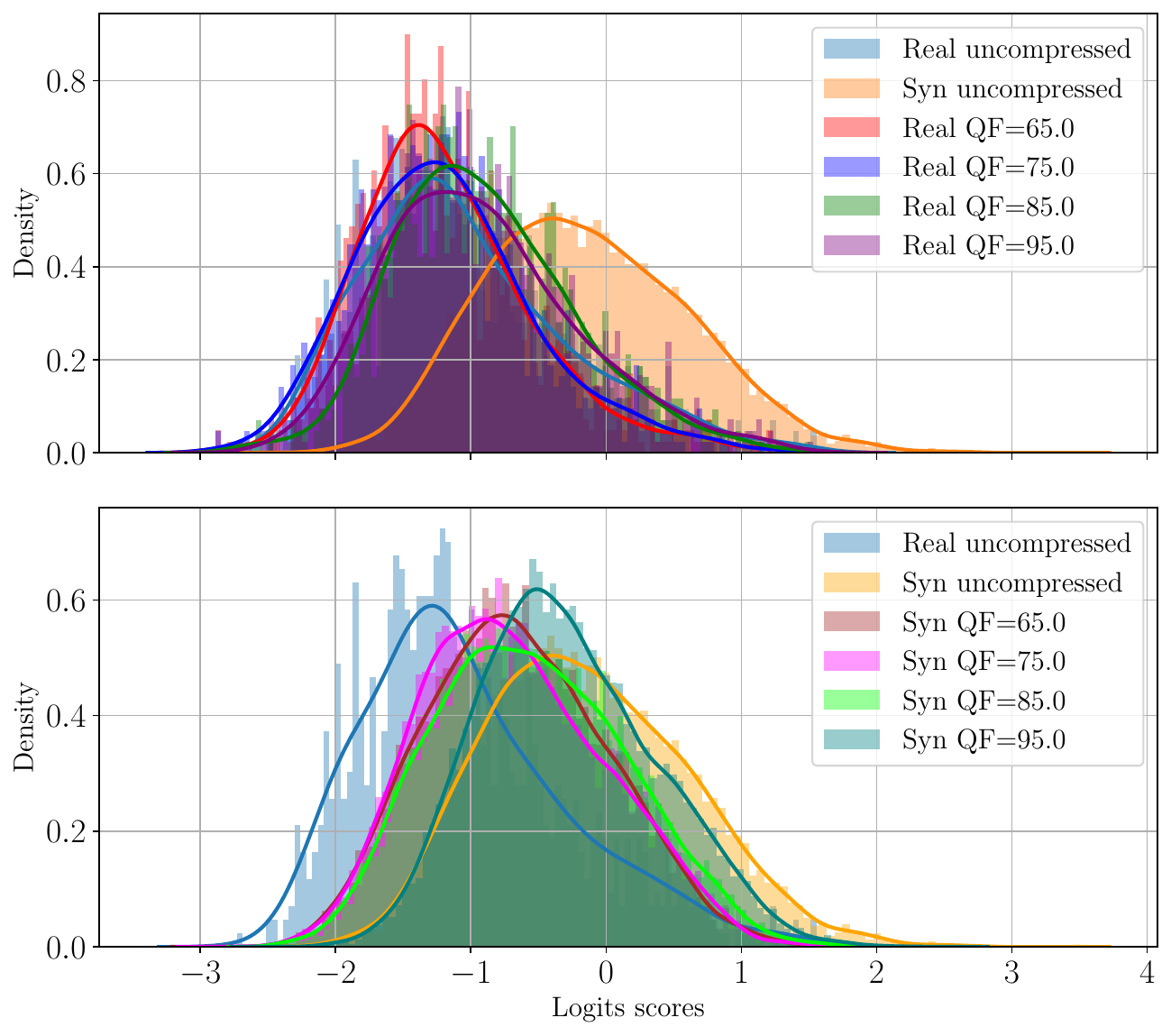}
            \label{fig:supp:clip2024a:raise:jpeg}}

        \caption{Scores distribution over the LSUN, LAION, and RAISE datasets of the \cite{cozzolino2023raising}-A detector.}
	\label{fig:supp:clip2024a:2}
\end{figure*}

\begin{figure*}[htb!]
	\centering
    
        \subfloat[\centering JPEG AI scores distribution over the Imagenet dataset.]{
            \includegraphics[width=\columnwidth]{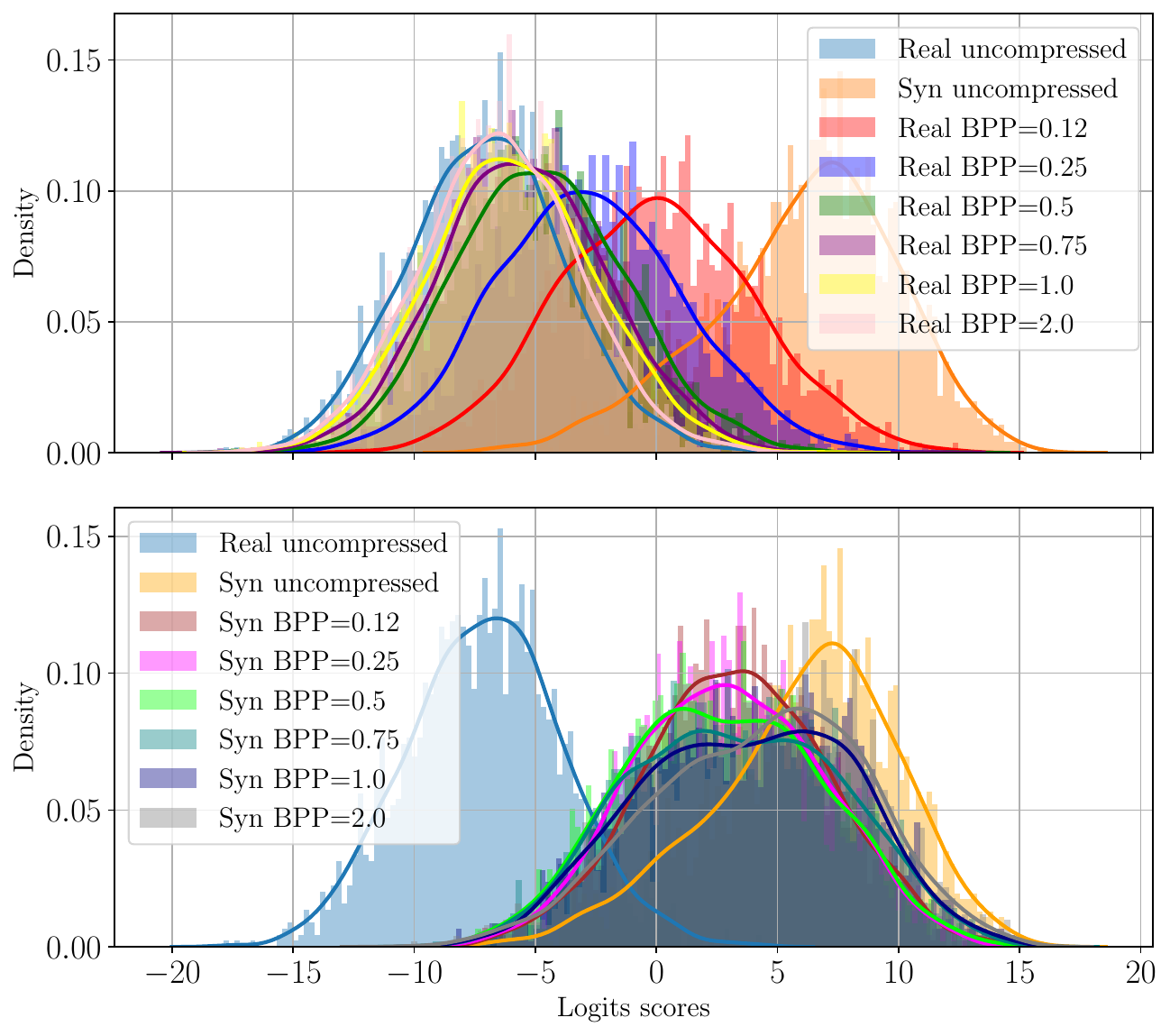}		\label{fig:supp:ohja2023:imagenet:jpegai}}
        \hfil
        \subfloat[\centering JPEG scores distribution over the Imagenet dataset.]{
            \includegraphics[width=\columnwidth]{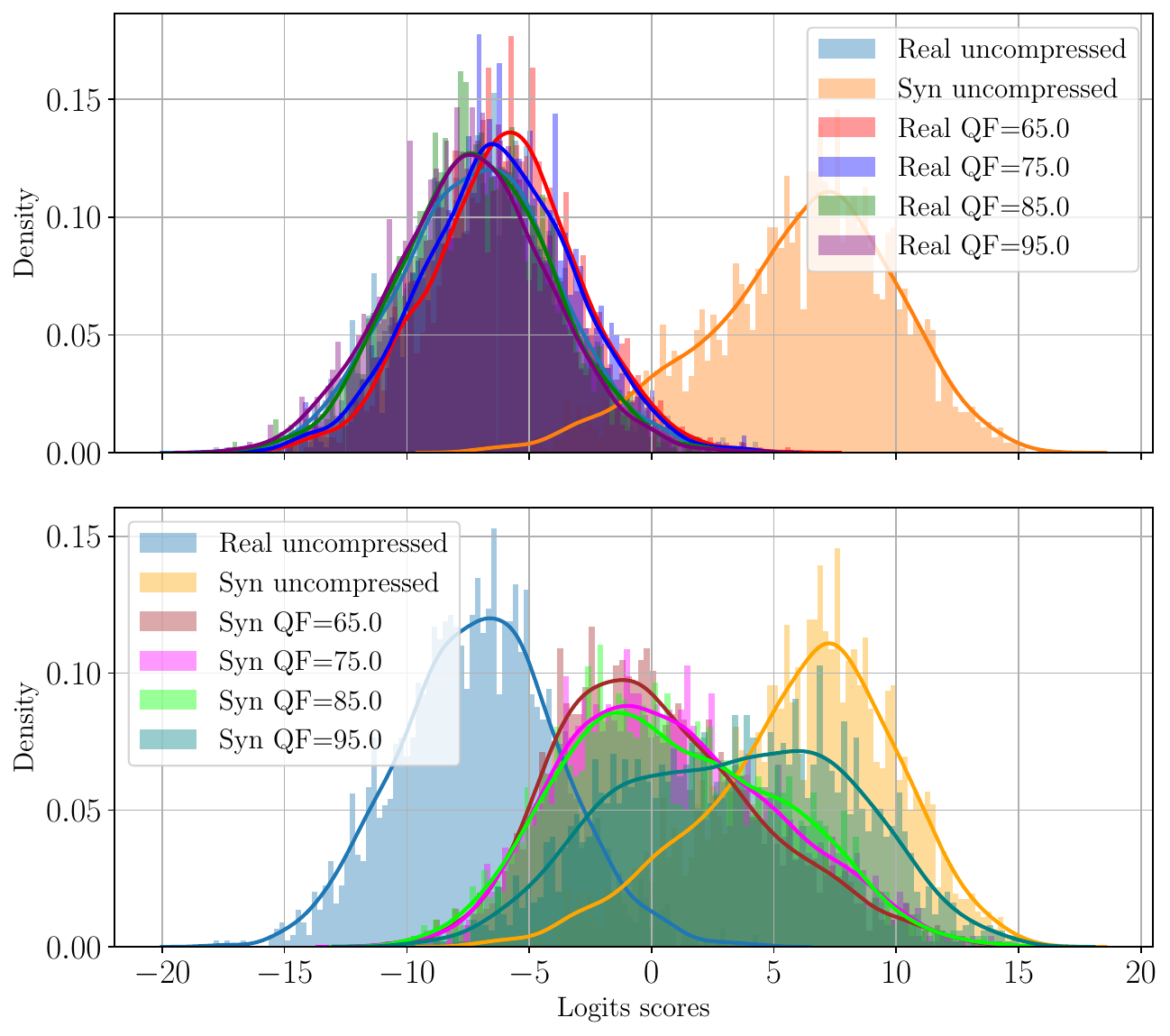}
            \label{fig:supp:ohja2023:imagenet:jpeg}}
        
        \subfloat[\centering JPEG AI scores distribution over the COCO dataset.]{
            \includegraphics[width=\columnwidth]{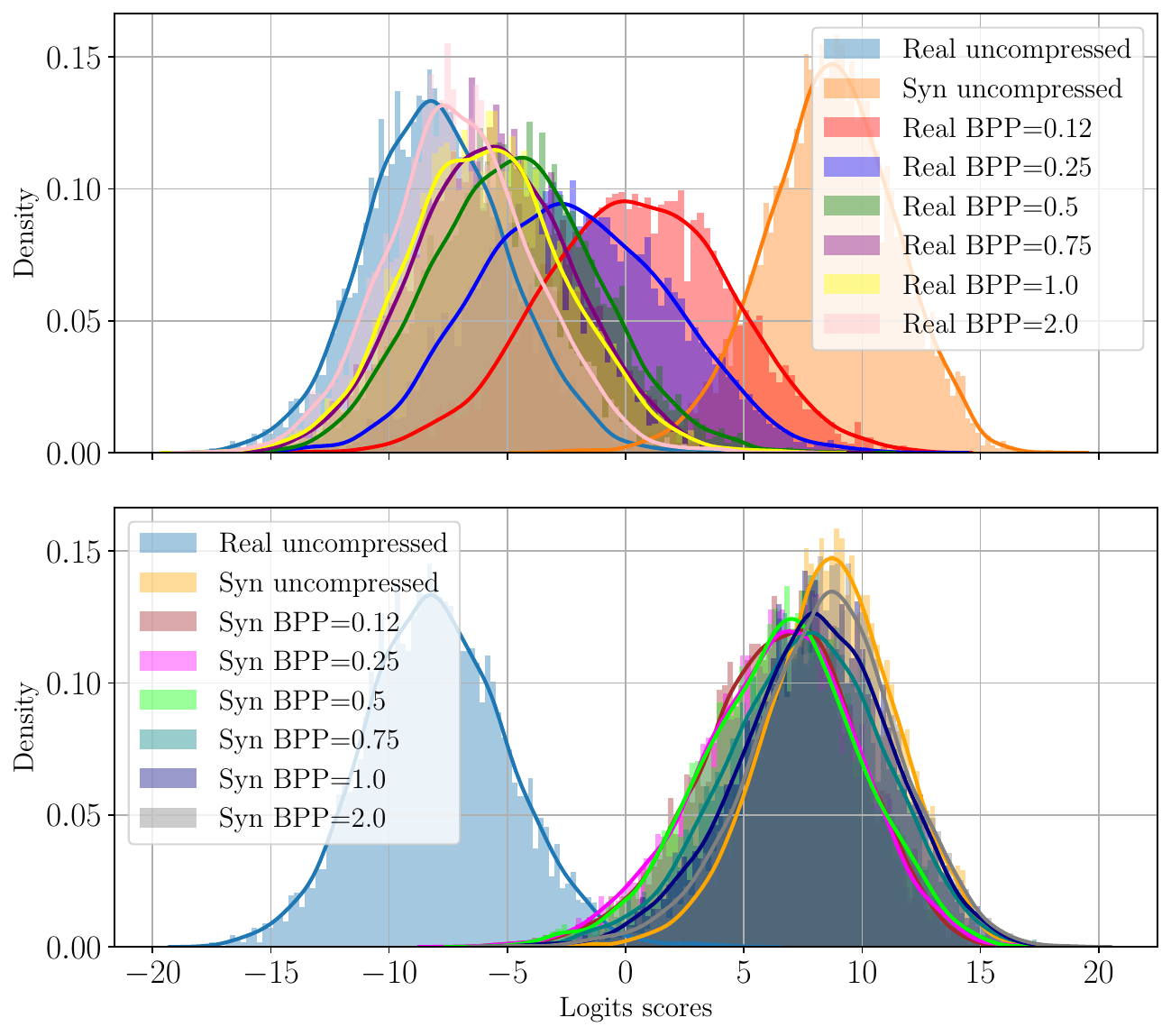}		\label{fig:supp:ohja2023:coco:jpegai}}
        \hfil
        \subfloat[\centering JPEG scores distribution over the COCO dataset.]{
            \includegraphics[width=\columnwidth]{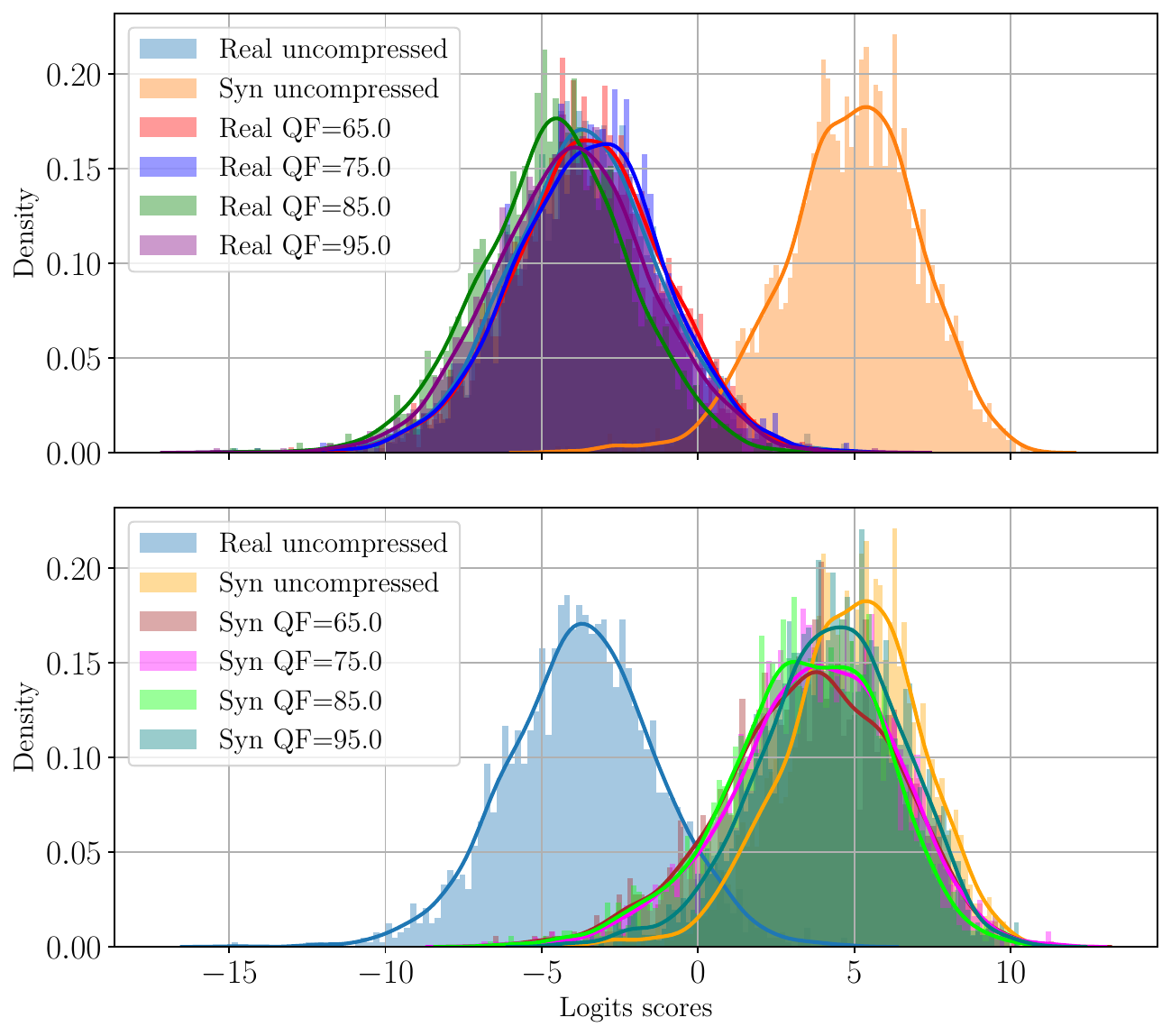}
            \label{fig:supp:ohja2023:coco:jpeg}}

        \subfloat[\centering JPEG AI scores distribution over the CelebA dataset.]{
            \includegraphics[width=\columnwidth]{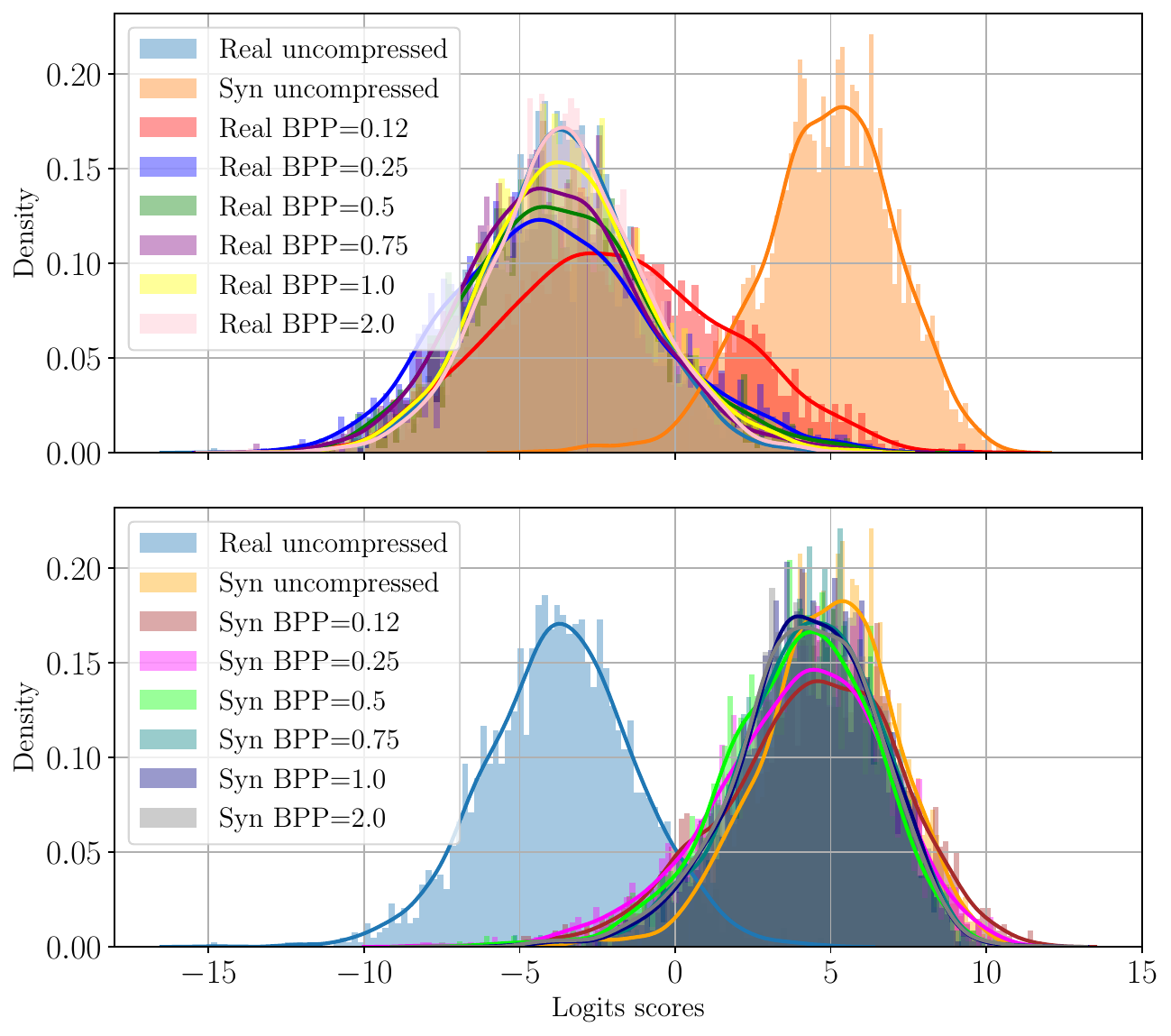}		\label{fig:supp:ohja2023:celeba:jpegai}}
        \hfil
        \subfloat[\centering JPEG scores distribution over the CelebA dataset.]{
            \includegraphics[width=\columnwidth]{figures/supp_material/Ojha2023_celeba_jpeg_dist.pdf}
            \label{fig:supp:ohja2023:celeba:jpeg}}

    \caption{Scores distribution over the Imagenet, COCO, and CelebA datasets of the \cite{Ojha_2023_CVPR} detector.}
	\label{fig:supp:ohja2023:1}
\end{figure*}

\begin{figure*}[htb!]

        \subfloat[\centering JPEG AI scores distribution over the LSUN dataset.]{
            \includegraphics[width=\columnwidth]{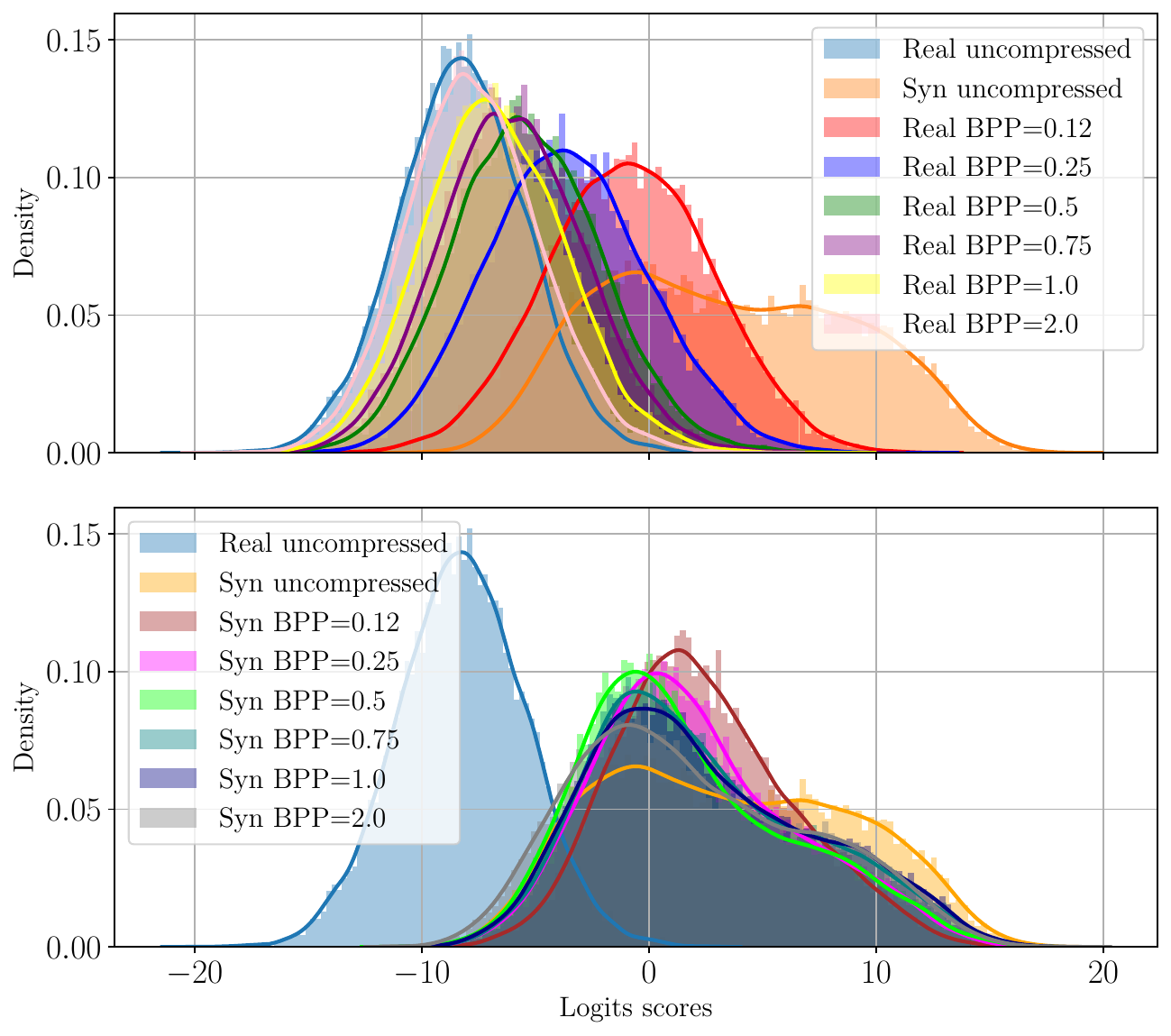}		\label{fig:supp:ohja2023:lsun:jpegai}}
        \hfil
        \subfloat[\centering JPEG scores distribution over the LSUN dataset.]{
            \includegraphics[width=\columnwidth]{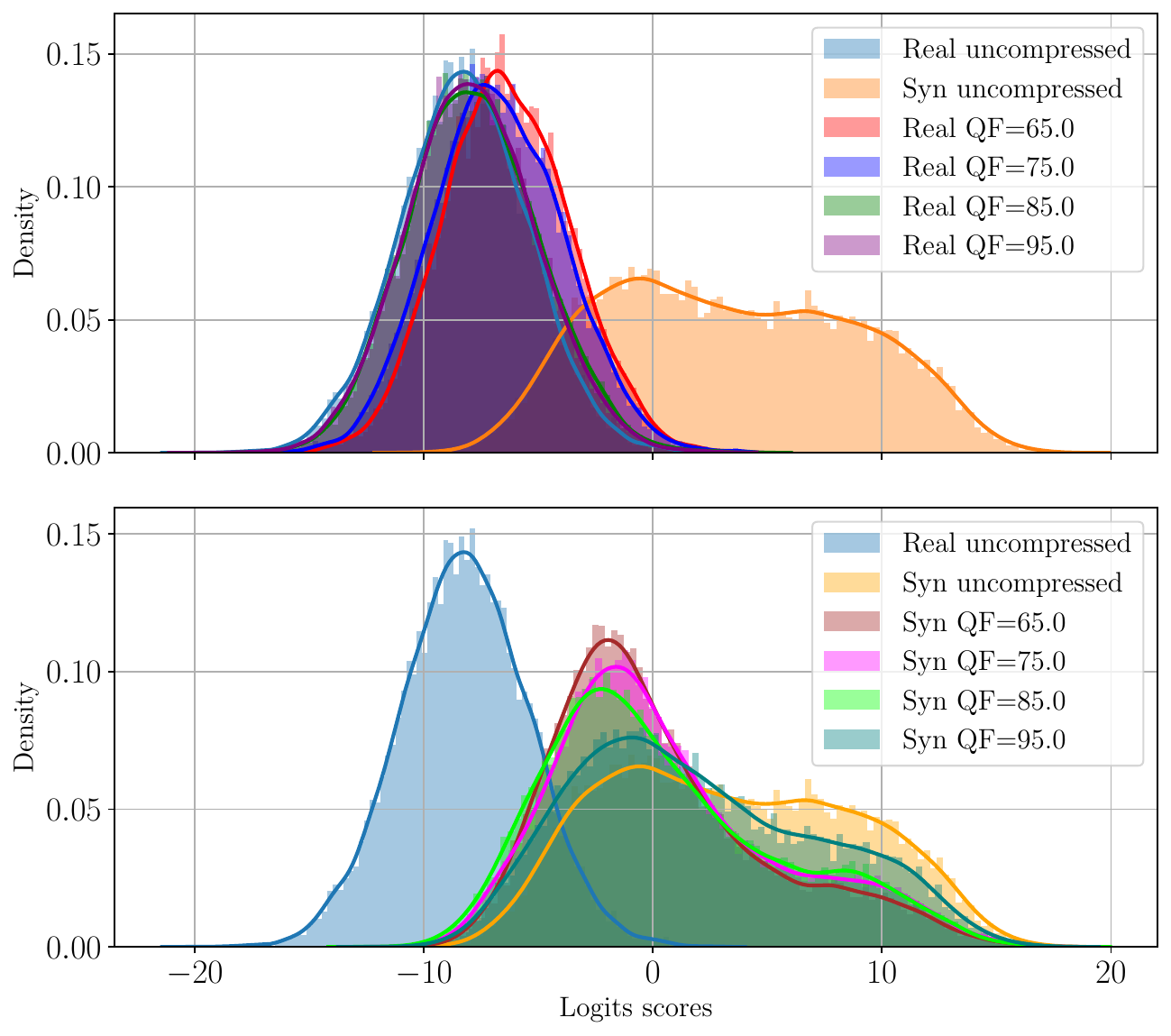}
            \label{fig:supp:ohja2023:lsun:jpeg}}

        \subfloat[\centering JPEG AI scores distribution over the LAION dataset.]{
            \includegraphics[width=\columnwidth]{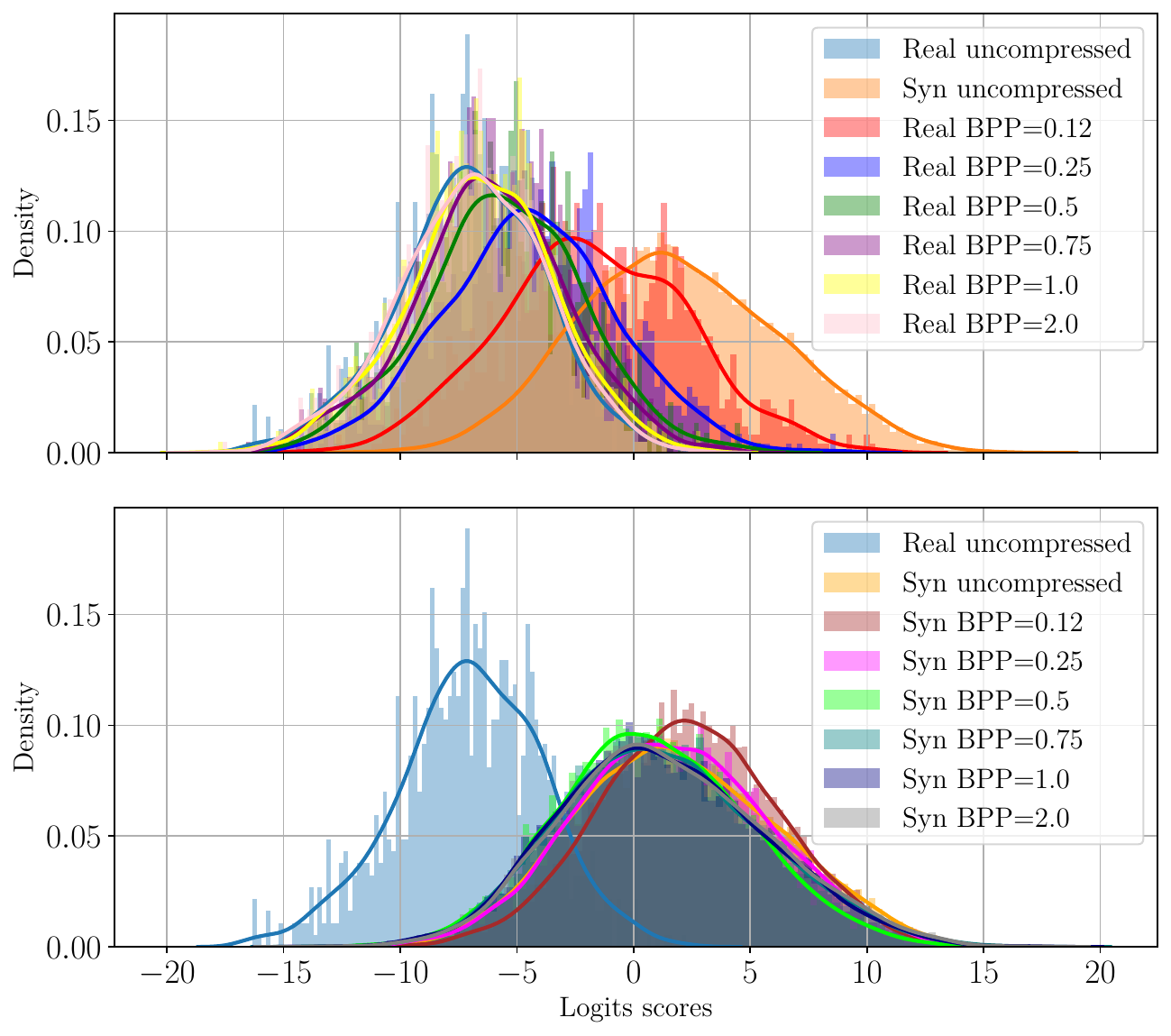}		\label{fig:supp:ohja2023:laion:jpegai}}
        \hfil
        \subfloat[\centering JPEG scores distribution over the LAION dataset.]{
            \includegraphics[width=\columnwidth]{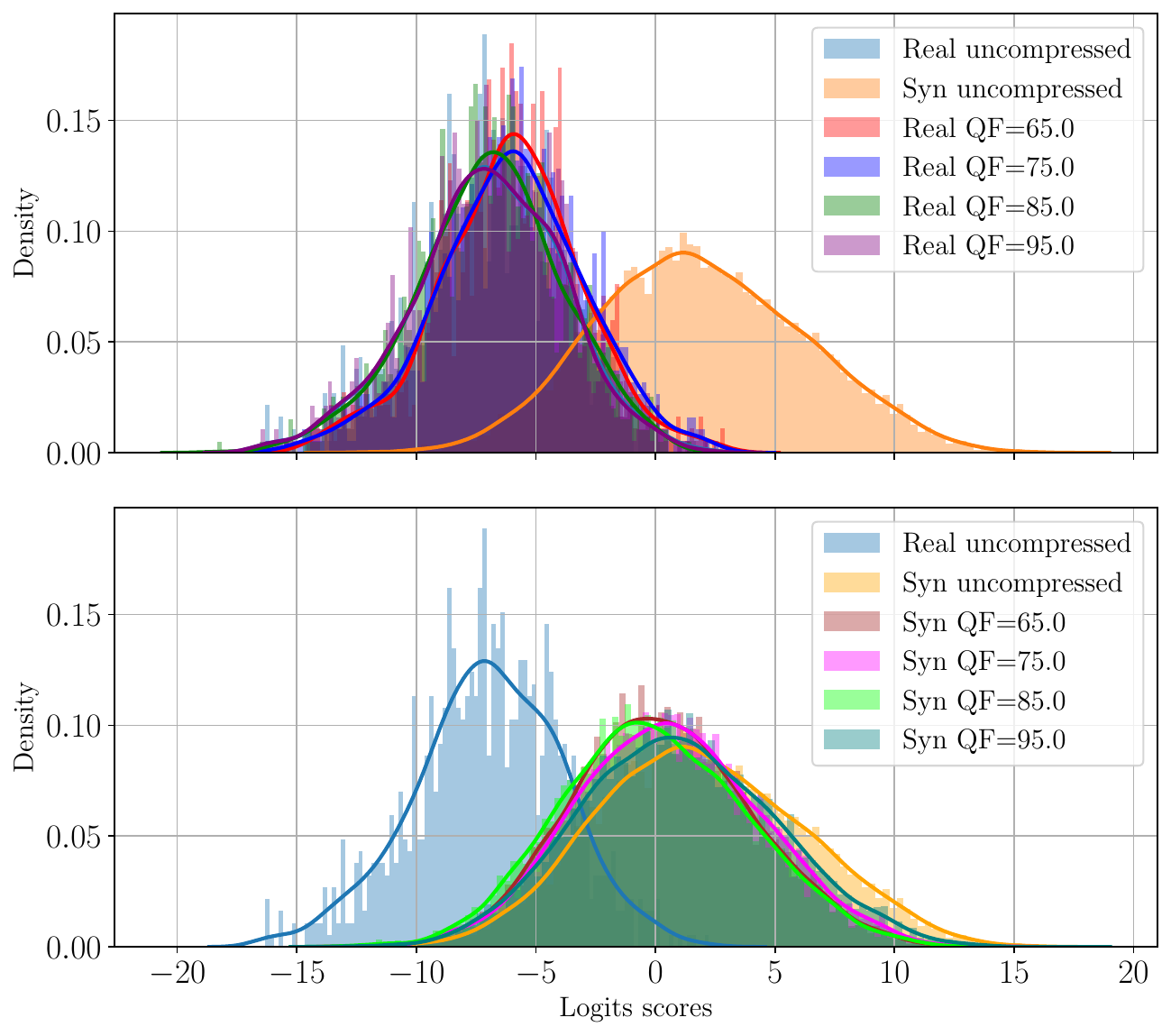}
            \label{fig:supp:ohja2023:laion:jpeg}}

        \caption{Scores distribution over the LSUN and LAION datasets of the \cite{Ojha_2023_CVPR} detector.}
	\label{fig:supp:ohja2023:2}
\end{figure*}

\begin{figure*}[htb!]
	\centering
    
        \subfloat[\centering JPEG AI scores distribution over the Imagenet dataset.]{
            \includegraphics[width=\columnwidth]{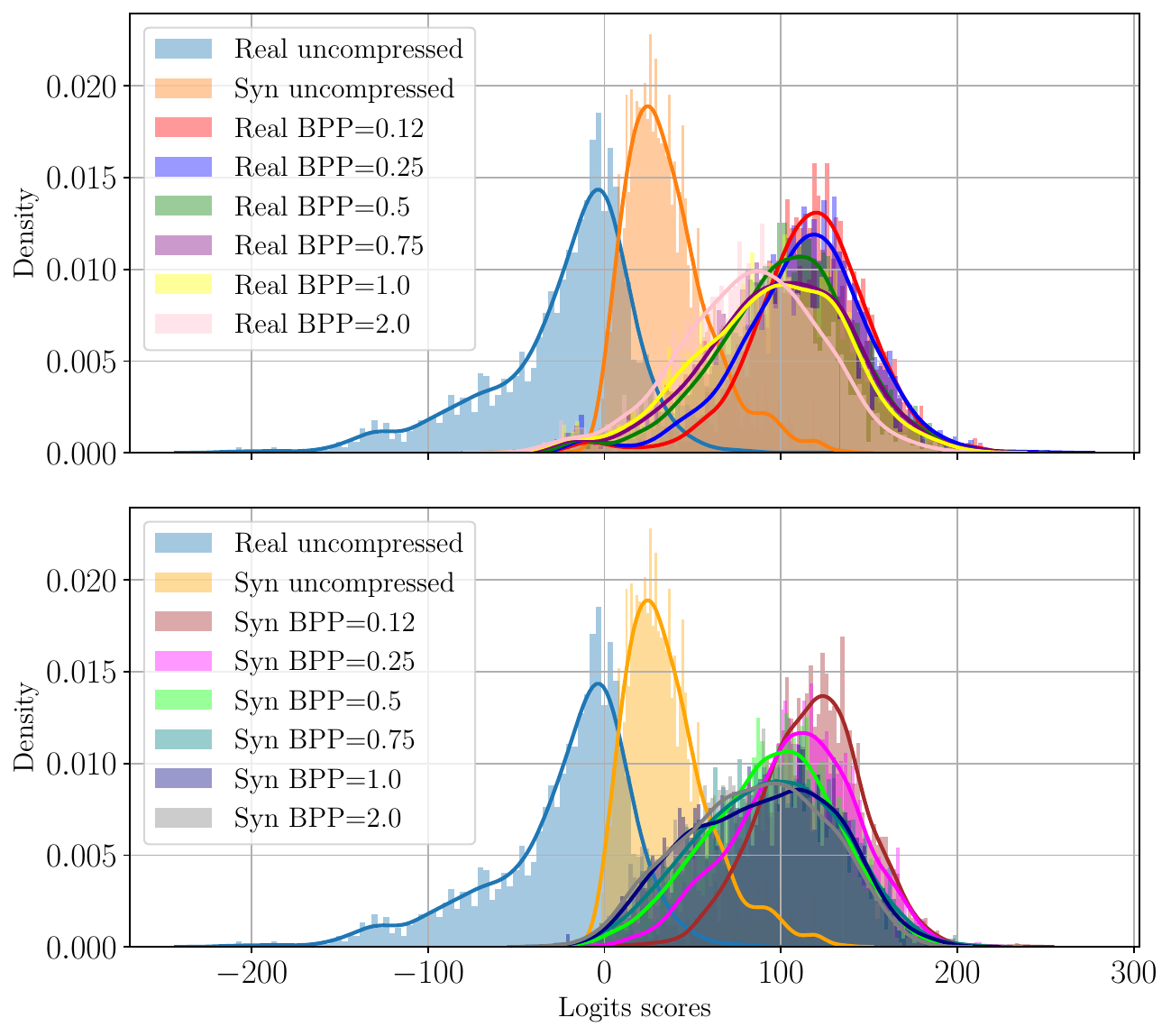}		\label{fig:supp:npr2023:imagenet:jpegai}}
        \hfil
        \subfloat[\centering JPEG scores distribution over the Imagenet dataset.]{
            \includegraphics[width=\columnwidth]{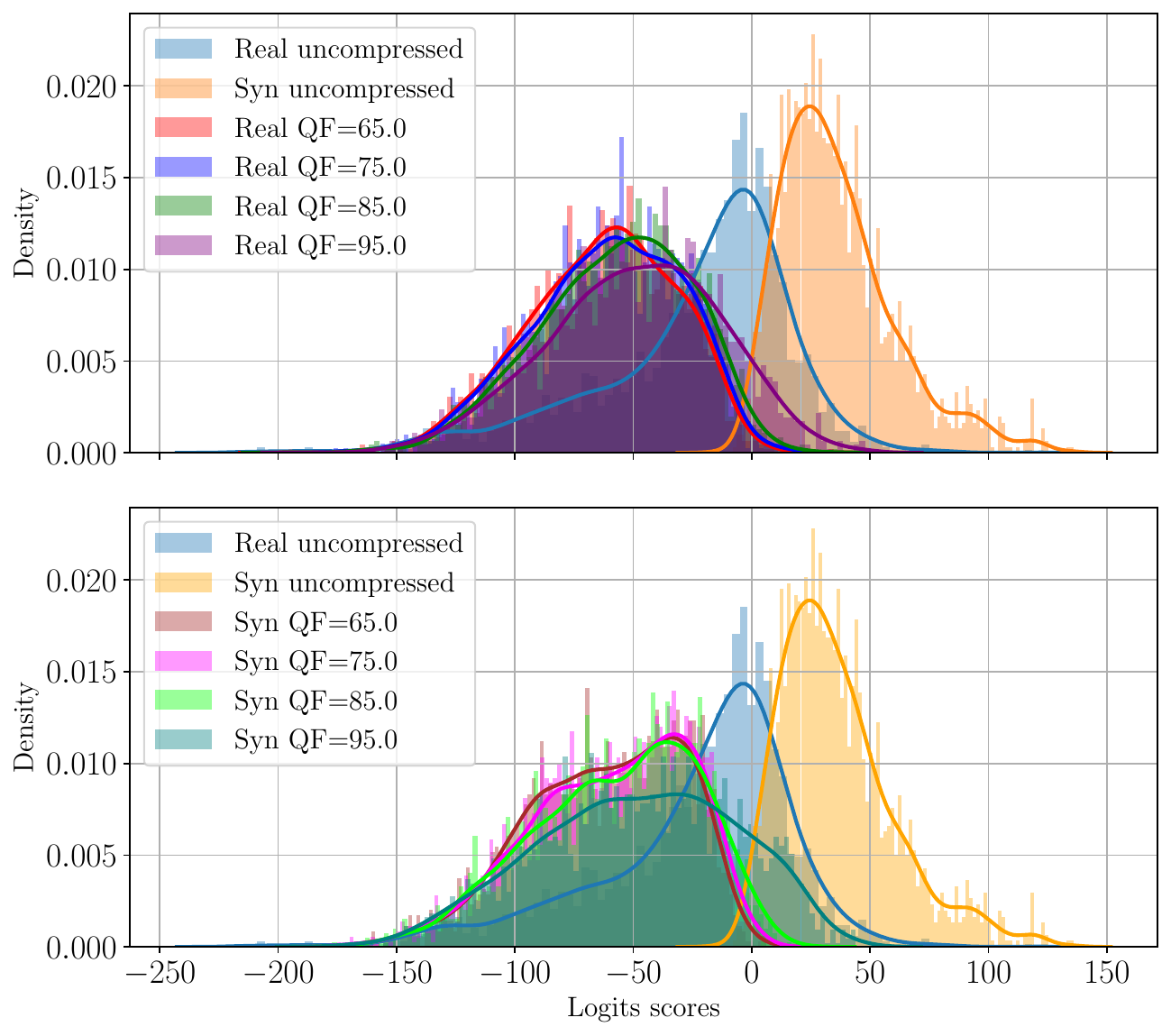}
            \label{fig:supp:npr2023:imagenet:jpeg}}
        
        \subfloat[\centering JPEG AI scores distribution over the COCO dataset.]{
            \includegraphics[width=\columnwidth]{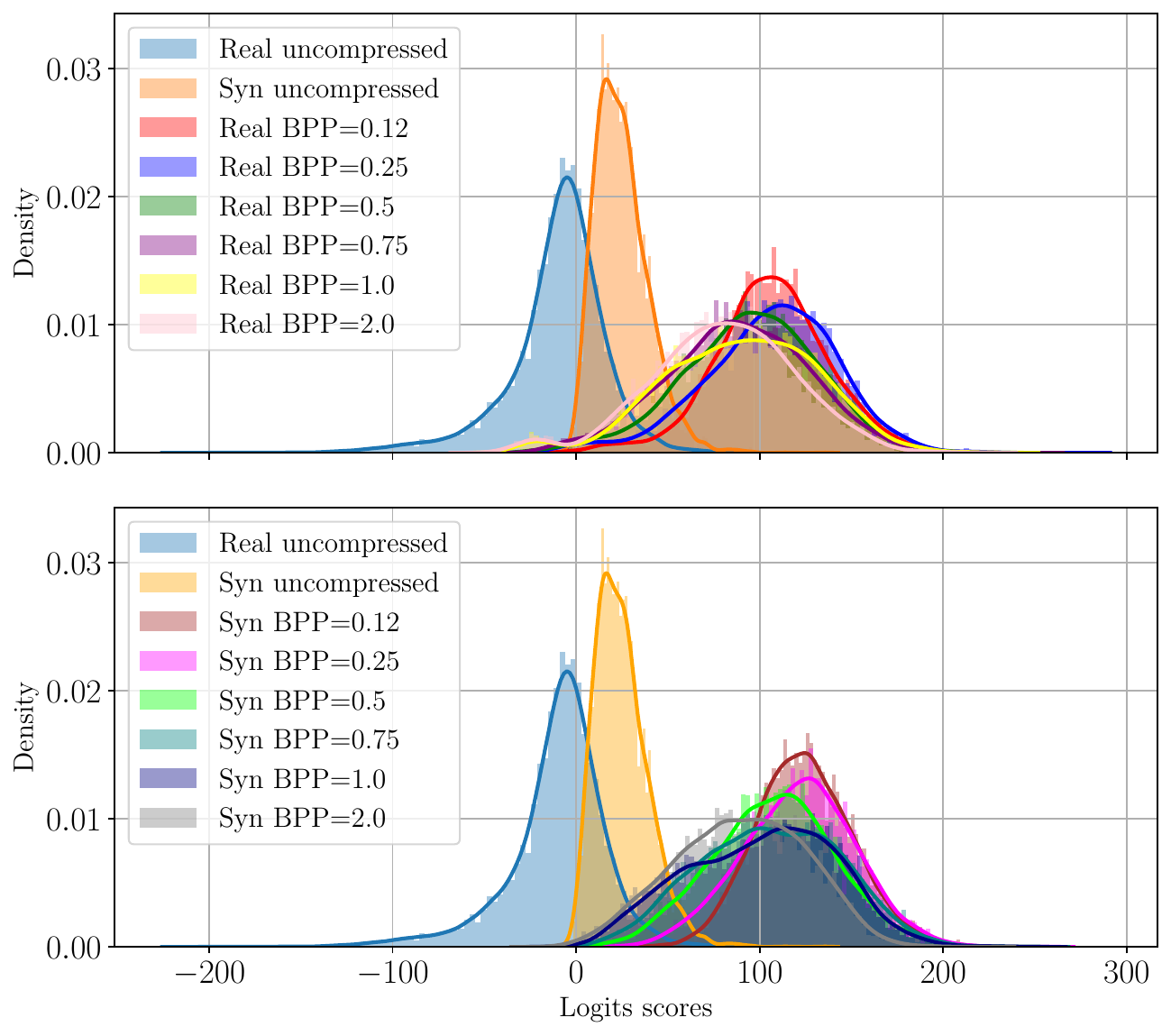}		\label{fig:supp:npr2023:coco:jpegai}}
        \hfil
        \subfloat[\centering JPEG scores distribution over the COCO dataset.]{
            \includegraphics[width=\columnwidth]{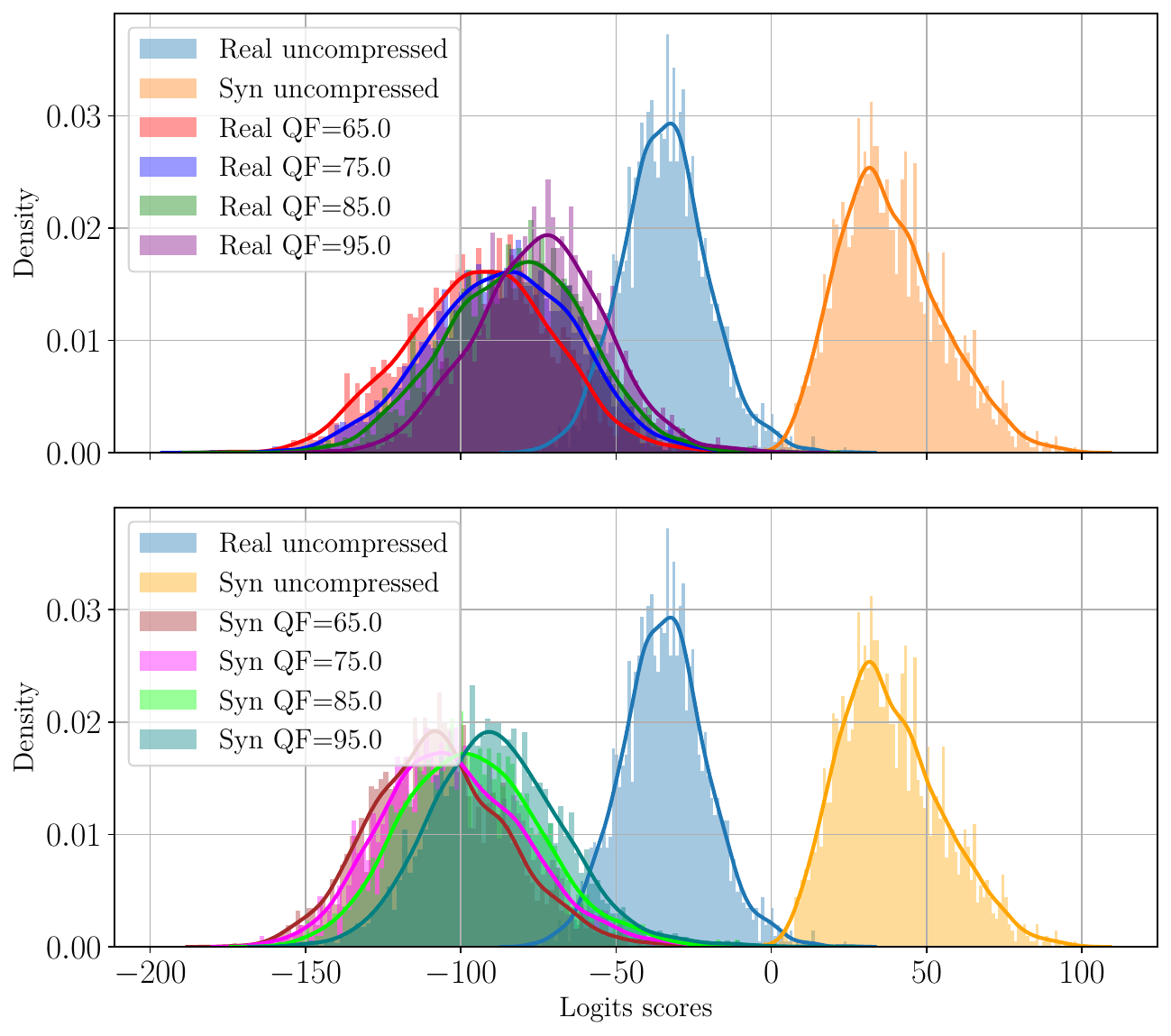}
            \label{fig:supp:npr2023:coco:jpeg}}

        \subfloat[\centering JPEG AI scores distribution over the CelebA dataset.]{
            \includegraphics[width=\columnwidth]{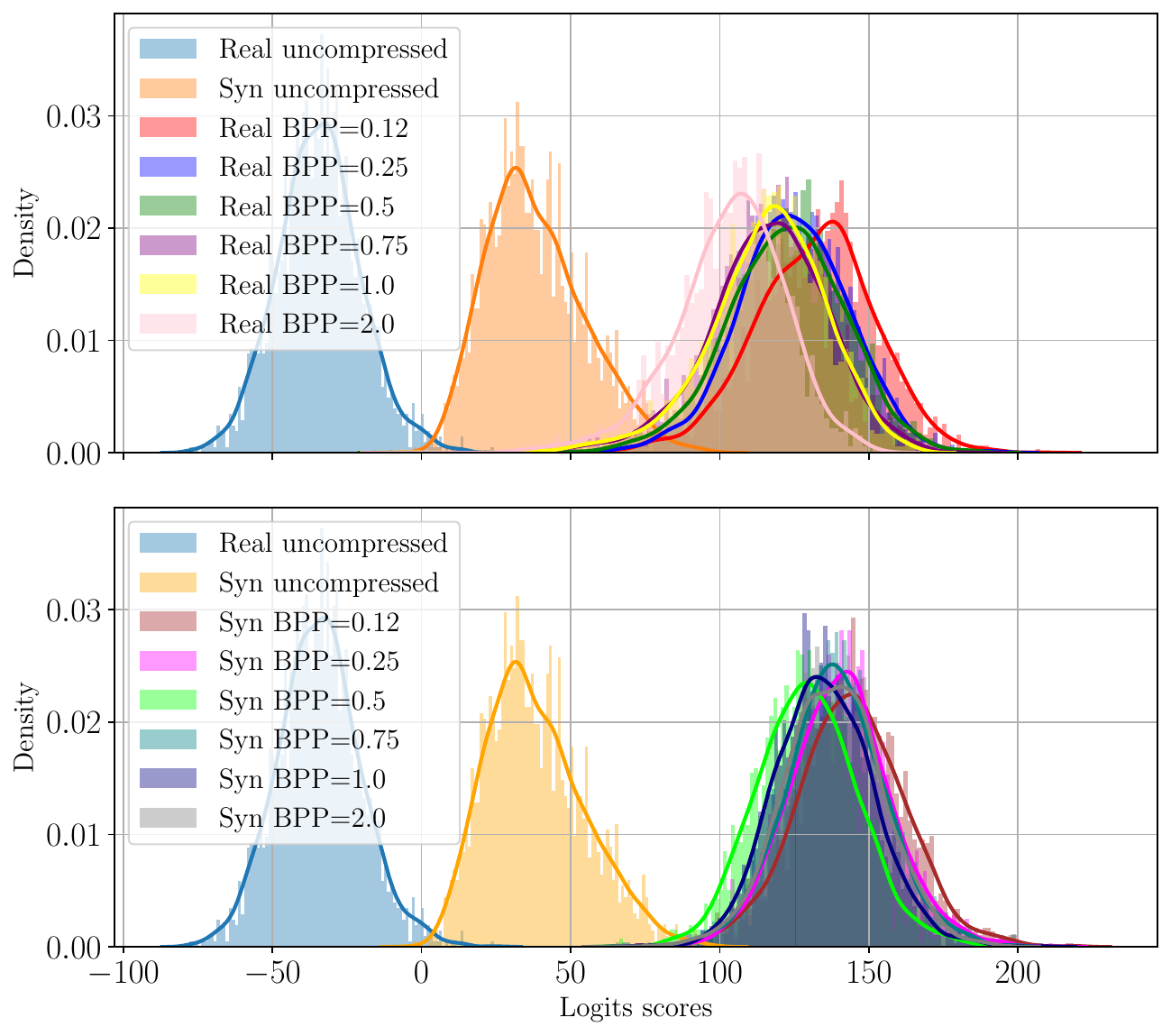}		\label{fig:supp:npr2023:celeba:jpegai}}
        \hfil
        \subfloat[\centering JPEG scores distribution over the CelebA dataset.]{
            \includegraphics[width=\columnwidth]{figures/supp_material/NPR_celeba_jpeg_dist.pdf}
            \label{fig:supp:npr2023:celeba:jpeg}}

    \caption{Scores distribution over the Imagenet, COCO, and CelebA datasets of the \cite{Tan_2024_CVPR} detector.}
	\label{fig:supp:npr2023:1}
\end{figure*}

\begin{figure*}[htb!]

        \subfloat[\centering JPEG AI scores distribution over the LSUN dataset.]{
            \includegraphics[width=\columnwidth]{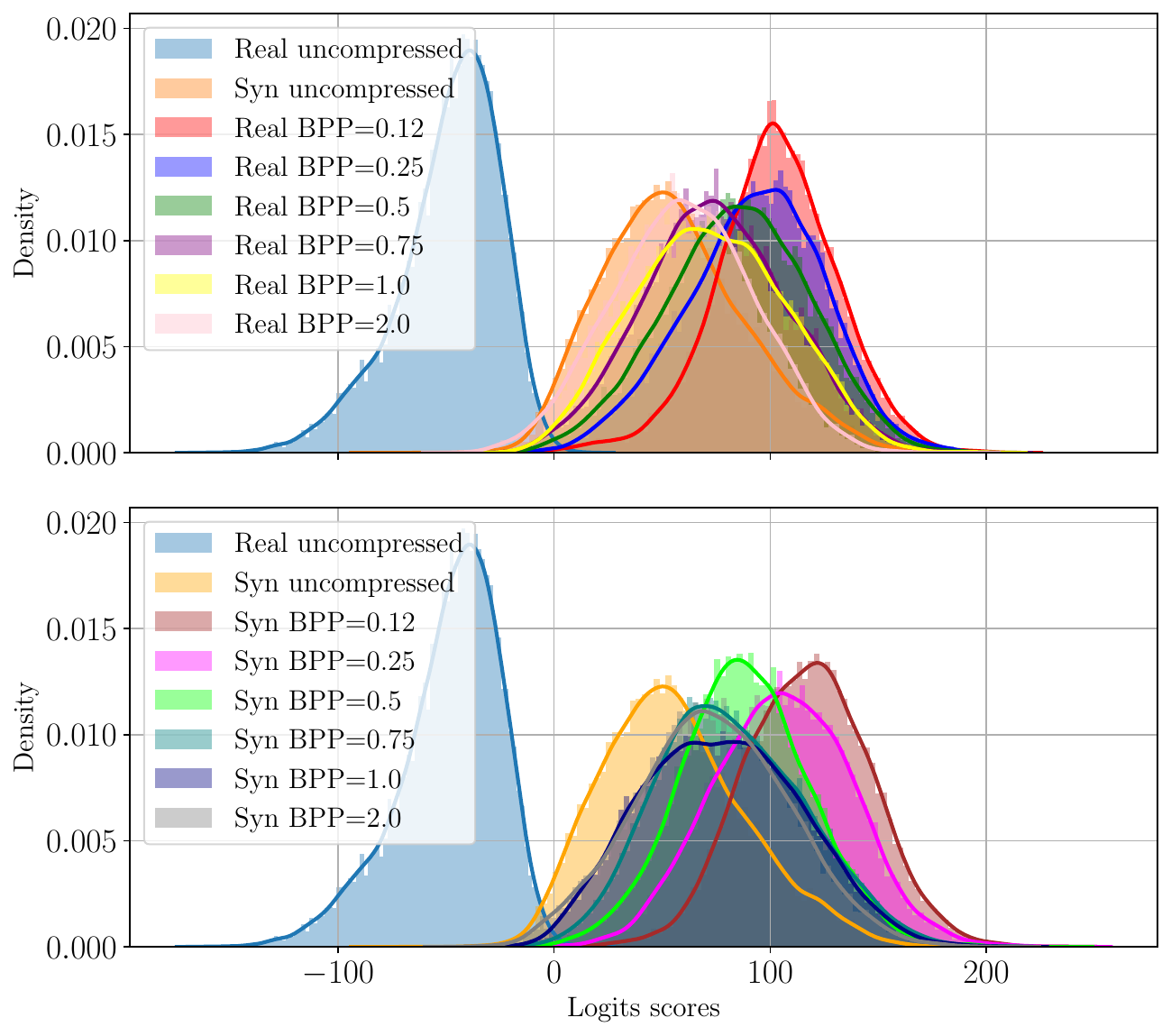}		\label{fig:supp:npr2023:lsun:jpegai}}
        \hfil
        \subfloat[\centering JPEG scores distribution over the LSUN dataset.]{
            \includegraphics[width=\columnwidth]{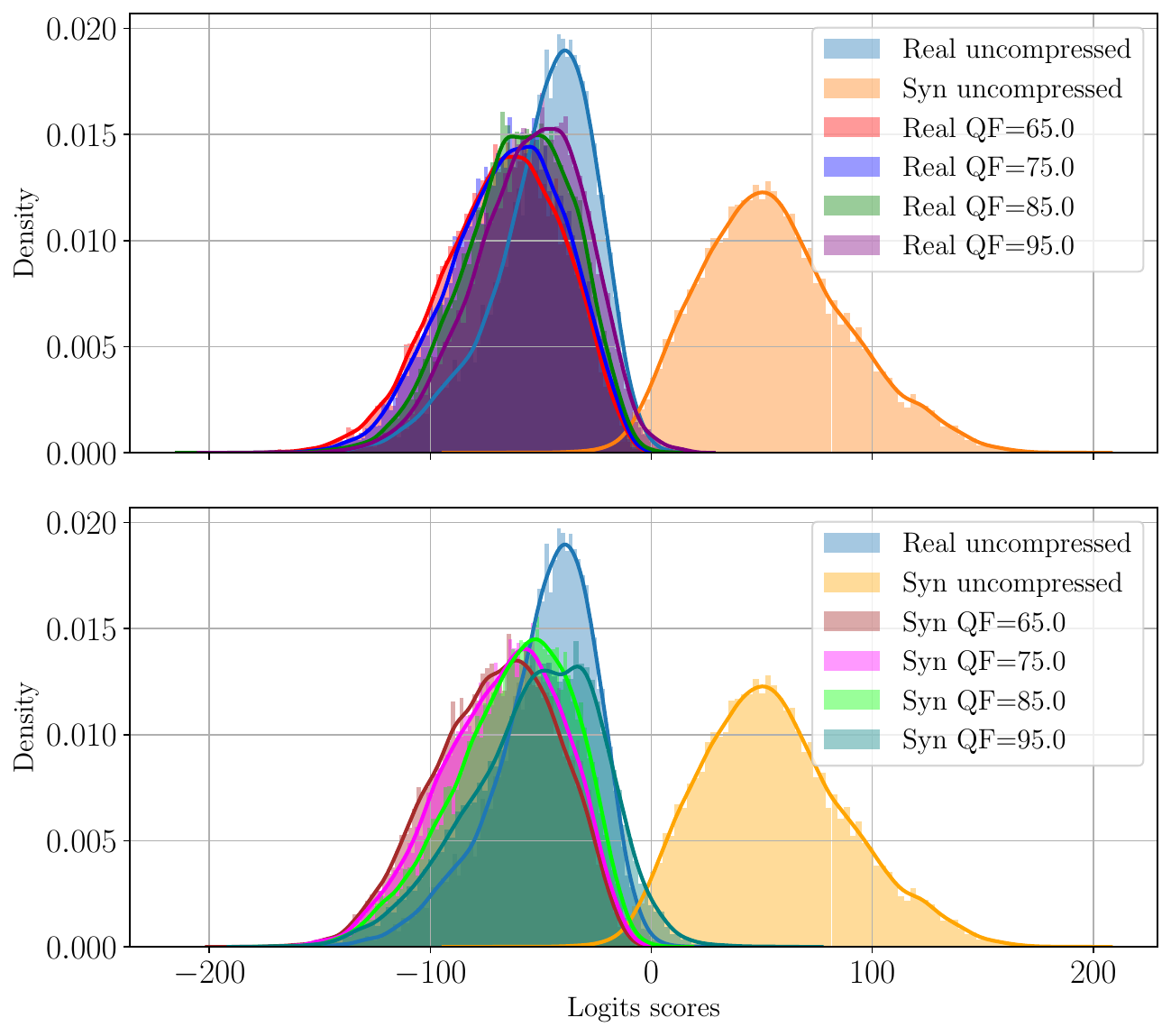}
            \label{fig:supp:npr2023:lsun:jpeg}}

        \subfloat[\centering JPEG AI scores distribution over the LAION dataset.]{
            \includegraphics[width=\columnwidth]{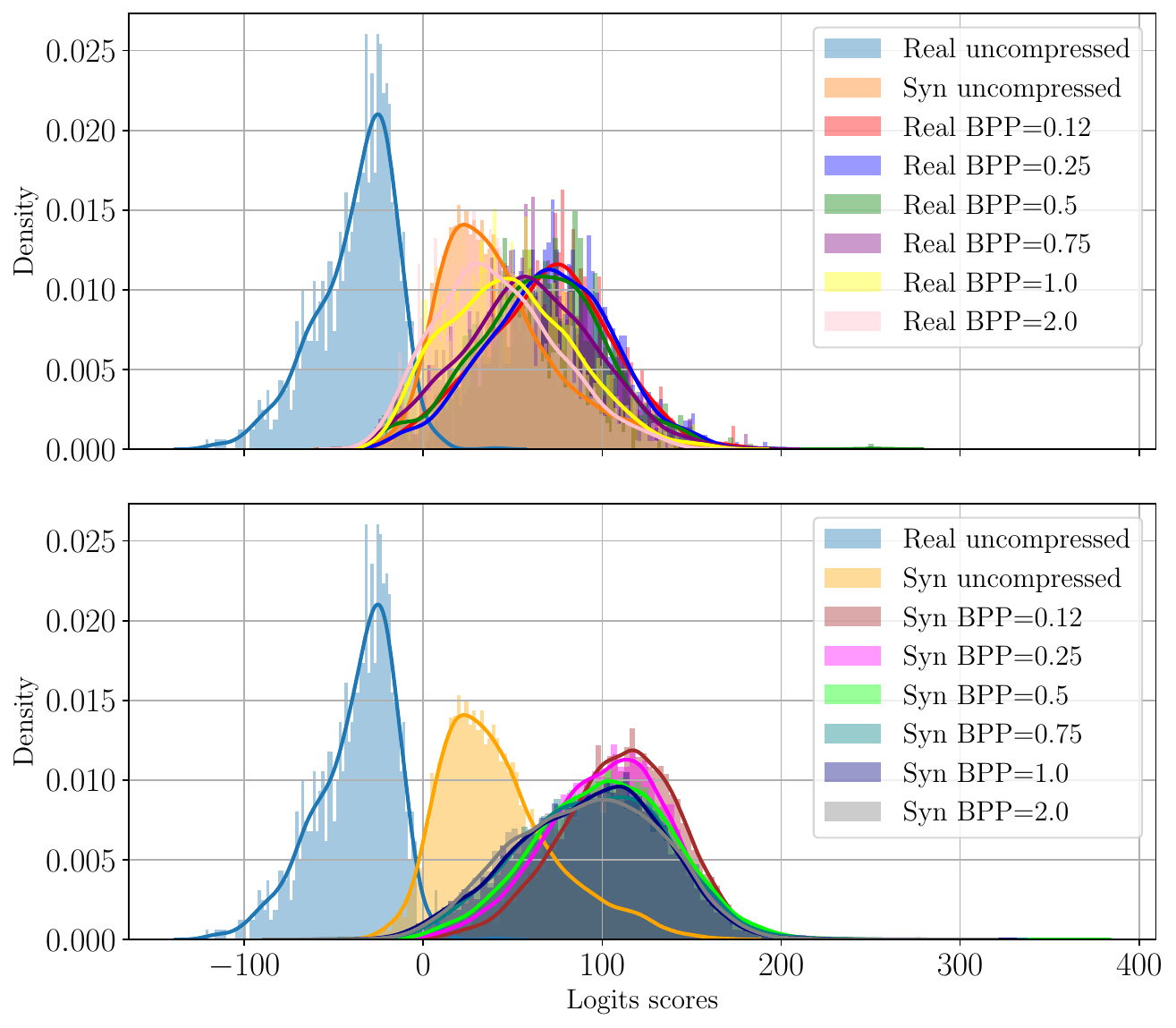}		\label{fig:supp:npr2023:laion:jpegai}}
        \hfil
        \subfloat[\centering JPEG scores distribution over the LAION dataset.]{
            \includegraphics[width=\columnwidth]{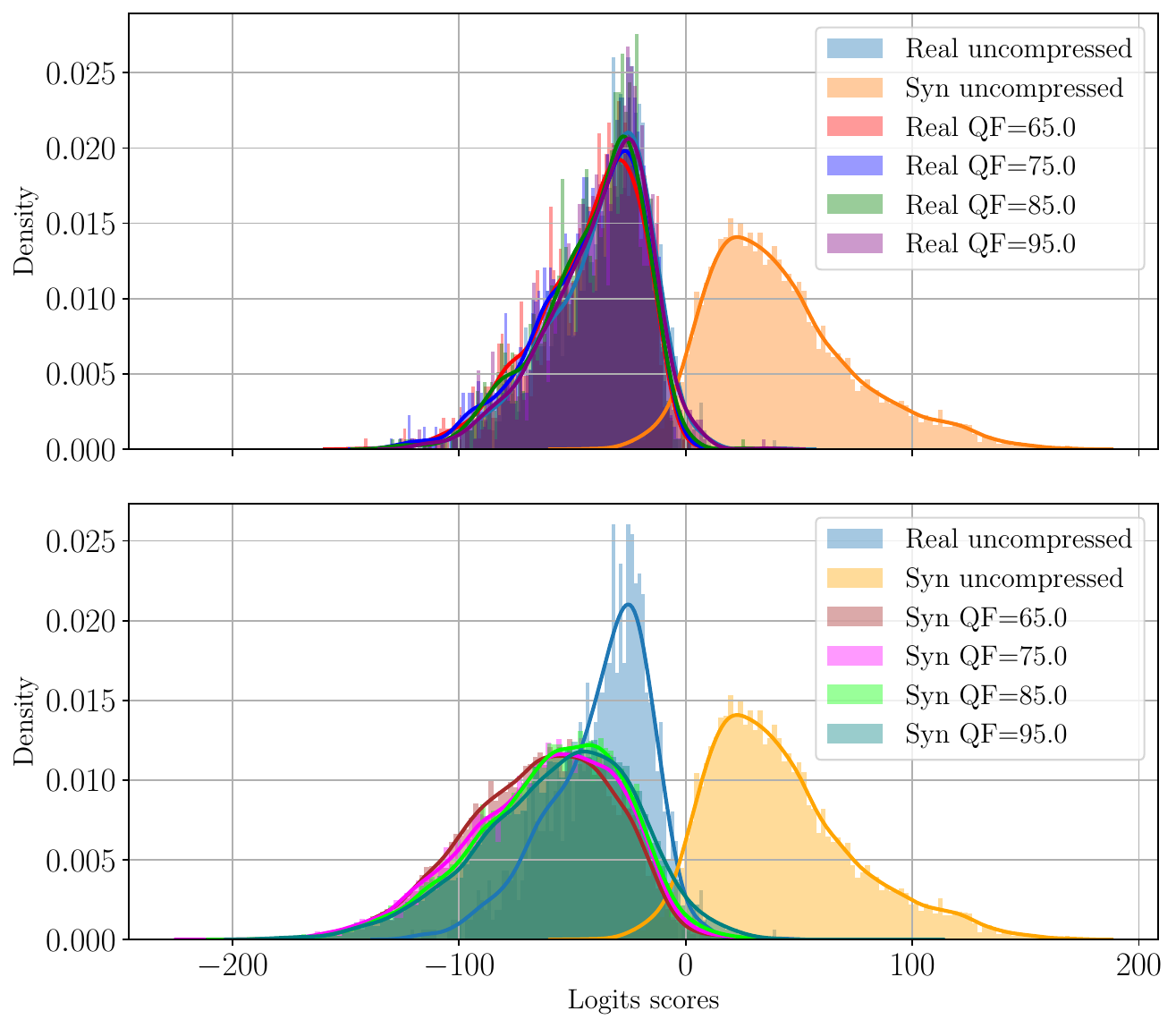}
            \label{fig:supp:npr2023:laion:jpeg}}

        \caption{Scores distribution over the LSUN and LAION datasets of the \cite{Tan_2024_CVPR} detector.}
	\label{fig:supp:npr2023:2}
\end{figure*}

\noindent \textbf{Scores distributions.} \cref{fig:supp:grag2021} to \cref{fig:supp:npr2023:2} report the scores distribution of the detectors \cite{gragnaniello2021}, \cite{cozzolino2023raising}-A, \cite{Ojha_2023_CVPR}, and \cite{Tan_2024_CVPR} while varying the JPEG AI compression ratio or the JPEG quality factor of the compressed images. We report one plot for each dataset on which we tested the detectors. Results are in line with the ones reported in \cref{fig:jpeg-ai_dist} and \cref{fig:jpeg_dist} of the main paper: while increasing the JPEG AI compression ratio creates a ``shift'' of the scores distribution towards the right, i.e., towards ``deepfake''-like scores \textcolor{blue}{above the $0$ threshold}, 
\textcolor{blue}{standard JPEG moves the distribution towards the left, i.e., towards ``pristine ''-like scores below the $0$ threshold. Such results are particularly relevant for the \cite{Tan_2024_CVPR} detector. This tool looks for \gls{npr}, i.e., the local interdependence between image
pixels caused by up-sampling operators. Such a trace seems particularly sensitive to JPEG compression. Indeed, we can see in \cref{fig:supp:npr2023:coco:jpeg} and \cref{fig:supp:npr2023:celeba:jpeg} how even the score distribution for pristine samples is moved below the distribution of pristine uncompressed images. However, this result should not surprise us, as it is well known in the forensic community that JPEG compression could conceal traces relative to the image history~\cite{Piva2013}.} 
We refer the reader to the code repository to visualize more distribution score plots.



\noindent \textbf{Complete analysis of JPEG AI compression ratio.} \cref{tab:jpegai_metrics_wang2020a} to \cref{tab:jpegai_metrics_npr} report the results of the experiments by varying the compression ratio of JPEG AI for each detector on every single dataset. For all tables, as commonly done in deepfake detection tasks, we evaluate the \gls{ba}, \gls{fpr}, and \gls{fnr} at the $0$ threshold. 



\begin{table*}[t]
\caption{\gls{auc}, \gls{ba}, \gls{fpr}, and \gls{fnr} for JPEG AI compression scenario for the \cite{wang2020}-A detector. 
}
\centering
\resizebox{\textwidth}{!}{
\begin{tabular}{l 
    S[table-format=1.2]@{\,/\,}S[table-format=2.2]@{\,/\,} S[table-format=1.2]@{\,/\,}S[table-format=1.2] |
    S[table-format=1.2]@{\,/\,}S[table-format=2.2]@{\,/\,}S[table-format=1.2]@{\,/\,}S[table-format=1.2]
    S[table-format=1.2]@{\,/\,}S[table-format=2.2]@{\,/\,}S[table-format=1.2]@{\,/\,}S[table-format=1.2]
    S[table-format=1.2]@{\,/\,}S[table-format=2.2]@{\,/\,}S[table-format=1.2]@{\,/\,}S[table-format=1.2]
    S[table-format=1.2]@{\,/\,}S[table-format=2.2]@{\,/\,}S[table-format=1.2]@{\,/\,}S[table-format=1.2]
    S[table-format=1.2]@{\,/\,}S[table-format=2.2]@{\,/\,}S[table-format=1.2]@{\,/\,}S[table-format=1.2]
    S[table-format=1.2]@{\,/\,}S[table-format=1.2]@{\,/\,}S[table-format=1.2]@{\,/\,}S[table-format=1.2]
}
\toprule
 & \multicolumn{4}{c |}{Original format} & \multicolumn{4}{c}{BPP = 0.12} & \multicolumn{4}{c}{BPP = 0.25} & 
 \multicolumn{4}{c}{BPP = 0.5} &
 \multicolumn{4}{c}{BPP = 0.75} & \multicolumn{4}{c}{BPP = 1.0} & \multicolumn{4}{c}{BPP = 2.0} \\
 \midrule
& \multicolumn{4}{c |}{$\textrm{AUC}$ /
$\textrm{BA}$ / $\textrm{FPR}$ / $\textrm{FNR}$} & \multicolumn{4}{c |}{$\textrm{AUC}$ /
$\textrm{BA}$ / $\textrm{FPR}$ / $\textrm{FNR}$} & \multicolumn{4}{c |}{$\textrm{AUC}$ /
$\textrm{BA}$ / $\textrm{FPR}$ / $\textrm{FNR}$} &
\multicolumn{4}{c |}{$\textrm{AUC}$ /
$\textrm{BA}$ / $\textrm{FPR}$ / $\textrm{FNR}$} &
\multicolumn{4}{c |}{$\textrm{AUC}$ /
$\textrm{BA}$ / $\textrm{FPR}$ / $\textrm{FNR}$} & \multicolumn{4}{c |}{$\textrm{AUC}$ /
$\textrm{BA}$ / $\textrm{FPR}$ / $\textrm{FNR}$} & \multicolumn{4}{c |}{$\textrm{AUC}$ /
$\textrm{BA}$ / $\textrm{FPR}$ / $\textrm{FNR}$} \\
\midrule
Imagenet & 0.88 & 0.70 & 0.07 & 0.53 & 0.58 & 0.55 & 0.59 & 0.32 & 0.66 & 0.61 & 0.46 & 0.33 & 0.73 & 0.66 & 0.24 & 0.45 & 0.78 & 0.68 & 0.17 & 0.48 & 0.81 & 0.68 & 0.11 & 0.54 & 0.86 & 0.64 & 0.04 & 0.68 \\
COCO & 0.92 & 0.79 & 0.07 & 0.35 & 0.78 & 0.67 & 0.52 & 0.15 & 0.83 & 0.73 & 0.41 & 0.14 & 0.89 & 0.81 & 0.17 & 0.20 & 0.92 & 0.84 & 0.10 & 0.23 & 0.92 & 0.82 & 0.10 & 0.26 & 0.93 & 0.75 & 0.04 & 0.46 \\
CelebA & 0.98 & 0.92 & 0.03 & 0.13 & 0.71 & 0.62 & 0.59 & 0.16 & 0.86 & 0.78 & 0.25 & 0.19 & 0.90 & 0.81 & 0.11 & 0.27 & 0.96 & 0.87 & 0.03 & 0.24 & 0.95 & 0.83 & 0.02 & 0.33 & 0.96 & 0.82 & 0.01 & 0.34 \\
LSUN & 0.99 & 0.89 & 0.00 & 0.22 & 0.72 & 0.63 & 0.55 & 0.19 & 0.83 & 0.75 & 0.27 & 0.23 & 0.88 & 0.78 & 0.10 & 0.34 & 0.93 & 0.81 & 0.05 & 0.33 & 0.95 & 0.82 & 0.02 & 0.34 & 0.98 & 0.80 & 0.00 & 0.40 \\
\bottomrule
\end{tabular}
}
\label{tab:jpegai_metrics_wang2020a}
\vspace{-10pt}
\end{table*}

\begin{table*}[t]
\caption{\gls{auc}, \gls{ba}, \gls{fpr}, and \gls{fnr} for JPEG AI compression scenario for the \cite{wang2020}-B detector. 
}
\centering
\resizebox{\textwidth}{!}{
\begin{tabular}{l 
    S[table-format=1.2]@{\,/\,}S[table-format=2.2]@{\,/\,} S[table-format=1.2]@{\,/\,}S[table-format=1.2] |
    S[table-format=1.2]@{\,/\,}S[table-format=2.2]@{\,/\,}S[table-format=1.2]@{\,/\,}S[table-format=1.2]
    S[table-format=1.2]@{\,/\,}S[table-format=2.2]@{\,/\,}S[table-format=1.2]@{\,/\,}S[table-format=1.2]
    S[table-format=1.2]@{\,/\,}S[table-format=2.2]@{\,/\,}S[table-format=1.2]@{\,/\,}S[table-format=1.2]
    S[table-format=1.2]@{\,/\,}S[table-format=2.2]@{\,/\,}S[table-format=1.2]@{\,/\,}S[table-format=1.2]
    S[table-format=1.2]@{\,/\,}S[table-format=2.2]@{\,/\,}S[table-format=1.2]@{\,/\,}S[table-format=1.2]
    S[table-format=1.2]@{\,/\,}S[table-format=1.2]@{\,/\,}S[table-format=1.2]@{\,/\,}S[table-format=1.2]
}
\toprule
 & \multicolumn{4}{c |}{Original format} & \multicolumn{4}{c}{BPP = 0.12} & \multicolumn{4}{c}{BPP = 0.25} & 
 \multicolumn{4}{c}{BPP = 0.5} &
 \multicolumn{4}{c}{BPP = 0.75} & \multicolumn{4}{c}{BPP = 1.0} & \multicolumn{4}{c}{BPP = 2.0} \\
 \midrule
& \multicolumn{4}{c |}{$\textrm{AUC}$ /
$\textrm{BA}$ / $\textrm{FPR}$ / $\textrm{FNR}$} & \multicolumn{4}{c |}{$\textrm{AUC}$ /
$\textrm{BA}$ / $\textrm{FPR}$ / $\textrm{FNR}$} & \multicolumn{4}{c |}{$\textrm{AUC}$ /
$\textrm{BA}$ / $\textrm{FPR}$ / $\textrm{FNR}$} &
\multicolumn{4}{c |}{$\textrm{AUC}$ /
$\textrm{BA}$ / $\textrm{FPR}$ / $\textrm{FNR}$} &
\multicolumn{4}{c |}{$\textrm{AUC}$ /
$\textrm{BA}$ / $\textrm{FPR}$ / $\textrm{FNR}$} & \multicolumn{4}{c |}{$\textrm{AUC}$ /
$\textrm{BA}$ / $\textrm{FPR}$ / $\textrm{FNR}$} & \multicolumn{4}{c |}{$\textrm{AUC}$ /
$\textrm{BA}$ / $\textrm{FPR}$ / $\textrm{FNR}$} \\
\midrule
Imagenet & 0.90 & 0.59 & 0.01 & 0.81 & 0.62 & 0.58 & 0.22 & 0.61 & 0.73 & 0.62 & 0.09 & 0.67 & 0.82 & 0.61 & 0.03 & 0.74 & 0.86 & 0.62 & 0.02 & 0.75 & 0.88 & 0.61 & 0.01 & 0.77 & 0.90 & 0.58 & 0.01 & 0.82 \\
COCO & 0.98 & 0.79 & 0.01 & 0.41 & 0.81 & 0.73 & 0.17 & 0.37 & 0.90 & 0.80 & 0.08 & 0.32 & 0.96 & 0.81 & 0.02 & 0.35 & 0.97 & 0.83 & 0.01 & 0.33 & 0.98 & 0.82 & 0.01 & 0.35 & 0.98 & 0.78 & 0.00 & 0.43 \\
CelebA & 0.95 & 0.81 & 0.01 & 0.37 & 0.71 & 0.60 & 0.06 & 0.74 & 0.83 & 0.62 & 0.02 & 0.74 & 0.90 & 0.71 & 0.01 & 0.56 & 0.93 & 0.74 & 0.01 & 0.50 & 0.93 & 0.76 & 0.01 & 0.48 & 0.94 & 0.76 & 0.00 & 0.48 \\
LSUN & 0.98 & 0.77 & 0.00 & 0.46 & 0.71 & 0.66 & 0.17 & 0.51 & 0.83 & 0.71 & 0.05 & 0.53 & 0.90 & 0.71 & 0.01 & 0.58 & 0.94 & 0.74 & 0.00 & 0.52 & 0.96 & 0.74 & 0.00 & 0.52 & 0.98 & 0.73 & 0.00 & 0.53 \\
\bottomrule
\end{tabular}
}
\label{tab:jpegai_metrics_wang2020b}
\vspace{-10pt}
\end{table*}

\begin{table*}[t]
\caption{\gls{auc}, \gls{ba}, \gls{fpr}, and \gls{fnr} for JPEG AI compression scenario for the \cite{gragnaniello2021} detector. 
}
\centering
\resizebox{\textwidth}{!}{
\begin{tabular}{l 
    S[table-format=1.2]@{\,/\,}S[table-format=2.2]@{\,/\,} S[table-format=1.2]@{\,/\,}S[table-format=1.2] |
    S[table-format=1.2]@{\,/\,}S[table-format=2.2]@{\,/\,}S[table-format=1.2]@{\,/\,}S[table-format=1.2]
    S[table-format=1.2]@{\,/\,}S[table-format=2.2]@{\,/\,}S[table-format=1.2]@{\,/\,}S[table-format=1.2]
    S[table-format=1.2]@{\,/\,}S[table-format=2.2]@{\,/\,}S[table-format=1.2]@{\,/\,}S[table-format=1.2]
    S[table-format=1.2]@{\,/\,}S[table-format=2.2]@{\,/\,}S[table-format=1.2]@{\,/\,}S[table-format=1.2]
    S[table-format=1.2]@{\,/\,}S[table-format=2.2]@{\,/\,}S[table-format=1.2]@{\,/\,}S[table-format=1.2]
    S[table-format=1.2]@{\,/\,}S[table-format=1.2]@{\,/\,}S[table-format=1.2]@{\,/\,}S[table-format=1.2]
}
\toprule
 & \multicolumn{4}{c |}{Original format} & \multicolumn{4}{c}{BPP = 0.12} & \multicolumn{4}{c}{BPP = 0.25} & 
 \multicolumn{4}{c}{BPP = 0.5} &
 \multicolumn{4}{c}{BPP = 0.75} & \multicolumn{4}{c}{BPP = 1.0} & \multicolumn{4}{c}{BPP = 2.0} \\
 \midrule
& \multicolumn{4}{c |}{$\textrm{AUC}$ /
$\textrm{BA}$ / $\textrm{FPR}$ / $\textrm{FNR}$} & \multicolumn{4}{c |}{$\textrm{AUC}$ /
$\textrm{BA}$ / $\textrm{FPR}$ / $\textrm{FNR}$} & \multicolumn{4}{c |}{$\textrm{AUC}$ /
$\textrm{BA}$ / $\textrm{FPR}$ / $\textrm{FNR}$} &
\multicolumn{4}{c |}{$\textrm{AUC}$ /
$\textrm{BA}$ / $\textrm{FPR}$ / $\textrm{FNR}$} &
\multicolumn{4}{c |}{$\textrm{AUC}$ /
$\textrm{BA}$ / $\textrm{FPR}$ / $\textrm{FNR}$} & \multicolumn{4}{c |}{$\textrm{AUC}$ /
$\textrm{BA}$ / $\textrm{FPR}$ / $\textrm{FNR}$} & \multicolumn{4}{c |}{$\textrm{AUC}$ /
$\textrm{BA}$ / $\textrm{FPR}$ / $\textrm{FNR}$} \\
\midrule
Imagenet & 0.89 & 0.68 & 0.01 & 0.64 & 0.45 & 0.48 & 0.95 & 0.09 & 0.51 & 0.47 & 0.83 & 0.23 & 0.72 & 0.66 & 0.36 & 0.32 & 0.80 & 0.70 & 0.14 & 0.45 & 0.84 & 0.70 & 0.06 & 0.54 & 0.89 & 0.68 & 0.01 & 0.64 \\
COCO & 0.96 & 0.67 & 0.01 & 0.66 & 0.56 & 0.52 & 0.94 & 0.02 & 0.68 & 0.55 & 0.82 & 0.07 & 0.83 & 0.74 & 0.33 & 0.20 & 0.89 & 0.77 & 0.08 & 0.39 & 0.90 & 0.73 & 0.04 & 0.50 & 0.95 & 0.67 & 0.01 & 0.65 \\
FFHQ & 0.93 & 0.80 & 0.06 & 0.33 & 0.58 & 0.50 & 0.99 & 0.00 & 0.56 & 0.52 & 0.94 & 0.02 & 0.81 & 0.76 & 0.44 & 0.04 & 0.75 & 0.70 & 0.45 & 0.14 & 0.91 & 0.84 & 0.15 & 0.18 & 0.88 & 0.72 & 0.06 & 0.50 \\
\bottomrule
\end{tabular}
}
\label{tab:jpegai_metrics_gragnaniello2021}
\vspace{-10pt}
\end{table*}

\begin{table*}[t]
\caption{\gls{auc}, \gls{ba}, \gls{fpr}, and \gls{fnr} for JPEG AI compression scenario for the \cite{Corvi_2023_ICASSP} detector.
}
\centering
\resizebox{\textwidth}{!}{
\begin{tabular}{l 
    S[table-format=1.2]@{\,/\,}S[table-format=2.2]@{\,/\,} S[table-format=1.2]@{\,/\,}S[table-format=1.2] |
    S[table-format=1.2]@{\,/\,}S[table-format=2.2]@{\,/\,}S[table-format=1.2]@{\,/\,}S[table-format=1.2]
    S[table-format=1.2]@{\,/\,}S[table-format=2.2]@{\,/\,}S[table-format=1.2]@{\,/\,}S[table-format=1.2]
    S[table-format=1.2]@{\,/\,}S[table-format=2.2]@{\,/\,}S[table-format=1.2]@{\,/\,}S[table-format=1.2]
    S[table-format=1.2]@{\,/\,}S[table-format=2.2]@{\,/\,}S[table-format=1.2]@{\,/\,}S[table-format=1.2]
    S[table-format=1.2]@{\,/\,}S[table-format=2.2]@{\,/\,}S[table-format=1.2]@{\,/\,}S[table-format=1.2]
    S[table-format=1.2]@{\,/\,}S[table-format=1.2]@{\,/\,}S[table-format=1.2]@{\,/\,}S[table-format=1.2]
}
\toprule
 & \multicolumn{4}{c |}{Original format} & \multicolumn{4}{c}{BPP = 0.12} & \multicolumn{4}{c}{BPP = 0.25} & 
 \multicolumn{4}{c}{BPP = 0.5} &
 \multicolumn{4}{c}{BPP = 0.75} & \multicolumn{4}{c}{BPP = 1.0} & \multicolumn{4}{c}{BPP = 2.0} \\
 \midrule
& \multicolumn{4}{c |}{$\textrm{AUC}$ /
$\textrm{BA}$ / $\textrm{FPR}$ / $\textrm{FNR}$} & \multicolumn{4}{c |}{$\textrm{AUC}$ /
$\textrm{BA}$ / $\textrm{FPR}$ / $\textrm{FNR}$} & \multicolumn{4}{c |}{$\textrm{AUC}$ /
$\textrm{BA}$ / $\textrm{FPR}$ / $\textrm{FNR}$} &
\multicolumn{4}{c |}{$\textrm{AUC}$ /
$\textrm{BA}$ / $\textrm{FPR}$ / $\textrm{FNR}$} &
\multicolumn{4}{c |}{$\textrm{AUC}$ /
$\textrm{BA}$ / $\textrm{FPR}$ / $\textrm{FNR}$} & \multicolumn{4}{c |}{$\textrm{AUC}$ /
$\textrm{BA}$ / $\textrm{FPR}$ / $\textrm{FNR}$} & \multicolumn{4}{c |}{$\textrm{AUC}$ /
$\textrm{BA}$ / $\textrm{FPR}$ / $\textrm{FNR}$} \\
\midrule
Imagenet & 0.87 & 0.70 & 0.01 & 0.59 & 0.48 & 0.50 & 0.99 & 0.01 & 0.55 & 0.52 & 0.66 & 0.31 & 0.69 & 0.64 & 0.05 & 0.68 & 0.77 & 0.66 & 0.01 & 0.67 & 0.81 & 0.67 & 0.01 & 0.65 & 0.85 & 0.68 & 0.01 & 0.63 \\
COCO & 0.96 & 0.74 & 0.00 & 0.52 & 0.47 & 0.49 & 0.99 & 0.03 & 0.69 & 0.58 & 0.59 & 0.24 & 0.80 & 0.74 & 0.02 & 0.50 & 0.87 & 0.72 & 0.00 & 0.55 & 0.91 & 0.72 & 0.00 & 0.57 & 0.95 & 0.72 & 0.00 & 0.56 \\
FFHQ & 0.89 & 0.87 & 0.00 & 0.27 & 0.79 & 0.52 & 0.93 & 0.03 & 0.75 & 0.68 & 0.32 & 0.32 & 0.84 & 0.58 & 0.02 & 0.82 & 0.84 & 0.62 & 0.00 & 0.76 & 0.89 & 0.67 & 0.00 & 0.66 & 0.90 & 0.75 & 0.00 & 0.51 \\
\bottomrule
\end{tabular}
}
\label{tab:jpegai_metrics_corvi2023}
\vspace{-10pt}
\end{table*}

\begin{table*}[t]
\caption{\gls{auc}, \gls{ba}, \gls{fpr}, and \gls{fnr} for JPEG AI compression scenario for the \cite{Ojha_2023_CVPR} detector. 
}
\centering
\resizebox{\textwidth}{!}{
\begin{tabular}{l 
    S[table-format=1.2]@{\,/\,}S[table-format=2.2]@{\,/\,} S[table-format=1.2]@{\,/\,}S[table-format=1.2] |
    S[table-format=1.2]@{\,/\,}S[table-format=2.2]@{\,/\,}S[table-format=1.2]@{\,/\,}S[table-format=1.2]
    S[table-format=1.2]@{\,/\,}S[table-format=2.2]@{\,/\,}S[table-format=1.2]@{\,/\,}S[table-format=1.2]
    S[table-format=1.2]@{\,/\,}S[table-format=2.2]@{\,/\,}S[table-format=1.2]@{\,/\,}S[table-format=1.2]
    S[table-format=1.2]@{\,/\,}S[table-format=2.2]@{\,/\,}S[table-format=1.2]@{\,/\,}S[table-format=1.2]
    S[table-format=1.2]@{\,/\,}S[table-format=2.2]@{\,/\,}S[table-format=1.2]@{\,/\,}S[table-format=1.2]
    S[table-format=1.2]@{\,/\,}S[table-format=1.2]@{\,/\,}S[table-format=1.2]@{\,/\,}S[table-format=1.2]
}
\toprule
 & \multicolumn{4}{c |}{Original format} & \multicolumn{4}{c}{BPP = 0.12} & \multicolumn{4}{c}{BPP = 0.25} & 
 \multicolumn{4}{c}{BPP = 0.5} &
 \multicolumn{4}{c}{BPP = 0.75} & \multicolumn{4}{c}{BPP = 1.0} & \multicolumn{4}{c}{BPP = 2.0} \\
 \midrule
& \multicolumn{4}{c |}{$\textrm{AUC}$ /
$\textrm{BA}$ / $\textrm{FPR}$ / $\textrm{FNR}$} & \multicolumn{4}{c |}{$\textrm{AUC}$ /
$\textrm{BA}$ / $\textrm{FPR}$ / $\textrm{FNR}$} & \multicolumn{4}{c |}{$\textrm{AUC}$ /
$\textrm{BA}$ / $\textrm{FPR}$ / $\textrm{FNR}$} &
\multicolumn{4}{c |}{$\textrm{AUC}$ /
$\textrm{BA}$ / $\textrm{FPR}$ / $\textrm{FNR}$} &
\multicolumn{4}{c |}{$\textrm{AUC}$ /
$\textrm{BA}$ / $\textrm{FPR}$ / $\textrm{FNR}$} & \multicolumn{4}{c |}{$\textrm{AUC}$ /
$\textrm{BA}$ / $\textrm{FPR}$ / $\textrm{FNR}$} & \multicolumn{4}{c |}{$\textrm{AUC}$ /
$\textrm{BA}$ / $\textrm{FPR}$ / $\textrm{FNR}$} \\
\midrule
Imagenet & 0.99 & 0.95 & 0.02 & 0.08 & 0.73 & 0.66 & 0.49 & 0.19 & 0.85 & 0.77 & 0.23 & 0.23 & 0.92 & 0.82 & 0.09 & 0.27 & 0.94 & 0.84 & 0.06 & 0.25 & 0.96 & 0.87 & 0.04 & 0.22 & 0.97 & 0.90 & 0.03 & 0.18 \\
COCO & 1.00 & 0.99 & 0.01 & 0.00 & 0.88 & 0.71 & 0.54 & 0.03 & 0.95 & 0.83 & 0.30 & 0.04 & 0.99 & 0.94 & 0.09 & 0.03 & 1.00 & 0.97 & 0.05 & 0.02 & 1.00 & 0.97 & 0.05 & 0.01 & 1.00 & 0.99 & 0.01 & 0.01 \\
CelebA & 0.99 & 0.96 & 0.07 & 0.02 & 0.91 & 0.82 & 0.30 & 0.07 & 0.96 & 0.89 & 0.12 & 0.09 & 0.97 & 0.91 & 0.11 & 0.06 & 0.98 & 0.93 & 0.09 & 0.04 & 0.98 & 0.93 & 0.10 & 0.04 & 0.98 & 0.93 & 0.08 & 0.05 \\
LSUN & 0.98 & 0.84 & 0.00 & 0.32 & 0.72 & 0.65 & 0.41 & 0.28 & 0.84 & 0.74 & 0.16 & 0.36 & 0.90 & 0.76 & 0.06 & 0.42 & 0.94 & 0.80 & 0.03 & 0.37 & 0.96 & 0.81 & 0.02 & 0.37 & 0.96 & 0.79 & 0.01 & 0.41 \\
LAION & 0.96 & 0.82 & 0.01 & 0.35 & 0.77 & 0.69 & 0.34 & 0.28 & 0.86 & 0.77 & 0.12 & 0.35 & 0.89 & 0.76 & 0.05 & 0.43 & 0.91 & 0.78 & 0.03 & 0.41 & 0.93 & 0.79 & 0.02 & 0.41 & 0.94 & 0.80 & 0.01 & 0.40 \\
\bottomrule
\end{tabular}
}
\label{tab:jpegai_metrics_ohja2023}
\vspace{-10pt}
\end{table*}

\begin{table*}[t]
\caption{\gls{auc}, \gls{ba}, \gls{fpr}, and \gls{fnr} for JPEG AI compression scenario for the \cite{cozzolino2023raising}-A detector.
}
\centering
\resizebox{\textwidth}{!}{
\begin{tabular}{l 
    S[table-format=1.2]@{\,/\,}S[table-format=2.2]@{\,/\,} S[table-format=1.2]@{\,/\,}S[table-format=1.2] |
    S[table-format=1.2]@{\,/\,}S[table-format=2.2]@{\,/\,}S[table-format=1.2]@{\,/\,}S[table-format=1.2]
    S[table-format=1.2]@{\,/\,}S[table-format=2.2]@{\,/\,}S[table-format=1.2]@{\,/\,}S[table-format=1.2]
    S[table-format=1.2]@{\,/\,}S[table-format=2.2]@{\,/\,}S[table-format=1.2]@{\,/\,}S[table-format=1.2]
    S[table-format=1.2]@{\,/\,}S[table-format=2.2]@{\,/\,}S[table-format=1.2]@{\,/\,}S[table-format=1.2]
    S[table-format=1.2]@{\,/\,}S[table-format=2.2]@{\,/\,}S[table-format=1.2]@{\,/\,}S[table-format=1.2]
    S[table-format=1.2]@{\,/\,}S[table-format=1.2]@{\,/\,}S[table-format=1.2]@{\,/\,}S[table-format=1.2]
}
\toprule
 & \multicolumn{4}{c |}{Original format} & \multicolumn{4}{c}{BPP = 0.12} & \multicolumn{4}{c}{BPP = 0.25} & 
 \multicolumn{4}{c}{BPP = 0.5} &
 \multicolumn{4}{c}{BPP = 0.75} & \multicolumn{4}{c}{BPP = 1.0} & \multicolumn{4}{c}{BPP = 2.0} \\
 \midrule
& \multicolumn{4}{c |}{$\textrm{AUC}$ /
$\textrm{BA}$ / $\textrm{FPR}$ / $\textrm{FNR}$} & \multicolumn{4}{c |}{$\textrm{AUC}$ /
$\textrm{BA}$ / $\textrm{FPR}$ / $\textrm{FNR}$} & \multicolumn{4}{c |}{$\textrm{AUC}$ /
$\textrm{BA}$ / $\textrm{FPR}$ / $\textrm{FNR}$} &
\multicolumn{4}{c |}{$\textrm{AUC}$ /
$\textrm{BA}$ / $\textrm{FPR}$ / $\textrm{FNR}$} &
\multicolumn{4}{c |}{$\textrm{AUC}$ /
$\textrm{BA}$ / $\textrm{FPR}$ / $\textrm{FNR}$} & \multicolumn{4}{c |}{$\textrm{AUC}$ /
$\textrm{BA}$ / $\textrm{FPR}$ / $\textrm{FNR}$} & \multicolumn{4}{c |}{$\textrm{AUC}$ /
$\textrm{BA}$ / $\textrm{FPR}$ / $\textrm{FNR}$} \\
\midrule
Imagenet & 0.90 & 0.70 & 0.57 & 0.04 & 0.59 & 0.50 & 0.99 & 0.00 & 0.69 & 0.52 & 0.94 & 0.01 & 0.75 & 0.56 & 0.87 & 0.02 & 0.79 & 0.59 & 0.79 & 0.02 & 0.81 & 0.62 & 0.74 & 0.03 & 0.85 & 0.63 & 0.71 & 0.03 \\
COCO & 0.91 & 0.74 & 0.44 & 0.09 & 0.50 & 0.47 & 0.99 & 0.06 & 0.59 & 0.48 & 0.95 & 0.09 & 0.69 & 0.52 & 0.86 & 0.10 & 0.74 & 0.59 & 0.75 & 0.07 & 0.81 & 0.65 & 0.63 & 0.08 & 0.83 & 0.67 & 0.57 & 0.09 \\
FFHQ & 0.96 & 0.90 & 0.09 & 0.10 & 0.95 & 0.90 & 0.07 & 0.14 & 0.96 & 0.91 & 0.04 & 0.13 & 0.94 & 0.88 & 0.10 & 0.14 & 0.95 & 0.90 & 0.07 & 0.14 & 0.95 & 0.90 & 0.07 & 0.14 & 0.97 & 0.91 & 0.06 & 0.11 \\
LSUN & 0.95 & 0.87 & 0.18 & 0.09 & 0.50 & 0.51 & 0.99 & 0.00 & 0.52 & 0.52 & 0.94 & 0.02 & 0.59 & 0.52 & 0.87 & 0.09 & 0.65 & 0.55 & 0.81 & 0.10 & 0.64 & 0.54 & 0.81 & 0.12 & 0.81 & 0.64 & 0.68 & 0.04 \\
LAION & 0.96 & 0.80 & 0.37 & 0.03 & 0.52 & 0.50 & 0.99 & 0.01 & 0.64 & 0.51 & 0.97 & 0.01 & 0.73 & 0.54 & 0.91 & 0.01 & 0.75 & 0.56 & 0.87 & 0.02 & 0.75 & 0.55 & 0.89 & 0.02 & 0.80 & 0.57 & 0.83 & 0.02 \\
RAISE & 0.80 & 0.65 & 0.11 & 0.60 & 0.68 & 0.65 & 0.23 & 0.47 & 0.73 & 0.65 & 0.11 & 0.59 & 0.58 & 0.55 & 0.25 & 0.64 & 0.79 & 0.64 & 0.09 & 0.62 & 0.80 & 0.64 & 0.10 & 0.62 & 0.79 & 0.64 & 0.12 & 0.61 \\
\bottomrule
\end{tabular}
}
\label{tab:jpegai_metrics_cozzolino2024A}
\vspace{-10pt}
\end{table*}

\begin{table*}[t]
\caption{\gls{auc}, \gls{ba}, \gls{fpr}, and \gls{fnr} for JPEG AI compression scenario for the \cite{cozzolino2023raising}-B detector. 
}
\centering
\resizebox{\textwidth}{!}{
\begin{tabular}{l 
    S[table-format=1.2]@{\,/\,}S[table-format=2.2]@{\,/\,} S[table-format=1.2]@{\,/\,}S[table-format=1.2] |
    S[table-format=1.2]@{\,/\,}S[table-format=2.2]@{\,/\,}S[table-format=1.2]@{\,/\,}S[table-format=1.2]
    S[table-format=1.2]@{\,/\,}S[table-format=2.2]@{\,/\,}S[table-format=1.2]@{\,/\,}S[table-format=1.2]
    S[table-format=1.2]@{\,/\,}S[table-format=2.2]@{\,/\,}S[table-format=1.2]@{\,/\,}S[table-format=1.2]
    S[table-format=1.2]@{\,/\,}S[table-format=2.2]@{\,/\,}S[table-format=1.2]@{\,/\,}S[table-format=1.2]
    S[table-format=1.2]@{\,/\,}S[table-format=2.2]@{\,/\,}S[table-format=1.2]@{\,/\,}S[table-format=1.2]
    S[table-format=1.2]@{\,/\,}S[table-format=1.2]@{\,/\,}S[table-format=1.2]@{\,/\,}S[table-format=1.2]
}
\toprule
 & \multicolumn{4}{c |}{Original format} & \multicolumn{4}{c}{BPP = 0.12} & \multicolumn{4}{c}{BPP = 0.25} & 
 \multicolumn{4}{c}{BPP = 0.5} &
 \multicolumn{4}{c}{BPP = 0.75} & \multicolumn{4}{c}{BPP = 1.0} & \multicolumn{4}{c}{BPP = 2.0} \\
 \midrule
& \multicolumn{4}{c |}{$\textrm{AUC}$ /
$\textrm{BA}$ / $\textrm{FPR}$ / $\textrm{FNR}$} & \multicolumn{4}{c |}{$\textrm{AUC}$ /
$\textrm{BA}$ / $\textrm{FPR}$ / $\textrm{FNR}$} & \multicolumn{4}{c |}{$\textrm{AUC}$ /
$\textrm{BA}$ / $\textrm{FPR}$ / $\textrm{FNR}$} &
\multicolumn{4}{c |}{$\textrm{AUC}$ /
$\textrm{BA}$ / $\textrm{FPR}$ / $\textrm{FNR}$} &
\multicolumn{4}{c |}{$\textrm{AUC}$ /
$\textrm{BA}$ / $\textrm{FPR}$ / $\textrm{FNR}$} & \multicolumn{4}{c |}{$\textrm{AUC}$ /
$\textrm{BA}$ / $\textrm{FPR}$ / $\textrm{FNR}$} & \multicolumn{4}{c |}{$\textrm{AUC}$ /
$\textrm{BA}$ / $\textrm{FPR}$ / $\textrm{FNR}$} \\
\midrule
Imagenet & 0.76 & 0.68 & 0.42 & 0.21 & 0.61 & 0.51 & 0.98 & 0.00 & 0.68 & 0.53 & 0.94 & 0.01 & 0.71 & 0.55 & 0.87 & 0.03 & 0.72 & 0.57 & 0.82 & 0.03 & 0.74 & 0.59 & 0.78 & 0.04 & 0.75 & 0.60 & 0.74 & 0.06 \\
COCO & 0.86 & 0.77 & 0.32 & 0.14 & 0.53 & 0.49 & 0.98 & 0.04 & 0.65 & 0.51 & 0.92 & 0.06 & 0.73 & 0.55 & 0.83 & 0.08 & 0.77 & 0.58 & 0.76 & 0.07 & 0.80 & 0.62 & 0.67 & 0.08 & 0.83 & 0.68 & 0.55 & 0.10 \\
FFHQ & 0.88 & 0.77 & 0.10 & 0.36 & 0.96 & 0.86 & 0.21 & 0.08 & 0.96 & 0.88 & 0.15 & 0.09 & 0.95 & 0.88 & 0.14 & 0.11 & 0.94 & 0.87 & 0.11 & 0.15 & 0.94 & 0.87 & 0.10 & 0.17 & 0.95 & 0.89 & 0.09 & 0.14 \\
LSUN & 0.79 & 0.68 & 0.49 & 0.14 & 0.52 & 0.51 & 0.98 & 0.01 & 0.56 & 0.52 & 0.93 & 0.03 & 0.62 & 0.53 & 0.85 & 0.08 & 0.67 & 0.55 & 0.80 & 0.09 & 0.66 & 0.55 & 0.79 & 0.11 & 0.72 & 0.59 & 0.73 & 0.08 \\
RAISE & 0.86 & 0.77 & 0.15 & 0.31 & 0.76 & 0.67 & 0.47 & 0.20 & 0.79 & 0.72 & 0.29 & 0.27 & 0.74 & 0.68 & 0.35 & 0.29 & 0.84 & 0.75 & 0.20 & 0.29 & 0.84 & 0.76 & 0.19 & 0.29 & 0.85 & 0.76 & 0.17 & 0.32 \\
\bottomrule
\end{tabular}
}
\label{tab:jpegai_metrics_cozzolino2024b}
\vspace{-10pt}
\end{table*}

\begin{table*}[t]
\caption{\gls{auc}, \gls{ba}, \gls{fpr}, and \gls{fnr} for JPEG AI compression scenario for the \cite{Mandelli2024} detector. 
}
\centering
\resizebox{\textwidth}{!}{
\begin{tabular}{l 
    S[table-format=1.2]@{\,/\,}S[table-format=2.2]@{\,/\,} S[table-format=1.2]@{\,/\,}S[table-format=1.2] |
    S[table-format=1.2]@{\,/\,}S[table-format=2.2]@{\,/\,}S[table-format=1.2]@{\,/\,}S[table-format=1.2]
    S[table-format=1.2]@{\,/\,}S[table-format=2.2]@{\,/\,}S[table-format=1.2]@{\,/\,}S[table-format=1.2]
    S[table-format=1.2]@{\,/\,}S[table-format=2.2]@{\,/\,}S[table-format=1.2]@{\,/\,}S[table-format=1.2]
    S[table-format=1.2]@{\,/\,}S[table-format=2.2]@{\,/\,}S[table-format=1.2]@{\,/\,}S[table-format=1.2]
    S[table-format=1.2]@{\,/\,}S[table-format=2.2]@{\,/\,}S[table-format=1.2]@{\,/\,}S[table-format=1.2]
    S[table-format=1.2]@{\,/\,}S[table-format=1.2]@{\,/\,}S[table-format=1.2]@{\,/\,}S[table-format=1.2]
}
\toprule
 & \multicolumn{4}{c |}{Original format} & \multicolumn{4}{c}{BPP = 0.12} & \multicolumn{4}{c}{BPP = 0.25} & 
 \multicolumn{4}{c}{BPP = 0.5} &
 \multicolumn{4}{c}{BPP = 0.75} & \multicolumn{4}{c}{BPP = 1.0} & \multicolumn{4}{c}{BPP = 2.0} \\
 \midrule
& \multicolumn{4}{c |}{$\textrm{AUC}$ /
$\textrm{BA}$ / $\textrm{FPR}$ / $\textrm{FNR}$} & \multicolumn{4}{c |}{$\textrm{AUC}$ /
$\textrm{BA}$ / $\textrm{FPR}$ / $\textrm{FNR}$} & \multicolumn{4}{c |}{$\textrm{AUC}$ /
$\textrm{BA}$ / $\textrm{FPR}$ / $\textrm{FNR}$} &
\multicolumn{4}{c |}{$\textrm{AUC}$ /
$\textrm{BA}$ / $\textrm{FPR}$ / $\textrm{FNR}$} &
\multicolumn{4}{c |}{$\textrm{AUC}$ /
$\textrm{BA}$ / $\textrm{FPR}$ / $\textrm{FNR}$} & \multicolumn{4}{c |}{$\textrm{AUC}$ /
$\textrm{BA}$ / $\textrm{FPR}$ / $\textrm{FNR}$} & \multicolumn{4}{c |}{$\textrm{AUC}$ /
$\textrm{BA}$ / $\textrm{FPR}$ / $\textrm{FNR}$} \\
\midrule
CelebA & 1.00 & 1.00 & 0.00 & 0.00 & 0.81 & 0.50 & 1.00 & 0.00 & 0.92 & 0.54 & 0.93 & 0.00 & 0.98 & 0.91 & 0.18 & 0.00 & 1.00 & 0.99 & 0.02 & 0.00 & 1.00 & 0.99 & 0.01 & 0.00 & 1.00 & 1.00 & 0.00 & 0.00 \\
FFHQ & 1.00 & 1.00 & 0.00 & 0.00 & 0.97 & 0.79 & 0.39 & 0.03 & 0.99 & 0.94 & 0.07 & 0.05 & 1.00 & 0.99 & 0.00 & 0.01 & 1.00 & 1.00 & 0.00 & 0.00 & 1.00 & 1.00 & 0.00 & 0.00 & 1.00 & 1.00 & 0.00 & 0.00 \\
Imagenet & 0.71 & 0.62 & 0.13 & 0.64 & 0.51 & 0.50 & 0.72 & 0.28 & 0.53 & 0.52 & 0.50 & 0.47 & 0.55 & 0.54 & 0.39 & 0.54 & 0.57 & 0.56 & 0.33 & 0.56 & 0.59 & 0.56 & 0.28 & 0.60 & 0.64 & 0.59 & 0.23 & 0.59 \\
LAION & 0.72 & 0.68 & 0.18 & 0.46 & 0.61 & 0.59 & 0.51 & 0.32 & 0.63 & 0.60 & 0.39 & 0.41 & 0.69 & 0.63 & 0.32 & 0.42 & 0.72 & 0.66 & 0.25 & 0.44 & 0.73 & 0.67 & 0.23 & 0.43 & 0.73 & 0.66 & 0.22 & 0.46 \\
LSUN & 0.83 & 0.76 & 0.12 & 0.37 & 0.50 & 0.51 & 0.82 & 0.16 & 0.61 & 0.59 & 0.55 & 0.27 & 0.58 & 0.55 & 0.50 & 0.40 & 0.63 & 0.59 & 0.38 & 0.44 & 0.63 & 0.60 & 0.32 & 0.49 & 0.73 & 0.68 & 0.20 & 0.44 \\
RAISE & 0.86 & 0.80 & 0.02 & 0.38 & 0.63 & 0.59 & 0.34 & 0.48 & 0.75 & 0.68 & 0.14 & 0.51 & 0.83 & 0.75 & 0.03 & 0.48 & 0.81 & 0.75 & 0.04 & 0.46 & 0.82 & 0.76 & 0.02 & 0.46 & 0.81 & 0.78 & 0.03 & 0.42 \\
\bottomrule
\end{tabular}
}
\label{tab:jpegai_metrics_mandelli}
\vspace{-10pt}
\end{table*}

\begin{table*}[t]
\caption{\gls{auc}, \gls{ba}, \gls{fpr}, and \gls{fnr} for JPEG AI compression scenario for the \cite{Tan_2024_CVPR} detector.
}
\centering
\resizebox{\textwidth}{!}{
\begin{tabular}{l 
    S[table-format=1.2]@{\,/\,}S[table-format=2.2]@{\,/\,} S[table-format=1.2]@{\,/\,}S[table-format=1.2] |
    S[table-format=1.2]@{\,/\,}S[table-format=2.2]@{\,/\,}S[table-format=1.2]@{\,/\,}S[table-format=1.2]
    S[table-format=1.2]@{\,/\,}S[table-format=2.2]@{\,/\,}S[table-format=1.2]@{\,/\,}S[table-format=1.2]
    S[table-format=1.2]@{\,/\,}S[table-format=2.2]@{\,/\,}S[table-format=1.2]@{\,/\,}S[table-format=1.2]
    S[table-format=1.2]@{\,/\,}S[table-format=2.2]@{\,/\,}S[table-format=1.2]@{\,/\,}S[table-format=1.2]
    S[table-format=1.2]@{\,/\,}S[table-format=2.2]@{\,/\,}S[table-format=1.2]@{\,/\,}S[table-format=1.2]
    S[table-format=1.2]@{\,/\,}S[table-format=1.2]@{\,/\,}S[table-format=1.2]@{\,/\,}S[table-format=1.2]
}
\toprule
 & \multicolumn{4}{c |}{Original format} & \multicolumn{4}{c}{BPP = 0.12} & \multicolumn{4}{c}{BPP = 0.25} & 
 \multicolumn{4}{c}{BPP = 0.5} &
 \multicolumn{4}{c}{BPP = 0.75} & \multicolumn{4}{c}{BPP = 1.0} & \multicolumn{4}{c}{BPP = 2.0} \\
 \midrule
& \multicolumn{4}{c |}{$\textrm{AUC}$ /
$\textrm{BA}$ / $\textrm{FPR}$ / $\textrm{FNR}$} & \multicolumn{4}{c |}{$\textrm{AUC}$ /
$\textrm{BA}$ / $\textrm{FPR}$ / $\textrm{FNR}$} & \multicolumn{4}{c |}{$\textrm{AUC}$ /
$\textrm{BA}$ / $\textrm{FPR}$ / $\textrm{FNR}$} &
\multicolumn{4}{c |}{$\textrm{AUC}$ /
$\textrm{BA}$ / $\textrm{FPR}$ / $\textrm{FNR}$} &
\multicolumn{4}{c |}{$\textrm{AUC}$ /
$\textrm{BA}$ / $\textrm{FPR}$ / $\textrm{FNR}$} & \multicolumn{4}{c |}{$\textrm{AUC}$ /
$\textrm{BA}$ / $\textrm{FPR}$ / $\textrm{FNR}$} & \multicolumn{4}{c |}{$\textrm{AUC}$ /
$\textrm{BA}$ / $\textrm{FPR}$ / $\textrm{FNR}$} \\
\midrule
Imagenet & 0.94 & 0.84 & 0.30 & 0.01 & 0.50 & 0.51 & 0.99 & 0.00 & 0.46 & 0.51 & 0.98 & 0.00 & 0.44 & 0.51 & 0.98 & 0.00 & 0.46 & 0.51 & 0.98 & 0.01 & 0.47 & 0.51 & 0.97 & 0.01 & 0.52 & 0.51 & 0.97 & 0.01 \\
COCO & 0.91 & 0.82 & 0.34 & 0.01 & 0.67 & 0.50 & 1.00 & 0.00 & 0.62 & 0.50 & 0.99 & 0.00 & 0.60 & 0.51 & 0.99 & 0.00 & 0.63 & 0.51 & 0.98 & 0.00 & 0.58 & 0.51 & 0.97 & 0.00 & 0.57 & 0.51 & 0.97 & 0.00 \\
CelebA & 1.00 & 0.99 & 0.01 & 0.00 & 0.67 & 0.50 & 1.00 & 0.00 & 0.74 & 0.50 & 1.00 & 0.00 & 0.60 & 0.50 & 1.00 & 0.00 & 0.81 & 0.50 & 1.00 & 0.00 & 0.77 & 0.50 & 1.00 & 0.00 & 0.92 & 0.50 & 1.00 & 0.00 \\
LSUN & 1.00 & 0.99 & 0.00 & 0.03 & 0.64 & 0.50 & 1.00 & 0.00 & 0.61 & 0.50 & 1.00 & 0.00 & 0.53 & 0.50 & 0.99 & 0.00 & 0.55 & 0.50 & 0.99 & 0.00 & 0.54 & 0.50 & 0.99 & 0.01 & 0.61 & 0.51 & 0.97 & 0.02 \\
LAION & 0.99 & 0.97 & 0.02 & 0.05 & 0.80 & 0.51 & 0.97 & 0.00 & 0.75 & 0.51 & 0.97 & 0.00 & 0.75 & 0.52 & 0.96 & 0.00 & 0.74 & 0.52 & 0.95 & 0.01 & 0.79 & 0.53 & 0.92 & 0.01 & 0.82 & 0.55 & 0.89 & 0.01 \\
\bottomrule
\end{tabular}
}
\label{tab:jpegai_metrics_npr}
\vspace{-10pt}
\end{table*}

\noindent \textbf{Comparison with standard JPEG.} \cref{tab:jpeg_metrics_wang2020a} to \cref{tab:jpeg_metrics_npr} report the results of the experiments varying the quality factor of JPEG for each detector on every single dataset. For all tables, as commonly done in deepfake detection tasks, we evaluate the \gls{ba}, \gls{fpr}, and \gls{fnr} at the $0$ threshold. 


\begin{table*}
\caption{\gls{auc}, \gls{ba}, \gls{fpr}, and \gls{fnr} for the JPEG compression scenario for the \cite{wang2020}-A detector.
}
\centering
\resizebox{\textwidth}{!}{
\begin{tabular}{l S[table-format=2.2]@{\,/\,} S[table-format=1.2]@{\,/\,}S[table-format=1.2]@{\,/\,}S[table-format=1.2] |
    S[table-format=2.2]@{\,/\,}S[table-format=1.2]@{\,/\,}S[table-format=1.2]@{\,/\,}S[table-format=1.2]
    S[table-format=2.2]@{\,/\,}S[table-format=1.2]@{\,/\,}S[table-format=1.2]@{\,/\,}S[table-format=1.2]
    S[table-format=2.2]@{\,/\,}S[table-format=1.2]@{\,/\,}S[table-format=1.2]@{\,/\,}S[table-format=1.2]
    S[table-format=2.2]@{\,/\,}S[table-format=1.2]@{\,/\,}S[table-format=1.2]@{\,/\,}S[table-format=1.2]
}
\toprule
 & \multicolumn{4}{c |}{Original format} & \multicolumn{4}{c}{QF = 65} & \multicolumn{4}{c}{BPP = 75} & 
 \multicolumn{4}{c}{QF = 85} &
 \multicolumn{4}{c}{QF = 95} \\
 \midrule
& \multicolumn{4}{c |}{$\textrm{AUC}$ /
$\textrm{BA}$ / $\textrm{FPR}$ / $\textrm{FNR}$} & \multicolumn{4}{c}{$\textrm{AUC}$ /
$\textrm{BA}$ / $\textrm{FPR}$ / $\textrm{FNR}$} & \multicolumn{4}{c}{$\textrm{AUC}$ /
$\textrm{BA}$ / $\textrm{FPR}$ / $\textrm{FNR}$} &
\multicolumn{4}{c}{$\textrm{AUC}$ /
$\textrm{BA}$ / $\textrm{FPR}$ / $\textrm{FNR}$} &
\multicolumn{4}{c}{$\textrm{AUC}$ /
$\textrm{BA}$ / $\textrm{FPR}$ / $\textrm{FNR}$} \\
\midrule
Imagenet & 0.88 & 0.70 & 0.07 & 0.53 & 0.88 & 0.60 & 0.01 & 0.79 & 0.88 & 0.60 & 0.01 & 0.79 & 0.89 & 0.60 & 0.01 & 0.80 & 0.89 & 0.62 & 0.01 & 0.74 \\
COCO & 0.92 & 0.79 & 0.07 & 0.35 & 0.96 & 0.70 & 0.00 & 0.59 & 0.96 & 0.69 & 0.00 & 0.61 & 0.96 & 0.68 & 0.00 & 0.64 & 0.96 & 0.70 & 0.01 & 0.60 \\
CelebA & 0.98 & 0.92 & 0.03 & 0.13 & 0.89 & 0.76 & 0.04 & 0.44 & 0.89 & 0.75 & 0.04 & 0.47 & 0.91 & 0.75 & 0.02 & 0.47 & 0.93 & 0.79 & 0.02 & 0.40 \\
LSUN & 0.99 & 0.89 & 0.00 & 0.22 & 0.95 & 0.72 & 0.00 & 0.56 & 0.96 & 0.74 & 0.00 & 0.53 & 0.97 & 0.75 & 0.00 & 0.50 & 0.99 & 0.79 & 0.00 & 0.42 \\

\bottomrule
\end{tabular}
}
\label{tab:jpeg_metrics_wang2020a}
\vspace{-10pt}
\end{table*}

\begin{table*}
\caption{\gls{auc}, \gls{ba}, \gls{fpr}, and \gls{fnr} for the JPEG compression scenario for the \cite{wang2020}-B detector.
}
\centering
\resizebox{\textwidth}{!}{
\begin{tabular}{l S[table-format=2.2]@{\,/\,} S[table-format=1.2]@{\,/\,}S[table-format=1.2]@{\,/\,}S[table-format=1.2] |
    S[table-format=2.2]@{\,/\,}S[table-format=1.2]@{\,/\,}S[table-format=1.2]@{\,/\,}S[table-format=1.2]
    S[table-format=2.2]@{\,/\,}S[table-format=1.2]@{\,/\,}S[table-format=1.2]@{\,/\,}S[table-format=1.2]
    S[table-format=2.2]@{\,/\,}S[table-format=1.2]@{\,/\,}S[table-format=1.2]@{\,/\,}S[table-format=1.2]
    S[table-format=2.2]@{\,/\,}S[table-format=1.2]@{\,/\,}S[table-format=1.2]@{\,/\,}S[table-format=1.2]
}
\toprule
 & \multicolumn{4}{c |}{Original format} & \multicolumn{4}{c}{QF = 65} & \multicolumn{4}{c}{BPP = 75} & 
 \multicolumn{4}{c}{QF = 85} &
 \multicolumn{4}{c}{QF = 95} \\
 \midrule
& \multicolumn{4}{c |}{$\textrm{AUC}$ /
$\textrm{BA}$ / $\textrm{FPR}$ / $\textrm{FNR}$} & \multicolumn{4}{c}{$\textrm{AUC}$ /
$\textrm{BA}$ / $\textrm{FPR}$ / $\textrm{FNR}$} & \multicolumn{4}{c}{$\textrm{AUC}$ /
$\textrm{BA}$ / $\textrm{FPR}$ / $\textrm{FNR}$} &
\multicolumn{4}{c}{$\textrm{AUC}$ /
$\textrm{BA}$ / $\textrm{FPR}$ / $\textrm{FNR}$} &
\multicolumn{4}{c}{$\textrm{AUC}$ /
$\textrm{BA}$ / $\textrm{FPR}$ / $\textrm{FNR}$} \\
\midrule
Imagenet & 0.90 & 0.59 & 0.01 & 0.81 & 0.89 & 0.57 & 0.01 & 0.86 & 0.89 & 0.57 & 0.00 & 0.86 & 0.90 & 0.57 & 0.01 & 0.86 & 0.90 & 0.57 & 0.01 & 0.85 \\
COCO & 0.98 & 0.79 & 0.01 & 0.41 & 0.99 & 0.75 & 0.00 & 0.49 & 0.99 & 0.75 & 0.00 & 0.50 & 0.99 & 0.75 & 0.00 & 0.51 & 0.99 & 0.76 & 0.00 & 0.48 \\
CelebA & 0.95 & 0.81 & 0.01 & 0.37 & 0.91 & 0.73 & 0.01 & 0.52 & 0.92 & 0.74 & 0.01 & 0.50 & 0.92 & 0.75 & 0.01 & 0.50 & 0.93 & 0.76 & 0.01 & 0.47 \\
LSUN & 0.98 & 0.77 & 0.00 & 0.46 & 0.96 & 0.69 & 0.00 & 0.61 & 0.96 & 0.70 & 0.00 & 0.59 & 0.97 & 0.71 & 0.00 & 0.58 & 0.98 & 0.73 & 0.00 & 0.54 \\

\bottomrule
\end{tabular}
}
\label{tab:jpeg_metrics_wang2020b}
\vspace{-10pt}
\end{table*}

\begin{table*}
\caption{\gls{auc}, \gls{ba}, \gls{fpr}, and \gls{fnr} for the JPEG compression scenario for the \cite{gragnaniello2021} detector. 
}
\centering
\resizebox{\textwidth}{!}{
\begin{tabular}{l S[table-format=2.2]@{\,/\,} S[table-format=1.2]@{\,/\,}S[table-format=1.2]@{\,/\,}S[table-format=1.2] |
    S[table-format=2.2]@{\,/\,}S[table-format=1.2]@{\,/\,}S[table-format=1.2]@{\,/\,}S[table-format=1.2]
    S[table-format=2.2]@{\,/\,}S[table-format=1.2]@{\,/\,}S[table-format=1.2]@{\,/\,}S[table-format=1.2]
    S[table-format=2.2]@{\,/\,}S[table-format=1.2]@{\,/\,}S[table-format=1.2]@{\,/\,}S[table-format=1.2]
    S[table-format=2.2]@{\,/\,}S[table-format=1.2]@{\,/\,}S[table-format=1.2]@{\,/\,}S[table-format=1.2]
}
\toprule
 & \multicolumn{4}{c |}{Original format} & \multicolumn{4}{c}{QF = 65} & \multicolumn{4}{c}{BPP = 75} & 
 \multicolumn{4}{c}{QF = 85} &
 \multicolumn{4}{c}{QF = 95} \\
 \midrule
& \multicolumn{4}{c |}{$\textrm{AUC}$ /
$\textrm{BA}$ / $\textrm{FPR}$ / $\textrm{FNR}$} & \multicolumn{4}{c}{$\textrm{AUC}$ /
$\textrm{BA}$ / $\textrm{FPR}$ / $\textrm{FNR}$} & \multicolumn{4}{c}{$\textrm{AUC}$ /
$\textrm{BA}$ / $\textrm{FPR}$ / $\textrm{FNR}$} &
\multicolumn{4}{c}{$\textrm{AUC}$ /
$\textrm{BA}$ / $\textrm{FPR}$ / $\textrm{FNR}$} &
\multicolumn{4}{c}{$\textrm{AUC}$ /
$\textrm{BA}$ / $\textrm{FPR}$ / $\textrm{FNR}$} \\
\midrule
Imagenet & 0.89 & 0.68 & 0.01 & 0.64 & 0.84 & 0.62 & 0.01 & 0.76 & 0.85 & 0.62 & 0.01 & 0.75 & 0.86 & 0.63 & 0.00 & 0.74 & 0.88 & 0.64 & 0.00 & 0.71 \\
COCO & 0.96 & 0.67 & 0.01 & 0.66 & 0.95 & 0.55 & 0.00 & 0.89 & 0.96 & 0.56 & 0.00 & 0.87 & 0.97 & 0.59 & 0.00 & 0.82 & 0.97 & 0.63 & 0.00 & 0.75 \\
FFHQ & 0.93 & 0.80 & 0.06 & 0.33 & 0.42 & 0.55 & 0.00 & 0.91 & 0.45 & 0.56 & 0.00 & 0.88 & 0.52 & 0.58 & 0.01 & 0.84 & 0.77 & 0.62 & 0.02 & 0.74 \\

\bottomrule
\end{tabular}
}
\label{tab:jpeg_metrics_gragnaniello2021}
\vspace{-10pt}
\end{table*}

\begin{table*}
\caption{\gls{auc}, \gls{ba}, \gls{fpr}, and \gls{fnr} for the JPEG compression scenario for the \cite{Corvi_2023_ICASSP} detector.
}
\centering
\resizebox{\textwidth}{!}{
\begin{tabular}{l S[table-format=2.2]@{\,/\,} S[table-format=1.2]@{\,/\,}S[table-format=1.2]@{\,/\,}S[table-format=1.2] |
    S[table-format=2.2]@{\,/\,}S[table-format=1.2]@{\,/\,}S[table-format=1.2]@{\,/\,}S[table-format=1.2]
    S[table-format=2.2]@{\,/\,}S[table-format=1.2]@{\,/\,}S[table-format=1.2]@{\,/\,}S[table-format=1.2]
    S[table-format=2.2]@{\,/\,}S[table-format=1.2]@{\,/\,}S[table-format=1.2]@{\,/\,}S[table-format=1.2]
    S[table-format=2.2]@{\,/\,}S[table-format=1.2]@{\,/\,}S[table-format=1.2]@{\,/\,}S[table-format=1.2]
}
\toprule
 & \multicolumn{4}{c |}{Original format} & \multicolumn{4}{c}{QF = 65} & \multicolumn{4}{c}{BPP = 75} & 
 \multicolumn{4}{c}{QF = 85} &
 \multicolumn{4}{c}{QF = 95} \\
 \midrule
& \multicolumn{4}{c |}{$\textrm{AUC}$ /
$\textrm{BA}$ / $\textrm{FPR}$ / $\textrm{FNR}$} & \multicolumn{4}{c}{$\textrm{AUC}$ /
$\textrm{BA}$ / $\textrm{FPR}$ / $\textrm{FNR}$} & \multicolumn{4}{c}{$\textrm{AUC}$ /
$\textrm{BA}$ / $\textrm{FPR}$ / $\textrm{FNR}$} &
\multicolumn{4}{c}{$\textrm{AUC}$ /
$\textrm{BA}$ / $\textrm{FPR}$ / $\textrm{FNR}$} &
\multicolumn{4}{c}{$\textrm{AUC}$ /
$\textrm{BA}$ / $\textrm{FPR}$ / $\textrm{FNR}$} \\
\midrule
Imagenet & 0.87 & 0.70 & 0.01 & 0.59 & 0.83 & 0.72 & 0.04 & 0.53 & 0.83 & 0.71 & 0.03 & 0.56 & 0.84 & 0.70 & 0.01 & 0.59 & 0.87 & 0.69 & 0.01 & 0.61 \\
COCO & 0.96 & 0.74 & 0.00 & 0.52 & 0.98 & 0.81 & 0.00 & 0.37 & 0.98 & 0.79 & 0.00 & 0.42 & 0.98 & 0.75 & 0.00 & 0.50 & 0.97 & 0.73 & 0.00 & 0.54 \\
FFHQ & 0.89 & 0.87 & 0.00 & 0.27 & 0.93 & 0.81 & 0.28 & 0.10 & 0.93 & 0.84 & 0.19 & 0.13 & 0.92 & 0.87 & 0.06 & 0.20 & 0.91 & 0.86 & 0.00 & 0.28 \\

\bottomrule
\end{tabular}
}
\label{tab:jpeg_metrics_corvi2023}
\vspace{-10pt}
\end{table*}

\begin{table*}
\caption{\gls{auc}, \gls{ba}, \gls{fpr}, and \gls{fnr} for the JPEG compression scenario for the \cite{Ojha_2023_CVPR} detector.
}
\centering
\resizebox{\textwidth}{!}{
\begin{tabular}{l S[table-format=2.2]@{\,/\,} S[table-format=1.2]@{\,/\,}S[table-format=1.2]@{\,/\,}S[table-format=1.2] |
    S[table-format=2.2]@{\,/\,}S[table-format=1.2]@{\,/\,}S[table-format=1.2]@{\,/\,}S[table-format=1.2]
    S[table-format=2.2]@{\,/\,}S[table-format=1.2]@{\,/\,}S[table-format=1.2]@{\,/\,}S[table-format=1.2]
    S[table-format=2.2]@{\,/\,}S[table-format=1.2]@{\,/\,}S[table-format=1.2]@{\,/\,}S[table-format=1.2]
    S[table-format=2.2]@{\,/\,}S[table-format=1.2]@{\,/\,}S[table-format=1.2]@{\,/\,}S[table-format=1.2]
}
\toprule
 & \multicolumn{4}{c |}{Original format} & \multicolumn{4}{c}{QF = 65} & \multicolumn{4}{c}{BPP = 75} & 
 \multicolumn{4}{c}{QF = 85} &
 \multicolumn{4}{c}{QF = 95} \\
 \midrule
& \multicolumn{4}{c |}{$\textrm{AUC}$ /
$\textrm{BA}$ / $\textrm{FPR}$ / $\textrm{FNR}$} & \multicolumn{4}{c}{$\textrm{AUC}$ /
$\textrm{BA}$ / $\textrm{FPR}$ / $\textrm{FNR}$} & \multicolumn{4}{c}{$\textrm{AUC}$ /
$\textrm{BA}$ / $\textrm{FPR}$ / $\textrm{FNR}$} &
\multicolumn{4}{c}{$\textrm{AUC}$ /
$\textrm{BA}$ / $\textrm{FPR}$ / $\textrm{FNR}$} &
\multicolumn{4}{c}{$\textrm{AUC}$ /
$\textrm{BA}$ / $\textrm{FPR}$ / $\textrm{FNR}$} \\
\midrule
Imagenet & 0.99 & 0.95 & 0.02 & 0.08 & 0.90 & 0.72 & 0.03 & 0.53 & 0.91 & 0.75 & 0.02 & 0.47 & 0.94 & 0.76 & 0.01 & 0.46 & 0.97 & 0.86 & 0.01 & 0.28 \\
COCO & 1.00 & 0.99 & 0.01 & 0.00 & 0.99 & 0.93 & 0.01 & 0.13 & 0.99 & 0.95 & 0.01 & 0.10 & 1.00 & 0.96 & 0.01 & 0.07 & 1.00 & 0.98 & 0.01 & 0.03 \\
CelebA & 0.99 & 0.96 & 0.07 & 0.02 & 0.97 & 0.91 & 0.07 & 0.10 & 0.97 & 0.92 & 0.07 & 0.09 & 0.99 & 0.94 & 0.03 & 0.10 & 0.99 & 0.95 & 0.06 & 0.04 \\
LSUN & 0.98 & 0.84 & 0.00 & 0.32 & 0.91 & 0.71 & 0.01 & 0.58 & 0.92 & 0.72 & 0.01 & 0.54 & 0.94 & 0.72 & 0.01 & 0.55 & 0.96 & 0.79 & 0.00 & 0.42 \\
LAION & 0.96 & 0.82 & 0.01 & 0.35 & 0.91 & 0.75 & 0.03 & 0.48 & 0.91 & 0.76 & 0.02 & 0.45 & 0.92 & 0.74 & 0.01 & 0.52 & 0.94 & 0.78 & 0.01 & 0.42 \\

\bottomrule
\end{tabular}
}
\label{tab:jpeg_metrics_ohja2023}
\vspace{-10pt}
\end{table*}

\begin{table*}
\caption{\gls{auc}, \gls{ba}, \gls{fpr}, and \gls{fnr} for the JPEG compression scenario for the \cite{cozzolino2023raising}-A detector. 
}
\centering
\resizebox{\textwidth}{!}{
\begin{tabular}{l S[table-format=2.2]@{\,/\,} S[table-format=1.2]@{\,/\,}S[table-format=1.2]@{\,/\,}S[table-format=1.2] |
    S[table-format=2.2]@{\,/\,}S[table-format=1.2]@{\,/\,}S[table-format=1.2]@{\,/\,}S[table-format=1.2]
    S[table-format=2.2]@{\,/\,}S[table-format=1.2]@{\,/\,}S[table-format=1.2]@{\,/\,}S[table-format=1.2]
    S[table-format=2.2]@{\,/\,}S[table-format=1.2]@{\,/\,}S[table-format=1.2]@{\,/\,}S[table-format=1.2]
    S[table-format=2.2]@{\,/\,}S[table-format=1.2]@{\,/\,}S[table-format=1.2]@{\,/\,}S[table-format=1.2]
}
\toprule
 & \multicolumn{4}{c |}{Original format} & \multicolumn{4}{c}{QF = 65} & \multicolumn{4}{c}{BPP = 75} & 
 \multicolumn{4}{c}{QF = 85} &
 \multicolumn{4}{c}{QF = 95} \\
 \midrule
& \multicolumn{4}{c |}{$\textrm{AUC}$ /
$\textrm{BA}$ / $\textrm{FPR}$ / $\textrm{FNR}$} & \multicolumn{4}{c}{$\textrm{AUC}$ /
$\textrm{BA}$ / $\textrm{FPR}$ / $\textrm{FNR}$} & \multicolumn{4}{c}{$\textrm{AUC}$ /
$\textrm{BA}$ / $\textrm{FPR}$ / $\textrm{FNR}$} &
\multicolumn{4}{c}{$\textrm{AUC}$ /
$\textrm{BA}$ / $\textrm{FPR}$ / $\textrm{FNR}$} &
\multicolumn{4}{c}{$\textrm{AUC}$ /
$\textrm{BA}$ / $\textrm{FPR}$ / $\textrm{FNR}$} \\
\midrule
Imagenet & 0.90 & 0.70 & 0.57 & 0.04 & 0.77 & 0.69 & 0.12 & 0.50 & 0.79 & 0.68 & 0.11 & 0.52 & 0.75 & 0.64 & 0.13 & 0.59 & 0.85 & 0.76 & 0.34 & 0.14 \\
COCO & 0.91 & 0.74 & 0.44 & 0.09 & 0.85 & 0.75 & 0.05 & 0.44 & 0.87 & 0.74 & 0.04 & 0.47 & 0.85 & 0.73 & 0.08 & 0.46 & 0.92 & 0.85 & 0.13 & 0.17 \\
FFHQ & 0.96 & 0.90 & 0.09 & 0.10 & 0.94 & 0.75 & 0.03 & 0.46 & 0.92 & 0.61 & 0.01 & 0.77 & 0.80 & 0.60 & 0.06 & 0.75 & 0.93 & 0.84 & 0.10 & 0.23 \\
LSUN & 0.95 & 0.87 & 0.18 & 0.09 & 0.73 & 0.61 & 0.07 & 0.72 & 0.74 & 0.63 & 0.07 & 0.67 & 0.72 & 0.63 & 0.10 & 0.64 & 0.90 & 0.82 & 0.20 & 0.16 \\
LAION & 0.96 & 0.80 & 0.37 & 0.03 & 0.80 & 0.71 & 0.24 & 0.33 & 0.84 & 0.74 & 0.14 & 0.37 & 0.77 & 0.69 & 0.21 & 0.41 & 0.90 & 0.78 & 0.37 & 0.08 \\
RAISE & 0.80 & 0.65 & 0.11 & 0.60 & 0.72 & 0.56 & 0.05 & 0.82 & 0.71 & 0.56 & 0.06 & 0.82 & 0.64 & 0.56 & 0.10 & 0.78 & 0.75 & 0.60 & 0.11 & 0.68 \\

\bottomrule
\end{tabular}
}
\label{tab:jpeg_metrics_cozzolino2023a}
\vspace{-10pt}
\end{table*}

\begin{table*}
\caption{\gls{auc}, \gls{ba}, \gls{fpr}, and \gls{fnr} for the JPEG compression scenario for the \cite{cozzolino2023raising}-B detector.
}
\centering
\resizebox{\textwidth}{!}{
\begin{tabular}{l S[table-format=2.2]@{\,/\,} S[table-format=1.2]@{\,/\,}S[table-format=1.2]@{\,/\,}S[table-format=1.2] |
    S[table-format=2.2]@{\,/\,}S[table-format=1.2]@{\,/\,}S[table-format=1.2]@{\,/\,}S[table-format=1.2]
    S[table-format=2.2]@{\,/\,}S[table-format=1.2]@{\,/\,}S[table-format=1.2]@{\,/\,}S[table-format=1.2]
    S[table-format=2.2]@{\,/\,}S[table-format=1.2]@{\,/\,}S[table-format=1.2]@{\,/\,}S[table-format=1.2]
    S[table-format=2.2]@{\,/\,}S[table-format=1.2]@{\,/\,}S[table-format=1.2]@{\,/\,}S[table-format=1.2]
}
\toprule
 & \multicolumn{4}{c |}{Original format} & \multicolumn{4}{c}{QF = 65} & \multicolumn{4}{c}{BPP = 75} & 
 \multicolumn{4}{c}{QF = 85} &
 \multicolumn{4}{c}{QF = 95} \\
 \midrule
& \multicolumn{4}{c |}{$\textrm{AUC}$ /
$\textrm{BA}$ / $\textrm{FPR}$ / $\textrm{FNR}$} & \multicolumn{4}{c}{$\textrm{AUC}$ /
$\textrm{BA}$ / $\textrm{FPR}$ / $\textrm{FNR}$} & \multicolumn{4}{c}{$\textrm{AUC}$ /
$\textrm{BA}$ / $\textrm{FPR}$ / $\textrm{FNR}$} &
\multicolumn{4}{c}{$\textrm{AUC}$ /
$\textrm{BA}$ / $\textrm{FPR}$ / $\textrm{FNR}$} &
\multicolumn{4}{c}{$\textrm{AUC}$ /
$\textrm{BA}$ / $\textrm{FPR}$ / $\textrm{FNR}$} \\
\midrule
Imagenet & 0.76 & 0.68 & 0.42 & 0.21 & 0.66 & 0.62 & 0.40 & 0.36 & 0.71 & 0.66 & 0.36 & 0.33 & 0.64 & 0.60 & 0.32 & 0.48 & 0.69 & 0.64 & 0.52 & 0.20 \\
COCO & 0.86 & 0.77 & 0.32 & 0.14 & 0.81 & 0.73 & 0.33 & 0.22 & 0.86 & 0.77 & 0.19 & 0.26 & 0.82 & 0.73 & 0.17 & 0.37 & 0.83 & 0.75 & 0.30 & 0.20 \\
FFHQ & 0.88 & 0.77 & 0.10 & 0.36 & 0.87 & 0.72 & 0.08 & 0.47 & 0.77 & 0.65 & 0.18 & 0.52 & 0.63 & 0.54 & 0.14 & 0.77 & 0.83 & 0.67 & 0.11 & 0.54 \\
LSUN & 0.79 & 0.68 & 0.49 & 0.14 & 0.70 & 0.64 & 0.36 & 0.35 & 0.70 & 0.65 & 0.39 & 0.30 & 0.67 & 0.62 & 0.38 & 0.37 & 0.80 & 0.70 & 0.48 & 0.12 \\
RAISE & 0.86 & 0.77 & 0.15 & 0.31 & 0.78 & 0.70 & 0.22 & 0.37 & 0.76 & 0.69 & 0.21 & 0.40 & 0.74 & 0.68 & 0.22 & 0.41 & 0.84 & 0.75 & 0.20 & 0.31 \\

\bottomrule
\end{tabular}
}
\label{tab:jpeg_metrics_cozzolino2023b}
\vspace{-10pt}
\end{table*}

\begin{table*}
\caption{\gls{auc}, \gls{ba}, \gls{fpr}, and \gls{fnr} for the JPEG compression scenario for the \cite{Mandelli2024} detector. 
}
\centering
\resizebox{\textwidth}{!}{
\begin{tabular}{l S[table-format=2.2]@{\,/\,} S[table-format=1.2]@{\,/\,}S[table-format=1.2]@{\,/\,}S[table-format=1.2] |
    S[table-format=2.2]@{\,/\,}S[table-format=1.2]@{\,/\,}S[table-format=1.2]@{\,/\,}S[table-format=1.2]
    S[table-format=2.2]@{\,/\,}S[table-format=1.2]@{\,/\,}S[table-format=1.2]@{\,/\,}S[table-format=1.2]
    S[table-format=2.2]@{\,/\,}S[table-format=1.2]@{\,/\,}S[table-format=1.2]@{\,/\,}S[table-format=1.2]
    S[table-format=2.2]@{\,/\,}S[table-format=1.2]@{\,/\,}S[table-format=1.2]@{\,/\,}S[table-format=1.2]
}
\toprule
 & \multicolumn{4}{c |}{Original format} & \multicolumn{4}{c}{QF = 65} & \multicolumn{4}{c}{BPP = 75} & 
 \multicolumn{4}{c}{QF = 85} &
 \multicolumn{4}{c}{QF = 95} \\
 \midrule
& \multicolumn{4}{c |}{$\textrm{AUC}$ /
$\textrm{BA}$ / $\textrm{FPR}$ / $\textrm{FNR}$} & \multicolumn{4}{c}{$\textrm{AUC}$ /
$\textrm{BA}$ / $\textrm{FPR}$ / $\textrm{FNR}$} & \multicolumn{4}{c}{$\textrm{AUC}$ /
$\textrm{BA}$ / $\textrm{FPR}$ / $\textrm{FNR}$} &
\multicolumn{4}{c}{$\textrm{AUC}$ /
$\textrm{BA}$ / $\textrm{FPR}$ / $\textrm{FNR}$} &
\multicolumn{4}{c}{$\textrm{AUC}$ /
$\textrm{BA}$ / $\textrm{FPR}$ / $\textrm{FNR}$} \\
\midrule
CelebA & 1.00 & 1.00 & 0.00 & 0.00 & 1.00 & 1.00 & 0.01 & 0.00 & 1.00 & 1.00 & 0.00 & 0.00 & 1.00 & 1.00 & 0.00 & 0.00 & 1.00 & 1.00 & 0.00 & 0.00 \\
FFHQ & 1.00 & 1.00 & 0.00 & 0.00 & 1.00 & 1.00 & 0.00 & 0.01 & 1.00 & 1.00 & 0.00 & 0.00 & 1.00 & 1.00 & 0.00 & 0.00 & 1.00 & 1.00 & 0.00 & 0.00 \\
Imagenet & 0.85 & 0.78 & 0.25 & 0.20 & 0.58 & 0.56 & 0.28 & 0.60 & 0.60 & 0.56 & 0.30 & 0.58 & 0.62 & 0.59 & 0.32 & 0.50 & 0.69 & 0.64 & 0.36 & 0.35 \\
LAION & 0.85 & 0.75 & 0.32 & 0.17 & 0.74 & 0.66 & 0.25 & 0.42 & 0.73 & 0.66 & 0.26 & 0.41 & 0.73 & 0.66 & 0.30 & 0.38 & 0.74 & 0.68 & 0.33 & 0.31 \\
LSUN & 0.90 & 0.81 & 0.23 & 0.15 & 0.65 & 0.61 & 0.24 & 0.54 & 0.66 & 0.61 & 0.29 & 0.49 & 0.69 & 0.64 & 0.32 & 0.41 & 0.74 & 0.68 & 0.32 & 0.32 \\
RAISE & 0.80 & 0.77 & 0.15 & 0.32 & 0.83 & 0.76 & 0.03 & 0.46 & 0.85 & 0.78 & 0.03 & 0.41 & 0.83 & 0.81 & 0.04 & 0.34 & 0.81 & 0.82 & 0.04 & 0.31 \\

\bottomrule
\end{tabular}
}
\label{tab:jpeg_metrics_mandelli2024}
\vspace{-10pt}
\end{table*}

\begin{table*}
\caption{\gls{auc}, \gls{ba}, \gls{fpr}, and \gls{fnr} for the JPEG compression scenario for the \cite{Tan_2024_CVPR} detector. 
}
\centering
\resizebox{\textwidth}{!}{
\begin{tabular}{l S[table-format=2.2]@{\,/\,} S[table-format=1.2]@{\,/\,}S[table-format=1.2]@{\,/\,}S[table-format=1.2] |
    S[table-format=2.2]@{\,/\,}S[table-format=1.2]@{\,/\,}S[table-format=1.2]@{\,/\,}S[table-format=1.2]
    S[table-format=2.2]@{\,/\,}S[table-format=1.2]@{\,/\,}S[table-format=1.2]@{\,/\,}S[table-format=1.2]
    S[table-format=2.2]@{\,/\,}S[table-format=1.2]@{\,/\,}S[table-format=1.2]@{\,/\,}S[table-format=1.2]
    S[table-format=2.2]@{\,/\,}S[table-format=1.2]@{\,/\,}S[table-format=1.2]@{\,/\,}S[table-format=1.2]
}
\toprule
 & \multicolumn{4}{c |}{Original format} & \multicolumn{4}{c}{QF = 65} & \multicolumn{4}{c}{BPP = 75} & 
 \multicolumn{4}{c}{QF = 85} &
 \multicolumn{4}{c}{QF = 95} \\
 \midrule
& \multicolumn{4}{c |}{$\textrm{AUC}$ /
$\textrm{BA}$ / $\textrm{FPR}$ / $\textrm{FNR}$} & \multicolumn{4}{c}{$\textrm{AUC}$ /
$\textrm{BA}$ / $\textrm{FPR}$ / $\textrm{FNR}$} & \multicolumn{4}{c}{$\textrm{AUC}$ /
$\textrm{BA}$ / $\textrm{FPR}$ / $\textrm{FNR}$} &
\multicolumn{4}{c}{$\textrm{AUC}$ /
$\textrm{BA}$ / $\textrm{FPR}$ / $\textrm{FNR}$} &
\multicolumn{4}{c}{$\textrm{AUC}$ /
$\textrm{BA}$ / $\textrm{FPR}$ / $\textrm{FNR}$} \\
\midrule
Imagenet & 0.94 & 0.84 & 0.30 & 0.01 & 0.52 & 0.50 & 0.01 & 1.00 & 0.52 & 0.50 & 0.01 & 1.00 & 0.51 & 0.50 & 0.01 & 0.98 & 0.51 & 0.53 & 0.08 & 0.86 \\
COCO & 0.91 & 0.82 & 0.34 & 0.01 & 0.42 & 0.50 & 0.00 & 1.00 & 0.42 & 0.50 & 0.00 & 1.00 & 0.42 & 0.50 & 0.00 & 1.00 & 0.42 & 0.49 & 0.05 & 0.96 \\
CelebA & 1.00 & 0.99 & 0.01 & 0.00 & 0.34 & 0.50 & 0.00 & 1.00 & 0.31 & 0.50 & 0.00 & 1.00 & 0.32 & 0.50 & 0.00 & 1.00 & 0.31 & 0.50 & 0.00 & 1.00 \\
LSUN & 1.00 & 0.99 & 0.00 & 0.03 & 0.47 & 0.50 & 0.00 & 1.00 & 0.48 & 0.50 & 0.00 & 1.00 & 0.49 & 0.50 & 0.00 & 1.00 & 0.53 & 0.51 & 0.01 & 0.98 \\
LAION & 0.99 & 0.97 & 0.02 & 0.05 & 0.30 & 0.50 & 0.00 & 0.99 & 0.31 & 0.50 & 0.00 & 0.99 & 0.33 & 0.51 & 0.00 & 0.99 & 0.34 & 0.51 & 0.02 & 0.97 \\

\bottomrule
\end{tabular}
}
\label{tab:jpeg_metrics_npr}
\vspace{-10pt}
\end{table*}

\section{Image splicing localization additional results}
\label{sec:supp:splicing}
\begin{figure}[htbp]
\centering
    \includegraphics[width=\columnwidth]{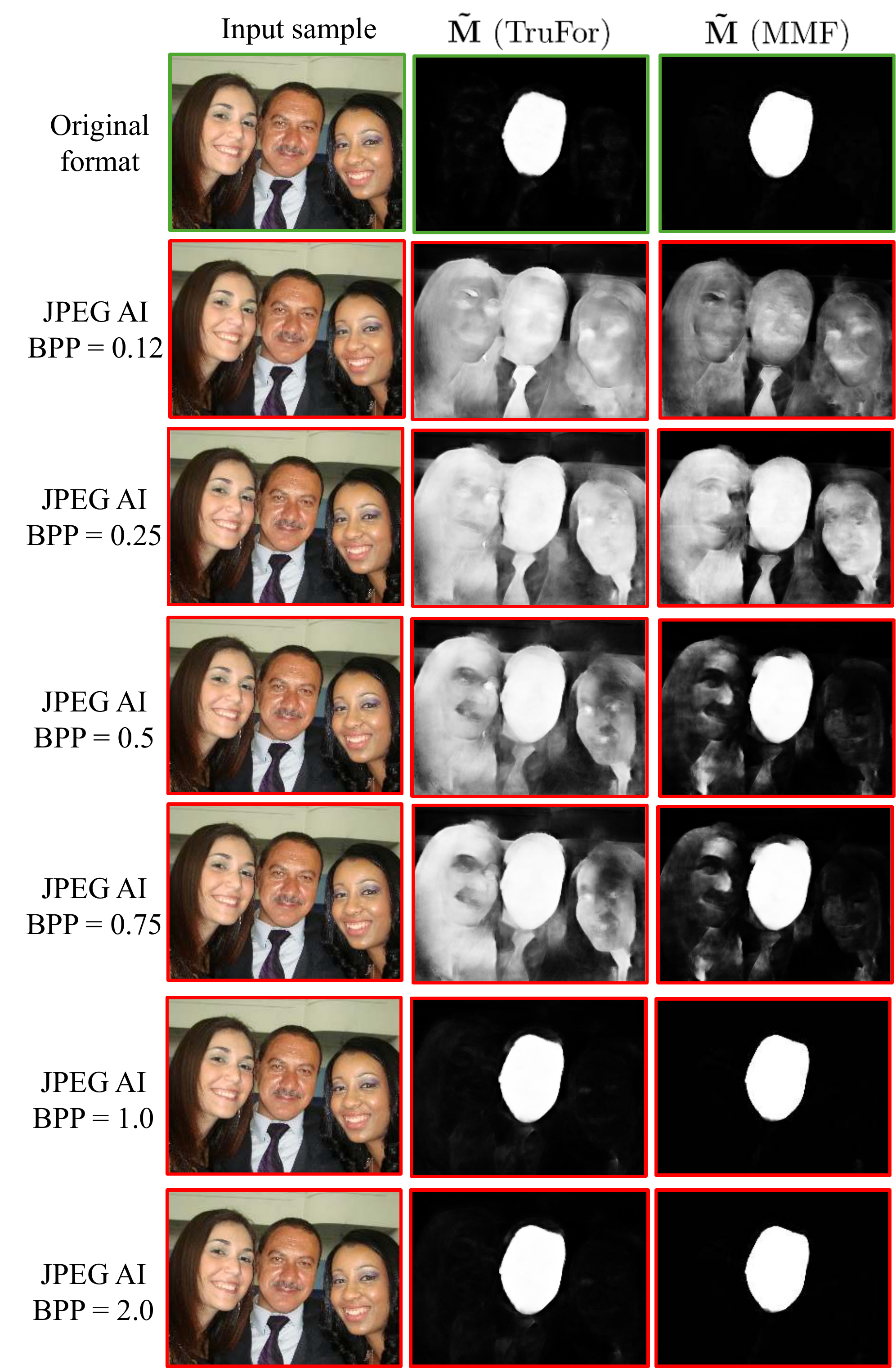}
    \caption{Output of the considered detectors for the same DS01 input under different JPEG AI compression settings.
    }
\label{fig:supp:splicing_1}
\vspace{-15pt}
\end{figure}

\begin{figure}
\centering
    \includegraphics[width=\columnwidth]{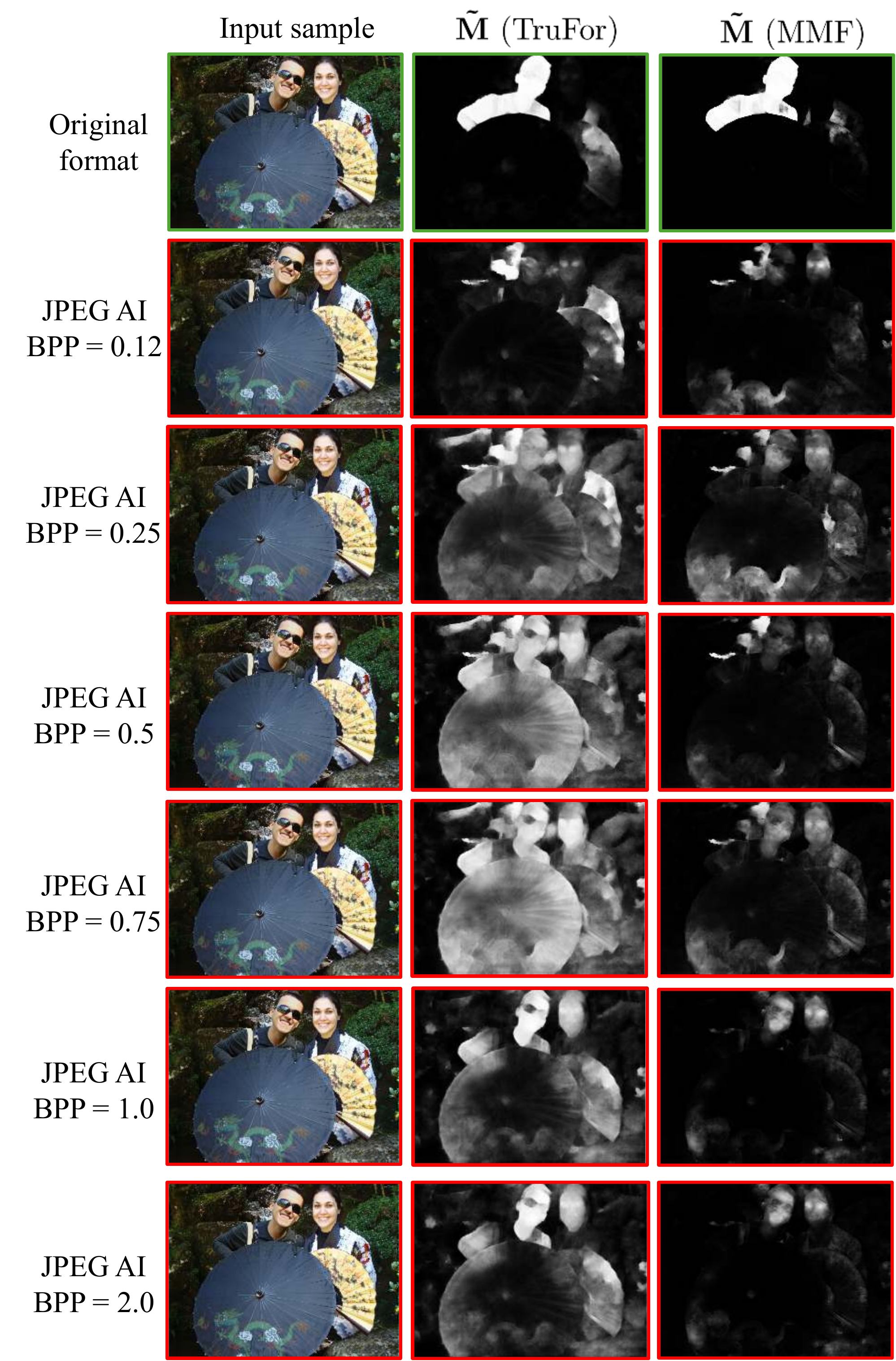}
    \caption{Output of the considered detectors for the same DS01 input under different JPEG AI compression settings.
    }
\label{fig:supp:splicing_2}
\vspace{-15pt}
\end{figure}

\begin{figure}
\centering
    \includegraphics[width=\columnwidth]{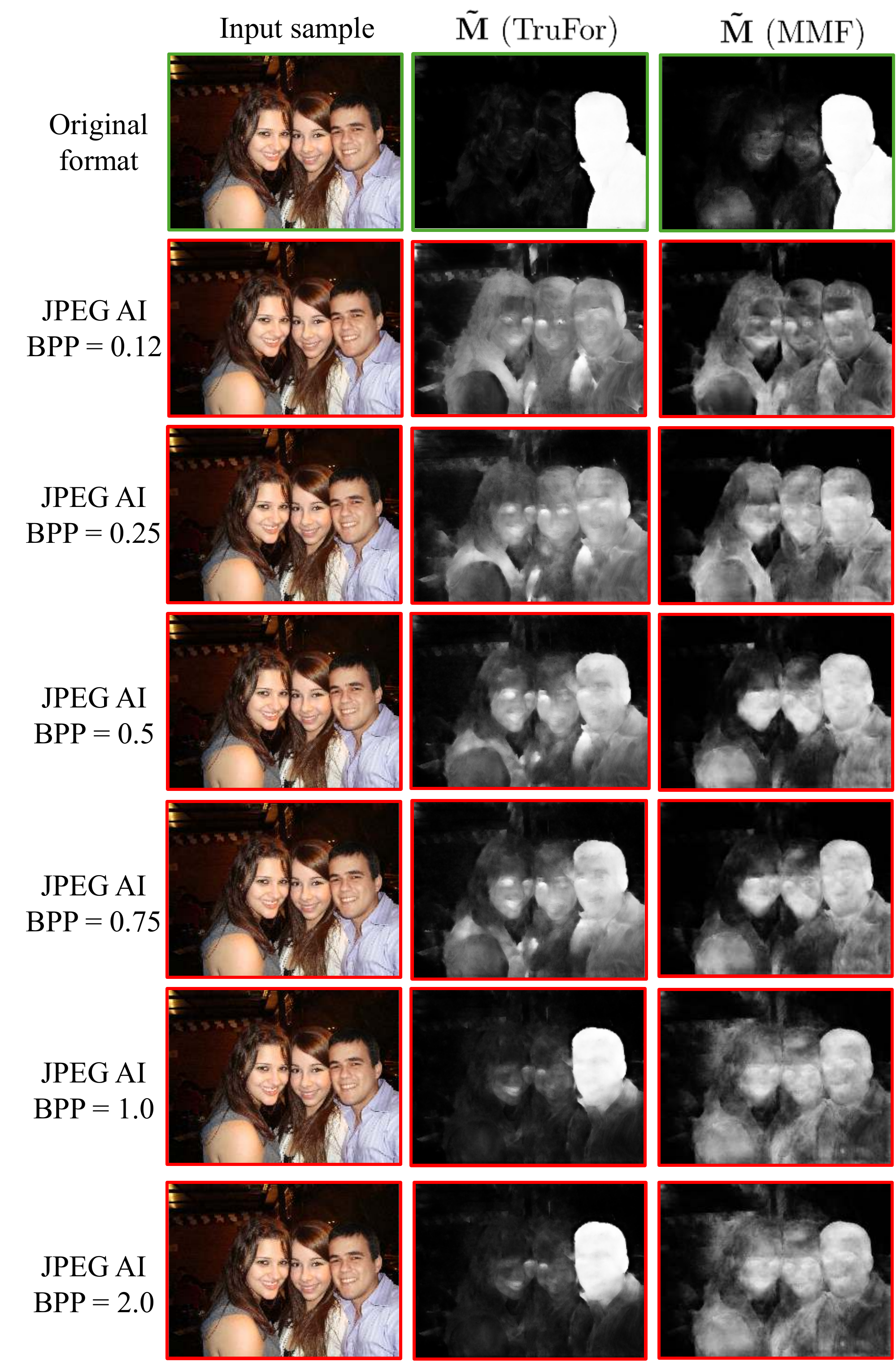}
    \caption{Output of the considered detectors for the same DS01 input under different JPEG AI compression settings.
    }
\label{fig:supp:splicing_3}
\vspace{-15pt}
\end{figure}

\begin{figure}
\centering
    \includegraphics[width=\columnwidth]{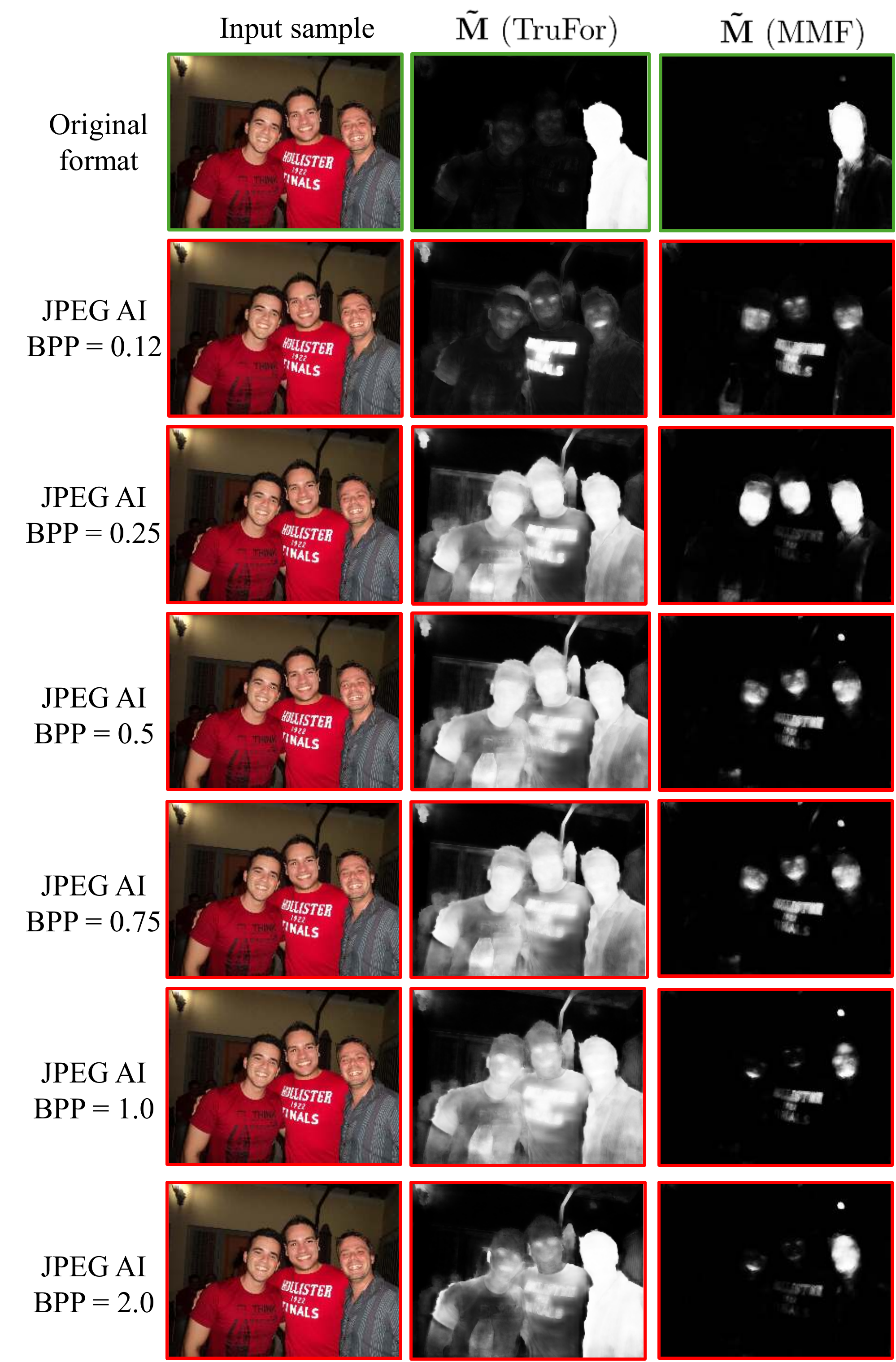}
    \caption{Output of the considered detectors for the same DS01 input under different JPEG AI compression settings.
    }
\label{fig:supp:splicing_4}
\vspace{-15pt}
\end{figure}

\cref{fig:supp:splicing_1} to \cref{fig:supp:splicing_4} report some examples of output maps $\mathbf{\Tilde{M}}$ for the TruFor and MMFusion detectors. In all plots, the detectors process the same DSO1 input image compressed with different JPEG AI \gls{bpp}. 

As we reported in \cref{sec:results} of the main paper, decreasing the JPEG AI compression ratio degrades the quality of the maps $\mathbf{\tilde{M}}$. All detectors do not correctly identify all the manipulated pixels in the tampered with area, while flagging many pristine pixels as manipulated. Whenever the \gls{bpp} values go over $1.0$, the output maps tend to match more closely the ones obtained when processing samples in their original format. This is the case, for instance, of~\cref{fig:supp:splicing_1}. 

However, it is interesting to notice that, for certain samples, increasing the \gls{bpp} during compression does not lead to a better $\mathbf{\tilde{M}}$. For example, the output of MMFusion in \cref{fig:supp:splicing_3} maintains a high number of false positives even for $\textrm{BPP}=1.0$ and $\textrm{BPP}=2.0$, and TruFor shows a similar behavior in \cref{fig:supp:splicing_4}. Therefore, while the overall tendency highlighted in \cref{tab:splicing} is clear, these examples seem to suggest that there is no guarantee that compressing images with high \gls{bpp} values will not interfere with the responses of the detectors. 

\section{Double JPEG AI compression} 
The study field relative to multiple JPEG compression is lively~\cite{Piva2013}. Therefore, in the following, we briefly analyze some of the implications of double JPEG AI compression. Please notice that this analysis is not exhaustive, and we plan to explore the more profound implications of double JPEG AI in future studies.

\begin{figure}[htb!]
	\centering
    
        \subfloat[\centering Single JPEG AI compression.]{
            \includegraphics[width=\columnwidth]{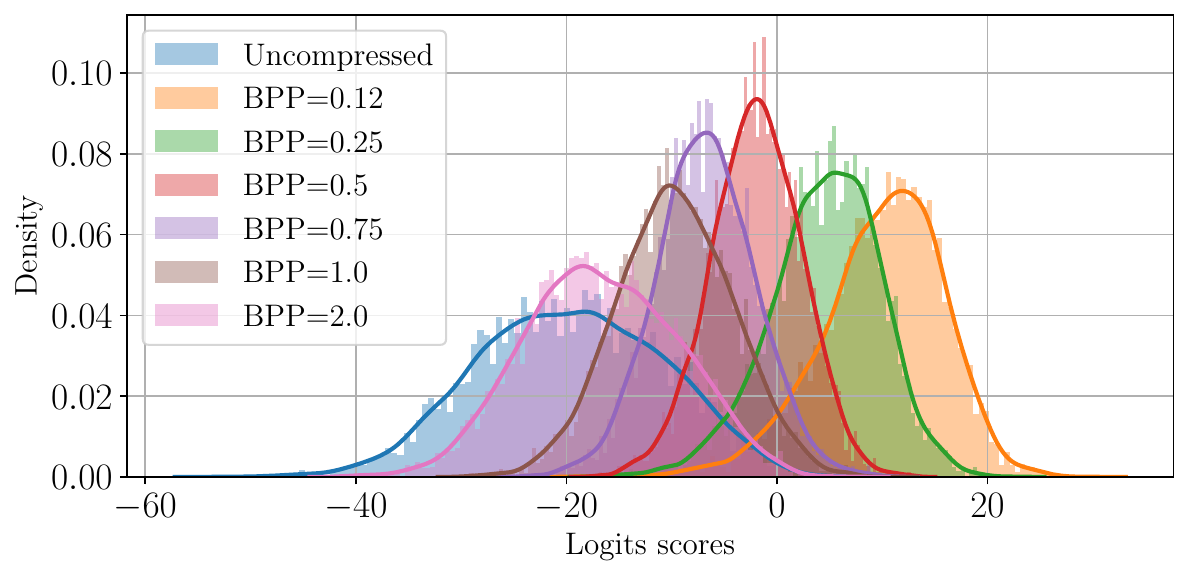}		
            \label{fig:supp:grag2021:djpeai:single}}
        
        \subfloat[\centering Double JPEG AI compression, starting \gls{bpp}$=0.12$.]{
            \includegraphics[width=\columnwidth]{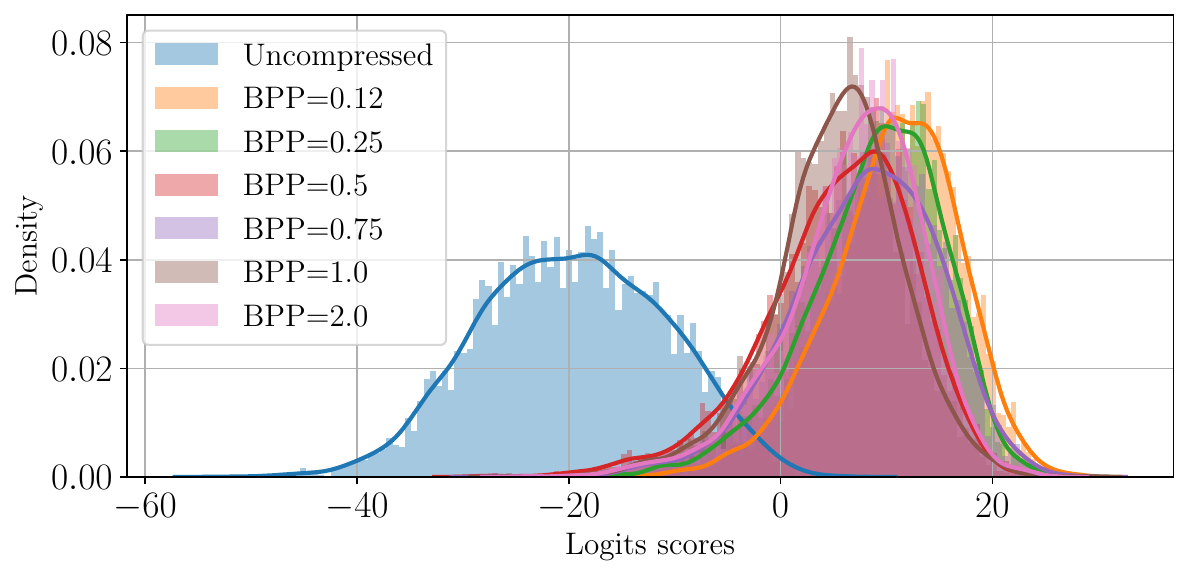}
            \label{fig:supp:grag2021:djpegai:source12}}
        
        \subfloat[\centering Double JPEG AI compression, starting \gls{bpp}$=2.0$.]{
            \includegraphics[width=\columnwidth]{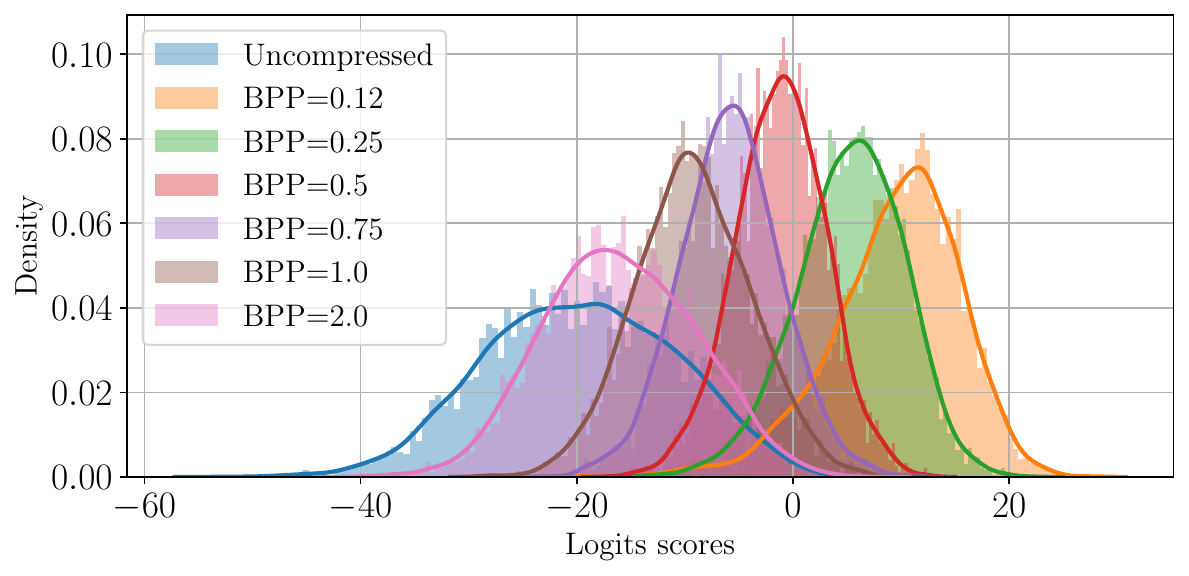}		\label{fig:supp:grag2021:djpegai:source20}}

        \caption{Scores distribution over COCO pristine samples of the \cite{gragnaniello2021} detector considering different JPEG AI compression stages.}
	\label{fig:supp:grag2021:djpegai}
\end{figure}

\noindent \textbf{Deepfake image detection. }\cref{fig:supp:grag2021:djpegai} reports the scores of the \cite{gragnaniello2021} detector extracted from pristine images compressed in the scenarios illustrated in \cref{sec:results:djpegai}, namely i) single JPEG AI compression, ii) double JPEG AI compression where the original-format images have been first compressed with the lowest \gls{bpp} considered, i.e., \gls{bpp}$ =0.12$, and then re-compressed at various \gls{bpp} values, and iii) the original-format images have been first compressed with the highest \gls{bpp} considered, i.e., \gls{bpp}$ =2.0$, and then re-compressed at various \gls{bpp} values. 

\cref{fig:supp:grag2021:djpegai:source12} reports the results for the first scenario. As we can quickly inspect, regardless of the \gls{bpp} value used in the second compression, the scores distribution of pristine samples will lie above the $0$ threshold in the area corresponding to the scores distribution of single compressed $0.12$ \gls{bpp} images. On the contrary, \cref{fig:supp:grag2021:djpegai:source20} shows that if we first compress the samples with a high \gls{bpp} value, the distribution of the scores will move accordingly to the \gls{bpp} value of the second compression operation, i.e., a $0.12$ \gls{bpp} after a $2.0$ \gls{bpp} compression will shift the scores in the same area as a single $0.12$ \gls{bpp} compression. 
These results suggest that JPEG AI behaves similarly to standard compression formats in the sense that, if a strong compression has been applied to the input samples, the compressed images will likely still bear forensic artifacts relative to the first compression stage regardless of the \gls{bpp} used in the second one.

\noindent \textbf{Image splicing localization. }In image splicing localization, the double compression usually does not interest the whole image. Indeed, it is rather common that only the spliced area indicated by the tampering mask $\mathbf{M}$ presents artifacts relative to a previous compression operation. This is because the source and target samples might have been saved with different formats or the same coding scheme but with one of the two images being processed with a more aggressive configuration.

We replicated this scenario by taking a spliced sample from an uncompressed dataset, i.e., DSO-1, and compressing this sample with all the \gls{bpp} values considered in our experiments. After that, we created several spliced images where the spliced area was JPEG AI compressed at various \gls{bpp}, and the surrounding content was uncompressed. We finally re-compressed all these images with \gls{bpp} equal to $0.12$ and $2.0$. In this way, we are creating a mismatch in the spatial distribution of compression traces due to the spliced area bearing traces of the first JPEG AI compression, similar to the scenario analyzed in double JPEG compression.

\begin{figure}
\centering
    \includegraphics[width=\columnwidth]{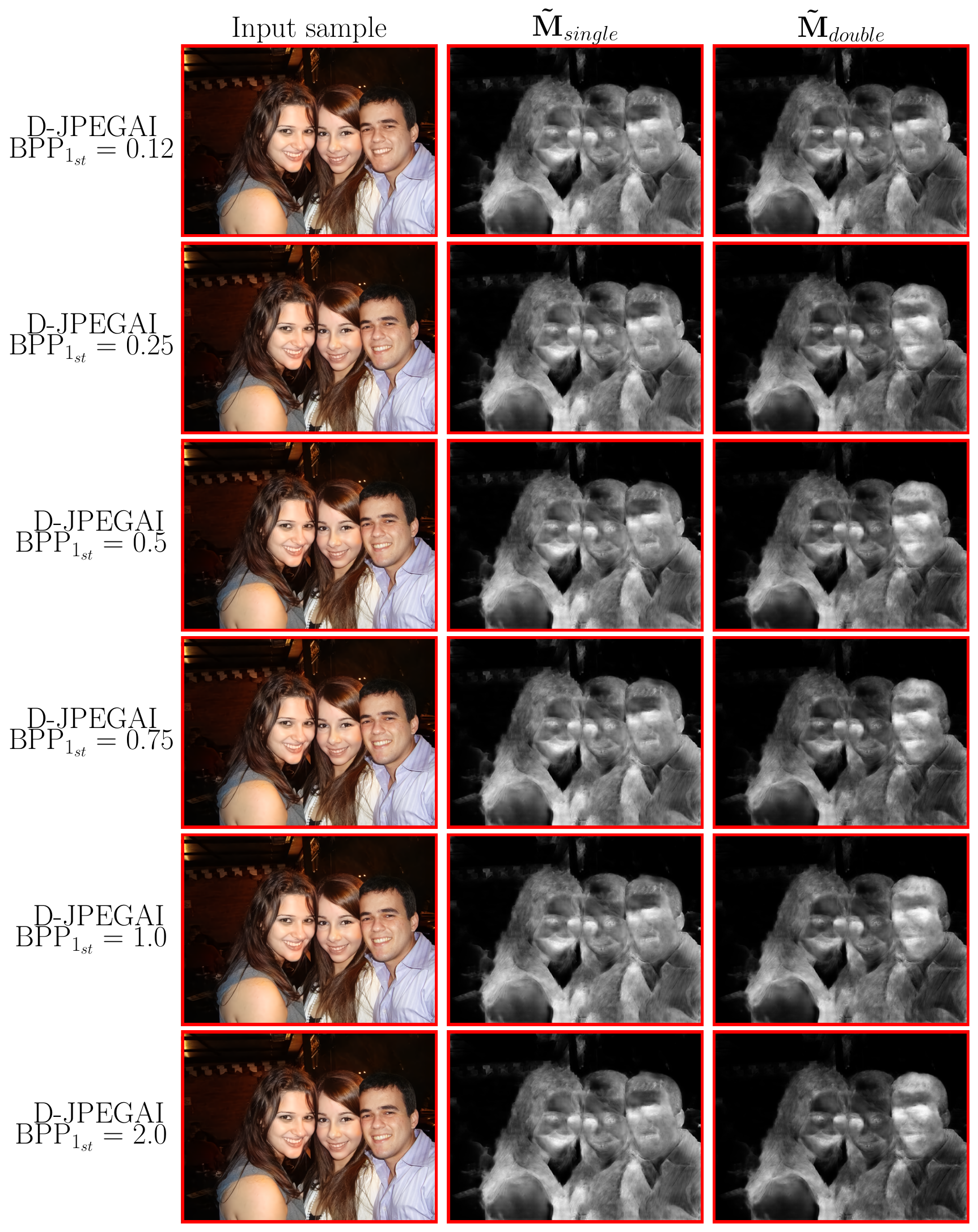}
    \caption{Output of MMFusion for DSO-1 double JPEG AI compressed images. The second compression stage is executed with \gls{bpp} equal to $0.12$. The first compression stage \gls{bpp} value varies from $0.12$ to $2.0$ from top to bottom.
    }
\label{fig:supp:djpegai012}
\vspace{-15pt}
\end{figure}

\begin{figure}
\centering
    \includegraphics[width=\columnwidth]{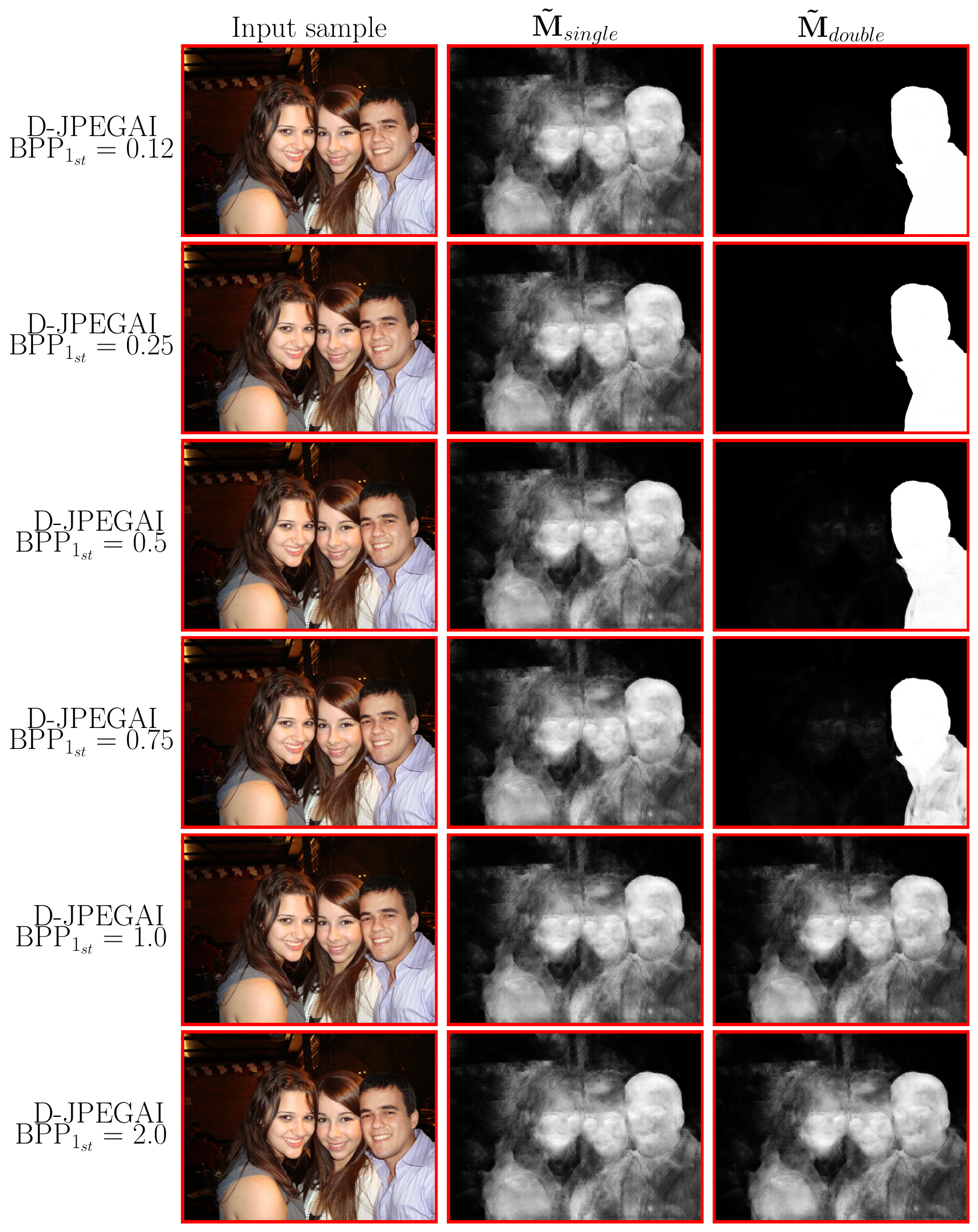}
    \caption{Output of MMFusion for DSO-1 double JPEG AI compressed images. The second compression stage is executed with \gls{bpp} equal to $2.0$. The first compression stage \gls{bpp} value varies from $0.12$ to $2.0$ from top to bottom.
    }
\label{fig:supp:djpegai20}
\vspace{-15pt}
\end{figure}

\cref{fig:supp:djpegai012} and \cref{fig:supp:djpegai20} report the results of our double JPEG AI splicing operation together with the mask estimated by MMFusion. We also include the results of single-compressing the original format spliced image for reference. As we can quickly inspect, we can draw similar conclusions to the deepfake image detection experiments. In \cref{fig:supp:djpegai20}, the images were saved with \gls{bpp}$=2.0$, i.e., with very high quality. In this scenario, if the spliced area has been initially compressed with low \gls{bpp}, MMFusion can detect this inconsistency and localize the spliced area with results much better than the reference single $2.0$ \gls{bpp} JPEG AI compression. On the other hand, in \cref{fig:supp:djpegai012}, the second compression stage at \gls{bpp}$=0.12$ likely erases all the traces relative to the first, hence making MMFusion deliver results very similar to a single JPEG AI compression.
These results suggest that the inconsistency in the spatial distribution of JPEG AI artifacts might become a very powerful forensic instrument.

\end{document}